\documentclass[reprint,nofootinbib]{revtex4}
\usepackage{graphicx}  % needed for figures
\usepackage{dcolumn}   % needed for some tables
\usepackage{bm}        % for math
\usepackage{amssymb}
\usepackage{hyperref}
\usepackage{multirow}
\usepackage{amsmath}
\usepackage{cases}
\usepackage{placeins}

\begin{document}
\title{Angular analysis of $B_s\, \to\, f_{2}'\,(1525)\,(\to K^+\,K^-)\,\mu^+ \,\mu^-$ decays as a probe to
lepton flavor universality violation}
\author{N~Rajeev${}^{1}$}
\email{rajeev\_rs@phy.nits.ac.in } 
\author{Niladribihari~Sahoo${}^{2}$}
\email{niladri.sahoo@warwick.ac.uk}
\author{Rupak~Dutta${}^{1}$}
\email{rupak@phy.nits.ac.in}
\affiliation{
\vspace{0.5cm}${}^1$National Institute of Technology Silchar, Silchar 788010, India\\ 
\vspace{0.5cm}${}^2$Department of Physics, University of Warwick, Coventry CV4 7AL, United Kingdom
}

\begin{abstract}
The flavor anomalies reported in $R_K$, $R_{K^*}$, $P_{5}^{\prime}$ and $\mathcal{B}(B_s\, \to\, \phi\, \mu^+\, \mu^-)$
indicate lepton flavor universality violation in $b \to s\,l^+\,l^-$ quark level transition decays. 
The deviation from the SM prediction reported in the underlying flavor observables currently 
stands at the level of $2.5\sigma$, $2.4\sigma$, $3.3\sigma$ and $3.7\sigma$, respectively.
In this context, we perform an angular analysis of the four-body differential 
decay of $B_s\, \to\, f_{2}^{\prime}\,(1525)\,(\to K^+\,K^-)\,\mu^+ \,\mu^-$  
in a model independent effective field theory framework.
The decay mode $B_s\, \to\, f_{2}^{\prime}(1525)\,l^+\,l^-$ undergoes similar $b \to s$ 
neutral current quark level transition and, in principle, can provide complementary information regarding lepton flavor universality 
violation in $b \to s\,l^+\,l^-$ quark level transition decays.
We give predictions of various physical 
observables such as the branching ratio,
the longitudinal polarization fraction, the forward-backward asymmetry, the angular observables
$P_1$, $P_2$, $P_{4}^{\prime}$, $P_{5}^{\prime}$ and also the lepton flavor sensitive observables such as
the ratio of branching ratio $R_{f_{2}^{\prime}}$, $Q_{F_L}$, $Q_{A_{FB}}$, $Q_{1}$, $Q_{2}$, $Q_{4}^{\prime}$, $Q_{5}^{\prime}$ for 
the $B_s\, \to\, f_{2}^{\prime}\,(1525)\,(\to K^+\,K^-)\,\mu^+ \,\mu^-$ decay mode in the standard model and in the presence of 
several 1D and 2D new physics scenarios. 
\end{abstract}

\maketitle

\section{Introduction}
\label{intro}
Exploring and identifying the Lorentz structure of possible new physics (NP) 
that lies beyond the standard model (SM) is of great importance particularly in semileptonic $B$ meson decays mediated via
$b \to s\,l^+\,l^-$ neutral current and $b \to c\,l\,\nu$ charged current interactions.
It is well known that the flavor sector could be an ideal platform to explore 
NP since it can provide possible indirect evidence of NP in the form of new interactions that
can, in principle, be very sensitive to the existing experiments. 
It is also well known that, apart from the flavor sector, existence of NP is also evident from several other phenomena such as  
the matter antimatter asymmetry of the universe, neutrino mass, dark matter, dark energy and so on.
In the recent years, several measurements have shown hints of lepton flavor universality violation~(LFUV)
in the semileptonic decays of $B$ mesons involving $b\, \to\, s\, l^+\,l^-$ ($l \in e,\mu$) neutral current and 
$b\, \to\, c\,l\,\nu$ ($l\in e/\mu,\tau$) charged current quark level transitions. 
Significant deviation from the SM expectation has been reported in various flavor observables such as $R_K$, $R_{K^*}$, 
$P_{5}^{\prime}$ in $B\, \to\, K^{(*)}\, l^+\,l^-$ decays; $\mathcal{B}(B_s\, \to\, \phi\, \mu^+\, \mu^-$);
$R_D$, $R_{D^*}$, $P^{\tau}_{D^*}$, $F_{L}^{D^*}$ in $B\, \to\, D^{(*)}\, l\,\nu$ decays and $R_{J/\Psi}$ in 
$B_c\, \to\, J/\Psi\, l\,\nu$ decays. Here we will focus mainly on the anomalies present in $B$ meson decays mediated via
$b \to s\,l^+\,l^-$ quark level transitions. 
The ratio of branching ratio $R_K$ and $R_{K^*}$ in $B\, \to\, (K\,,K^*)\, l^+\,l^-$ decays are defined as
\begin{equation}
 R_{K^{(*)}} = \frac{\mathcal{B}(B\, \to\, K^{(*)}\, \mu^+\,\mu^-)}{\mathcal{B}(B\, \to\, K^{(*)}\, e^+\,e^-)}\,.
\end{equation}
After the Rencontres  de  Moriond,  2019, the current status of several observables pertaining to $b \to s\,l^+\,l^-$ quark level
transition decays is as follows:
the measurement of $R_K$ from the combined data of both Run 1 and Run 2 of LHCb reports $R_K=0.846^{+0.060}_{-0.054}$ (stat)
$^{+0.016}_{-0.014}$ (syst)~\cite{Aaij:2019wad} in the central $q^2$ region~($1 \le q^2 \le 6\,{\rm GeV^2}$), 
where $q^2$ is the invariant mass-squared of the dilepton. The deviation from the SM value of 
$R_K \sim 1$~\cite{Bordone:2016gaq,Hiller:2003js} is observed to be at the level of $\sim2.5\sigma$. 
Similarly, $R_{K^*}$ was measured in two different $q^2$ bins by two different experiments: in the 
low $q^2$ bin~($0.045  \le q^2 \le 1.1\,{\rm GeV^2}$), LHCb reports
$R_{K^*}=0.660^{+0.110}_{-0.070}$ (stat) $\pm 0.024$ (syst)~\cite{Aaij:2017vbb,Aaij:2020nrf} and Belle reports 
$R_{K^*}=0.52^{+0.36}_{-0.26}$ (stat) $\pm 0.05$ (syst)~\cite{Abdesselam:2019wac}
and in the central $q^2$ bin~($1.1 \le q^2 \le 6\,{\rm GeV^2}$), LHCb 
reports $R_{K^*}=0.685^{+0.113}_{-0.069}$ (stat) $\pm 0.047$ (syst)~\cite{Aaij:2017vbb,Aaij:2020nrf} and Belle reports
$R_{K^*}=0.96^{+0.45}_{-0.29}$ (stat) $\pm 0.11$ (syst)~\cite{Abdesselam:2019wac}. These measurements differ from the SM prediction of 
$R_{K^*}\sim1$
~\cite{Bordone:2016gaq,Hiller:2003js} at the level of $\sim2.4\sigma$.
In addition to $R_K$ and $R_{K^*}$, deviation from the SM expectation is also observed in the measurements of the 
angular distributions of $B\, \to\, K^{*}\, \mu^+\,\mu^-$, particularly in $P_{5}^{\prime}$~\cite{DescotesGenon:2012zf,Descotes-Genon:2013vna}.
The ATLAS~\cite{Aaboud:2018krd} and LHCb~\cite{Aaij:2013qta,Aaij:2015oid} collaborations measured 
$P_{5}^{\prime}$ in the bin $q^2 \in [4,6]$ GeV$^2$ and they differ by $\sim 3.3\sigma$~\cite{Aebischer:2018iyb} from the SM expectation
\cite{Descotes-Genon:2013vna}. Similarly,
the CMS~\cite{CMS Collaboration} measurement in $q^2 \in [4.3,6]$ GeV$^2$ and the Belle~\cite{Abdesselam:2016llu} measurement in $q^2 \in [4.3,8]$ GeV$^2$ differ by $1\sigma$ 
and $2.1\sigma$, respectively from the SM expectations~\cite{DescotesGenon:2012zf,Descotes-Genon:2014uoa}. Moreover, the measured value
of the branching ratio $\mathcal{B}(B_s\, \to\, \phi\, \mu^+\, \mu^-)$~\cite{Aaij:2013aln,Aaij:2015esa} is found to deviate at the level of 
$\sim 3.7\sigma$ from the
SM expectations~\cite{Aebischer:2018iyb,Straub:2015ica}. In Table~\ref{status} we report the current status of $R_K$, $R_{K^*}$, 
$P_{5}^{\prime}$ and $\mathcal{B}(B_s\, \to\, \phi\, \mu^+\, \mu^-)$.
At present, the dedicated ongoing $B$ factory programs at Belle II 
and LHCb emerge as promising platforms that can either confirm or refute the existence of NP in $b \to s\,l^+\,l^-$ transition decays.

\begin{table}[htbp]
\centering
\setlength{\tabcolsep}{5pt} % Default value: 6pt
\renewcommand{\arraystretch}{1.5} % Default value: 1
\begin{tabular}{|c|c|c|c|c|}
\hline
       & $q^2$ bins & Theoretical predictions & Experimental measurements & Deviation\\
\hline
\hline
$R_K$  & [1.0, 6.0] & $1 \pm 0.01$~\cite{Bordone:2016gaq,Hiller:2003js} & $0.846^{+0.060}_{-0.054}$ (stat) $^{+0.016}_{-0.014}$ (syst)~\cite{Aaij:2019wad} & $\sim2.5\sigma$ \\
\hline
\multirow{4}{*}{$R_{K^*}$} & \multirow{2}{*}{[0.045, 1.1]} & $1 \pm 0.01$~\cite{Bordone:2016gaq,Hiller:2003js} & $0.660^{+0.110}_{-0.070}$ (stat) $\pm 0.024$ (syst)~\cite{Aaij:2017vbb,Aaij:2020nrf} & \multirow{4}{*}{$\sim2.4\sigma$}\\ 
                                                        &  & $1 \pm 0.01$~\cite{Bordone:2016gaq,Hiller:2003js} & $0.52^{+0.36}_{-0.26}$ (stat) $\pm 0.05$ (syst)~\cite{Abdesselam:2019wac} & \\ \cline{2-4}
                           & \multirow{2}{*}{[1.1, 6.0]}   & $1 \pm 0.01$~\cite{Bordone:2016gaq,Hiller:2003js} & $0.685^{+0.113}_{-0.069}$ (stat) $\pm 0.047$ (syst)~\cite{Aaij:2017vbb,Aaij:2020nrf} & \\
                                                       &   & $1 \pm 0.01$~\cite{Bordone:2016gaq,Hiller:2003js} & $0.96^{+0.45}_{-0.29}$ (stat) $\pm 0.11$ (syst)~\cite{Abdesselam:2019wac} & \\
\hline  
\multirow{3}{*}{$P_{5}^{\prime}$} & \multirow{1}{*}{[4.0, 6.0]} & $-0.757 \pm 0.074$~\cite{Descotes-Genon:2013vna} & $-0.21 \pm 0.15$~\cite{Aaboud:2018krd,Aaij:2013qta,Aaij:2015oid} & $\sim3.3\sigma$\\ \cline{2-5}
                                  & \multirow{1}{*}{[4.3, 6.0]} & $-0.774^{+0.0.061+0.087}_{-0.059-0.093}$~\cite{DescotesGenon:2012zf} & $-0.96^{+0.22}_{-0.21}$ (stat) $\pm 0.16$ (syst)~\cite{CMS Collaboration} & $\sim1.0\sigma$\\  \cline{2-5}
                                  & \multirow{1}{*}{[4.0, 8.0]} & $-0.881 \pm 0.082$~\cite{Descotes-Genon:2014uoa} & $-0.267^{+0.275}_{-0.269}$ (stat) $\pm 0.049$ (syst)~\cite{Abdesselam:2016llu} & $\sim2.1\sigma$\\
\hline
{$\mathcal{B}(B_s\, \to\, \phi\, \mu^+\, \mu^-$)} & {[1.0, 6.0]} & $(5.39 \pm 0.66)\times 10^{-8}$~\cite{Aebischer:2018iyb,Straub:2015ica} & $(2.57 \pm 0.37)\times 10^{-8}$~\cite{Aaij:2013aln,Aaij:2015esa} & $\sim3.7\sigma$\\
\hline 
\hline
\end{tabular}
\caption{Current status of $R_K$ and $R_{K^*}$ and $P_{5}^{\prime}$ in $B\, \to\, K^{(*)}\, l^+\,l^-$ and the branching ratio of
$\mathcal{B}(B_s\, \to\, \phi\, \mu^+\, \mu^-)$}
\label{status}
\end{table}
Our main aim is to study the impact of NP on $B_s\, \to\, f_{2}^{\prime}\,(1525)\,\mu^+ \,\mu^-$ decay observables in a model independent 
effective theory formalism.
The $B_s\, \to\, f_{2}^{\prime}\,(1525)\,\mu^+ \,\mu^-$ decay mode has received less attention both from the theoretical and the experimental 
side and it has not been discussed earlier in detail.
Although, in Ref.~\cite{Li:2010ra}, the authors discussed the SM results for both the $\mu$ mode and $\tau$ mode of 
$B_s\, \to\, f_{2}^{\prime}\,(1525)\,l^+ \,l^-$ along with the $B\, \to\, K_{2}^{*}(1430)\,l^+ \,l^-$ decays, 
more emphasis was given to $B\, \to\, K_{2}^{*}$ rather than $B_s\, \to\, f_{2}^{\prime}$ decays.  
Also the branching ratio of $f_{2}^{\prime}$ decaying into $ K^+\,K^-$ was not considered in their numerical analysis. 
In Ref.~\cite{Li:2010ra}, the authors also discussed the impact of NP on several observables coming from two different NP models such as the 
vector-like quark model and the family non-universal $Z^{\prime}$ model.
Similarly, there are ample number of literatures discussing the $B\, \to\, K_{2}^{*}(1430)\,l^+ \,l^-$ decays
~\cite{Ahmed:2012zzc,RaiChoudhury:2006bnu,Hatanaka:2009gb,Hatanaka:2009sj,Junaid:2012gz,Lu:2011jm,Aliev:2011gc,Das:2018orb} mediated via same
$b \to s\,l^+\,l^-$ quark level transition.

So far we don't have many experimental results on electroweak penguin decays involving spin 2 particles. The experimental techniques used for
$B_s\, \to\, \phi\,l^+ \,l^-$ can be adjusted to $B_s\, \to\, f_{2}^{\prime}\,(1525)\,l^+ \,l^-$ decay as well because both $\phi$ and 
$f_{2}^{\prime}(1525)$ decay to a pair of charged kaons which are easily detected by the LHCb detector. Since the dominating structures in 
$K^+K^-$ spectrum are the P wave $\phi(1020)$ and there are several possible resonances around 1500~MeV/$c^2$, it is natural to look at this 
regime to study. Further, the presence of D waves in this mass region yields a richer spectrum for exploring interesting angular observables. 

Although there are other resonances like $f_2(1270)$ and $f_0(1500)$ between $\phi(1020)$ and $f_2^{\prime}(1525)$, they have smaller branching fraction of 5\% or less into $K^+K^-$ final state and very unlikely to have large rates. Hence, $f_2^{\prime}(1525)$ is the best option after $\phi(1020)$~\cite{Aaij:2011ac}. 
This decay in muonic mode can be observed with the currently available data and we expect around 200 events for this mode.
The currently available data is statistically limited for performing angular analysis, 
 the branching fraction measurement is only possible claiming the first observation of this decay in muonic mode.
 With more data in Run 3, the measurement of angular observables is possible. 
 In the published Run 1 angular analysis of $B_s \to \phi \mu^+ \mu^-$ decays~\cite{Aaij:2015esa}, we see 10-20\% statistical 
 uncertainties across all angular observables. So far no results have been published with the full Run 1 and Run 2 data.
 We expect the statistical uncertainty in case of full Run 1 + Run 2 analysis to go down by a factor of 40\%, 
 that means 5-10\% total statistical uncertainty for angular observables in case of $B_s \to \phi \mu^+ \mu^-$ 
 decays with full Run 1 + Run 2 data. For $B_s \to f_2^{\prime}(1525) \mu^+ \mu^-$, we expect the statistical 
 uncertainty to be 3 times more with respect to $B_s \to \phi \mu^+ \mu^-$ decays i.e, 15-30\% statistical error.
 
The present paper is organized as follows: in Section~\ref{theory}, we start with a brief overview of the 
effective Hamiltonian for $b\, \to\, s\, l^+\, l^-$ quark level transition decays in the presence of new vector and axial vector NP operators.
A brief discussion of $B_s \to f_{2}^{\prime}$ hadronic matrix elements followed
by the angular distribution and the transversity amplitudes for $B_s \to f_{2}^{\prime}(1525)(\to K^+\,K^-)\,\mu^+ \,\mu^-$
decays are also reported. Finally we write down the decay distribution and expressions for several lepton flavor universal (LFU) observables. 
In Section~\ref{results}, 
we report our results that are obtained in the SM and in several NP scenarios. We conclude with a brief summary of our results in 
Section~\ref{conclusion}.

\section{Theoretical framework}
\label{theory}
\subsection{Effective Hamiltonian}

The effective Hamiltonian for $b\, \to\, s\, l^+\, l^-$ quark level transition decays in the presence of new vector and axial vector NP 
operators is written as
\begin{eqnarray}
\label{eff_ham}
 \mathcal{H}_{eff} &=& - \frac{G_F}{\sqrt{2}}\, V_{tb}\, V_{ts}^{*}\, \frac{\alpha_e}{4\, \pi} \Bigg[ 
 {C}_{9}^{eff}\, \bar{s}\, \gamma^{\mu}\, P_L\, b\, \bar{l}\, \gamma_{\mu}\, l\, +\, 
 {C}_{10}^{eff}\, \bar{s}\, \gamma^{\mu}\, P_L\, b\, \bar{l}\, \gamma_{\mu}\, \gamma_{5}\, l\, -\, 
 \frac{2\,m_b}{q^2}{C}_{7}^{eff}\, \bar{s}\, i\, q_{\nu}\, \sigma^{\mu \nu}\, P_R\, b\, \bar{l}\, \gamma_{\mu}\, l\, +\,
 \nonumber \\ &&
 {C}_{9}^{NP}\, \bar{s}\, \gamma^{\mu}\, P_L\, b\, \bar{l}\, \gamma_{\mu}\, l\, +\, 
 {C}_{10}^{NP}\, \bar{s}\, \gamma^{\mu}\, P_L\, b\, \bar{l}\, \gamma_{\mu}\, \gamma_{5}\, l\, +\,
 {C}_{9}^{\prime}\, \bar{s}\, \gamma^{\mu}\, P_R\, b\, \bar{l}\, \gamma_{\mu}\, l\, +\,
 {C}_{10}^{\prime}\, \bar{s}\, \gamma^{\mu}\, P_R\, b\, \bar{l}\, \gamma_{\mu}\, \gamma_{5}\, l\, \Bigg]\,,  
\end{eqnarray}
where $G_F$ is the Fermi coupling constant, $\alpha_e$ is the fine structure constant,
$V_{tb}$ and $V_{ts}$ are the corresponding Cabibbo Kobayashi Maskawa (CKM) matrix elements and $P_{L,R}=(1\mp\gamma_5)/2$.
The factorizable loop terms are incorporated within the effective Wilson coefficients (WCs) ${C}_{7}^{eff}$ and ${C}_{9}^{eff}$ 
as~\cite{Buras:1994dj}
\begin{eqnarray}
 {C}_{7}^{eff} &=& {C}_7 - \frac{{C}_5}{3} - {C}_6 \nonumber \\ 
 {C}_{9}^{eff} &=& {C}_{9}(\mu)\,+\,h(\hat{m}_c, \hat{s})\,{C}_{0}\, -\,\frac{1}{2}\,h(1,\hat{s})
 (4 {C}_{3}\,+\,4 {C}_{4}\,+\,3 {C}_{5}\,+\,{C}_{6})\, \nonumber \\ &&
 -\, \frac{1}{2}\,h(0,\hat{s}) ({C}_{3}\,+\, 3 {C}_{4})\, +\, \frac{2}{9}(3 {C}_{3}\,+ {C}_{4}\,+\,3 {C}_{5}\,+\,{C}_{6})\,,
\end{eqnarray}
where $\hat{s}=q^2/m_{b}^2$, $\hat{m}_c=m_c/m_b$ and
${C}_0=3 {C}_1\, +\, {C}_2\, +\, 3{C}_3\, +\,{C}_4\, +\, 3{C}_5\, +\, {C}_6$.
Similarly, the auxiliary functions are defined as
\begin{equation}
 h(z,\hat{s})=-\frac{8}{9}\ln \frac{m_b}{\mu}\,-\,\frac{8}{9}\ln z\, +\, \frac{8}{27}\, +\, \frac{4}{9}x\,-\,\frac{2}{9}(2+x)
 |1-x|^{1/2}\begin{cases}
             \ln \lvert \frac{\sqrt{1-x}+1}{\sqrt{1-x}-1}\rvert -i \pi\,, & \text{for $x \equiv \frac{4z^2}{\hat{s}}<1$} \\
             2 \arctan \frac{1}{\sqrt{x-1}}, & \text{for $x \equiv \frac{4z^2}{\hat{s}}>1$}
            \end{cases}
\end{equation}
\begin{equation}
 h(0,\hat{s})=-\frac{8}{9}\ln \frac{m_b}{\mu}\,-\,\frac{4}{9}\ln \hat{s}\, +\, \frac{8}{27}\, +\, \frac{4}{9} i \pi.
\end{equation}
The additional terms in the ${C}_{9}^{eff}$ describe the short distance contributions from the four-quark operators
which lie away from the $c\bar{c}$ resonance region. Similarly, the long distance contributions which include the resonant state
from $b\,\to\,c\,\bar{c}\,s$ which further annihilate into a lepton pair are excluded in the present analysis.
Hence, we only concentrate on the regions from $q^2\in[0.045,0.98]$ and $q^2\in[1.1,6.0]$\footnote{The $q^2\in[0.98,1.1]$ GeV$^2$
is excluded because of $\phi(1020)\, \to\, \mu^+\,\mu^-$ decays.} GeV$^2$.
It not necessary that the non-local effects are accounted only for the resonant states but also they are much important even 
below the charmonium contribution. This has been studied in detail in Refs.~\cite{Khodjamirian:2010vf,Khodjamirian:2012rm,
Bobeth:2017vxj,Gubernari:2020eft}. 
The authors in~\cite{Khodjamirian:2010vf} reports that, due to the virtual photon propagator, the non-factorizable contributions to 
$\Delta C_9$ are enhanced at small $q^2$ i.e., at $q^2>>4m_{l}^2$.
The factarizable soft gluon part $\Delta C_9(q^2)$ plays an important role in $B\to K^*$ decays.
The charm loop corrections $\Delta C_9(q^2)$ which almost reach upto $20\%$ of $C_9$ at $1\le q^2 \le 4$ GeV$^2$
in $B\to K^* ll$ decays and similarly, they will not exceed more than $5\%$ in the 
$B\to K ll$ decays. These non-factorizable contributions significantly affect the differential width
and the forward backward asymmetry in $B\to K^*$ decays. The zero crossing of the forward backward asymmetry
will be effected significantly in discriminating the new physics contributions.
Similarly, the non-local contributions in $B\to K^*$ and $B_s\to \phi$ decays have also been discussed very recently in 
\cite{Gubernari:2020eft}.
The authors in this particular paper proposed a modified analytic parameterization of non-local matrix
elements.
These effects infact enter the decay amplitudes in the form of non-perturbative non-local matrix
elements which are difficult to calculate with controlled uncertainties. 
The recalculations of beyond the OPE 
contributions involve the light cone sum rule and the full set of $B$ meson light cone distribution amplitudes.
Since, in most of the theoretical papers which try to address the LFU violation in $b \to s\,l^+\,l^-$ decays,
the hadronic non local effects are being neglected. Hence, we do not consider these corrections in our present analysis.

The new physics WCs in the effective Hamiltonian such as ${C}_{9,10}^{NP}$ and ${C}_{9,10}^{\prime}$ include the effects coming 
from the new vector and axial vector NP couplings. In SM, all these new WCs are considered to be zero.
In principle, one can have the new scalar, pseudoscalar and tensor NP WCs but they are severely constrained by $B_s\,\to\,\mu^+\,\mu^-$
and $b\,\to\,s\,\gamma$ measurements~\cite{Alok:2010zd,Alok:2011gv,Bardhan:2017xcc}.
The values for each WC obtained in the leading logarithmic approximation at the energy scale $\mu=m_{b,pole}$
are reported in Table~\ref{tab_wc}. Similarly, the values of each new WCs are obtained from the global fits of 
Ref.~\cite{Alok:2019ufo}.

\subsection{Spin $2$ polarization tensor and $B_s \to f_{2}^{\prime}$ hadronic matrix elements}
\subsubsection{Spin $2$ polarization tensor}
A spin $2$ polarization tensor $\epsilon^{\mu\nu}(n)$, where $n\in \pm2, \pm1, 0$, can be constructed via spin 1 polarization vector 
~\cite{Li:2010ra,Berger:2000wt,Wang:2010ni}. For the ${f_{2}^{\prime}}$ meson having the four momentum 
$(|\vec{p}_{f_{2}^{\prime}}|,0,0,E_{f_{2}^{\prime}})$,
where, $\vec{p}_{f_{2}^{\prime}}$ and $E_{f_{2}^{\prime}}$ are the momentum and energy of ${f_{2}^{\prime}}$
in the $B_s$ meson rest frame, the explicit structure of polarization tensor $\epsilon^{\mu\nu}(n)$
in the ordinary coordinate frame are constructed out of a massive vector state by the use of an appropriate
Clebsch-Gordan coefficients. Those are
\begin{eqnarray}
 \epsilon_{\mu\nu}(\pm2)&=&\epsilon_{\mu}(\pm)\, \epsilon_{\nu}(\pm) \nonumber \\
 \epsilon_{\mu\nu}(\pm1)&=&\frac{1}{\sqrt{2}}\bigg[\epsilon_{\mu}(\pm)\, \epsilon_{\nu}(0)+\epsilon_{\nu}(\pm)\, \epsilon_{\mu}(0)\bigg]
 \nonumber \\
 \epsilon_{\mu\nu}(0)&=&\frac{1}{\sqrt{6}}\bigg[\epsilon_{\mu}(+)\, \epsilon_{\nu}(-)+\epsilon_{\nu}(+)\, \epsilon_{\mu}(-)\bigg]
 +\, \sqrt{\frac{2}{3}}\, \epsilon_{\mu}(0)\,\epsilon_{\nu}(0)\,,
\end{eqnarray}
where
\begin{equation}
 \epsilon_{\mu}(0)=\frac{1}{m_{f_{2}^{\prime}}}(|\vec{p}_{f_{2}^{\prime}}|,0,0,E_{f_{2}^{\prime}}), \hspace{1cm}
 \epsilon_{\mu}(\pm)=\frac{1}{\sqrt{2}}(0,\mp 1,-i,0)\,.
\end{equation}
In the $B_s\, \to\, f_{2}^{\prime}(1525)\,l^+\,l^-$ decay, the $n=\pm2$ helicity states of the ${f_{2}^{\prime}}$ are not aware of
the two leptons that are obtained in the final state. Hence, it would be convenient to introduce a new polarization
vector $\epsilon_{T_{\mu}}(h)$ as
\begin{equation}
 \epsilon_{T_{\mu}}(h)=\frac{1}{m_b}\epsilon_{\mu\nu}(h)P_{B_s}^{\nu}\,,
\end{equation}
where $P_{B_s}$ is the four momentum of $B_s$ meson. The polarization vector $\epsilon_{T_{\mu}}(h)$ satisfies the
following equations~\cite{Li:2010ra}.
\begin{eqnarray}
 \epsilon_{T_{\mu}}(\pm2) &=& 0, \\ \nonumber
 \epsilon_{T_{\mu}}(\pm1) &=& \frac{1}{m_{B_s}}\frac{1}{\sqrt{2}}\,\epsilon(0)\cdotp P_{B_s}\epsilon_{\mu}(\pm)
 = \frac{\sqrt{\lambda}}{\sqrt{8}\,m_{B_s}m_{f_{2}^{\prime}}}\,\epsilon_{\mu}(\pm), \\ \nonumber
 \epsilon_{T_{\mu}}(0) &=& \frac{1}{m_{B_s}}\sqrt{\frac{2}{3}}\,\epsilon(0)\cdotp P_{B_s}\epsilon_{\mu}(0)
 =\frac{\sqrt{\lambda}}{\sqrt{6}\,m_{B_s}m_{f_{2}^{\prime}}}\,\epsilon_{\mu}(0)\,.
\end{eqnarray}

\subsubsection{$B_s \to f_{2}^{\prime}$ hadronic matrix elements}
Normally, for calculating the $B_s \to f_2^{\prime}(1525)$ form factors, the $f_2^{\prime}(1525)$ is treated to
be stable and contributes as a single particle. It means, it has a simple pole at
$k^2=m_{f_2^{\prime}(1525)}^2$. If there are any further higher order states, the hadronic representation goes
beyond the single pole. The purpose of implementing the narrow width limit can be done by calculating the sum rules
for the $K^+ K^-$ states.
Recently, there have been few improvements discussed in the theoretical predictions as of the local and non-local form factors are concerned.
In the propagation of the intermediate strange meson, the width effect of $f_{2}^{\prime}$ could be important.
This is because of the fact that the $\Gamma_{f_{2}^{\prime}} \gg \Gamma_{\phi}$. Hence, the narrow width approximation
may not work well for the $f_{2}^{\prime}$ case unlike $\phi$.
The finite-width effects which lead to 10\% corrections to the form factors which further lead to 20\% corrections to the 
branching fractions in the case of $\rho$ and $K^*$ are studied in detail in Refs.~\cite{Cheng:2017smj,Descotes-Genon:2019bud}.
Nevertheless, we do not consider these effects in the present analysis. Although, the corrections 
tend to increase the discrepancy between the SM predictions of the branching fractions and the corresponding
LHCb measurements, the normalized angular observables such as $P'_5$ and other LFU sensitive ratios which mainly depend on 
the form factors are insensitive to the finite width corrections. 
Moreover, the global fit results including all these corrections are still awaited~\cite{Descotes-Genon:2019bud,Virto:2021pmw}.

In general, the $B_s \to f_{2}^{\prime}$ hadronic matrix elements can be parameterized in terms of several form factors as follows 
~\cite{Li:2010ra,Hatanaka:2009gb,Hatanaka:2009sj,Yang:2010qd,Wang:2010ni}:
\begin{eqnarray}
 \langle\, f_{2}^{\prime}(P_{f_{2}^{\prime}}, \epsilon)|\bar{s}\gamma^{\mu}b|\bar{B}_s (P_{B_s})\,\rangle &=&
 -\frac{2V(q^2)}{m_{B_s}+m_{f_{2}^{\prime}}} \epsilon^{\mu \nu \rho \sigma}\, \epsilon_{T_{\nu}}^{*}\,P_{{B_s}\rho}P_{{f_{2}^{\prime}}\sigma} \nonumber \\ 
 \langle\, f_{2}^{\prime}(P_{f_{2}^{\prime}}, \epsilon)|\bar{s}\gamma^{\mu}\gamma_{5}b|\bar{B}_s (P_{B_s})\,\rangle &=&
 2 i\, m_{f_{2}^{\prime}}A_{0}(q^2)\frac{\epsilon_{T}^{*} \ldotp q}{q^2}\, q^{\mu}\, +\, i(m_{B_s}+m_{f_{2}^{\prime}})A_1 (q^2)
 \bigg[ \epsilon_{T_{\mu}}^{*}-\frac{\epsilon_{T}^{*} \ldotp q}{q^2}\, q^{\mu} \bigg]\, \nonumber \\ &&
 -\, iA_2 (q^2) \frac{\epsilon_{T}^{*} \ldotp q}{m_{B_s}+m_{f_{2}^{\prime}}}\, \bigg[ P^{\mu} - \frac{m_{B_s}^2 + m_{f_{2}^{\prime}}^2}{q^2}q^{\mu} \bigg] \nonumber \\
 \langle\, f_{2}^{\prime}(P_{f_{2}^{\prime}}, \epsilon)|\bar{s}\sigma^{\mu\nu}q_{\nu}b|\bar{B}_s (P_{B_s})\,\rangle &=&
 -2\,i\,T_1 (q^2)\, \epsilon^{\mu \nu \rho \sigma}\,\epsilon_{T_{\nu}}^{*}\,P_{{B_s}\rho}P_{{f_{2}^{\prime}}\sigma} \nonumber \\ 
 \langle\, f_{2}^{\prime}(P_{f_{2}^{\prime}}, \epsilon)|\bar{s}\sigma^{\mu\nu}\gamma_{5}q_{\nu}b|\bar{B}_s (P_{B_s})\,\rangle &=&
T_2(q^2)\, \bigg[(m_{B_s}^2 + m_{f_{2}^{\prime}}^2)\, \epsilon_{T_{\mu}}\,\epsilon_{T}^{*} \ldotp q\, P^{\mu}\bigg]\,
+\, T_3 (q^2)\, \epsilon_{T}^{*} \ldotp q\, \bigg[q^{\mu}-\frac{q^2}{m_{B_s}^2 + m_{f_{2}^{\prime}}^2}\,P^{\mu}\bigg]\,,
\end{eqnarray}
where $P_{B_s}$ and $P_{f_{2}^{\prime}}$ are the four momenta of $B_s$ meson and
$f_{2}^{\prime}$, respectively and $q=P_{B_s}-P_{f_{2}^{\prime}}$. In general, the $B_s \to f_{2}^{\prime}$
transition form factors are non-perturbative in nature and they can be calculated using several non-perturbative 
approaches. We follow Ref.~\cite{Wang:2010ni} and write the $B_s \to f_{2}^{\prime}$ transition form factors as
\begin{equation}
\label{ffdef}
 F(q^2)=\frac{F(0)}{(1\,-\,q^2/m_{B_s}^2)\,\left[1\,-\,a(q^2/m_{B_s}^2)\,+\,b(q^2/m_{B_s}^2)^2\right]}\,,
\end{equation}
where $F$ denotes $A_0$, $A_1$, $V$, $T_1$, $T_2$ and $T_3$, respectively. Similarly, $A_2$ is related to
$A_0$ and $A_1$ by
\begin{equation}
 A_2 (q^2)= \frac{m_{B_s}+m_{f_{2}^{\prime}}}{m_{B_s}^2-q^2}\,\left[(m_{B_s}+m_{f_{2}^{\prime}})A_1(q^2)\,-\,
 2\,m_{f_{2}^{\prime}}\,A_0(q^2)\right]
\end{equation}
The numerical entries of the $B_s \to f_{2}^{\prime}$ form factors at the maximum recoil point and the two 
fitted parameters $a$ and $b$ are reported in Table~\ref{tab_ff}. 

The $B\to T$ form factors contain one more pole structure in the $q^2$ distribution and they are expected
to be sharper than the $B\to V$ form factors.
But the parametrization of $B\to T$ form factors is analogous to $B\to V$ form factors and the only difference is the replacement of
$\epsilon$ by $\epsilon_T$. 
This can be easily related when we mark the pole at $q^2=0$ and we get the relation 
$2\,m_T A_0 (0)= (m_{B_s} + m_T)A_1 (0)-(m_{B_s} - m_T)A_2 (0)$ which has a similar relation as of $B\to V$ case.

The Lorentz structures of the wave functions and the $B$ decay form factors involving the vector and 
tensor mesons have great similarities. Hence this allows to obtain the factorization formulas of $B\to T$
form factors from $B\to V$ ones. Further, the two set of $B\to V$ and $B\to T$ form factors have
the same signs and related $q^2$ dependancy. This is because the light cone distribution amplitudes of the 
tensor mesons and the vector mesons have similar shapes in the dominant region of the pQCD approach.
In the pQCD, the factorization formula is given by~\cite{Wang:2010ni},
\begin{eqnarray}
\label{fact}
 \mathcal{M} &=& \int_{0}^{1} dx_1 dx_2 \int d^2 \vec{b}_1 d^2 \vec{b}_2\, \phi_B (x_1, \vec{b}_1, P_B, t)\, 
 T_x (x_1, x_2, \vec{b}_1, \vec{b}_1, t)\, \phi_2 (x_2, \vec{b}_2, P_2, t)\, S_t (x_2)\, exp[-S_B (t)-S_2 (t)]
\end{eqnarray}
This has been generalized to number of transition form factors for various final state mesons including scalar, vector, pseudoscalar
and axial-vector mesons. The correspondence between vector and tensor mesons are obtained in a comparative way. Both LCDAs of the tensor 
meson and $B\to T$ form factors coincide with the quantities involving a vector meson as
\begin{equation}
 \phi_{V}^{(i)} \leftrightarrow \phi_{T}^{(i)}, \hspace{1cm}
 F^{B\to T} \leftrightarrow F^{B\to V}
\end{equation}
where $F$ and $\phi_{V,T}^{(i)}$ represent the $B\to (T,V)$ form factors and LCDA respectively. 
The polarization vector $\epsilon$ is replaced by $\epsilon_{\bullet}$ and $\epsilon_T$ 
respectively in the LCDAs and in the transition form factors. As a result the $B \to T$ form factors are factorized as
\begin{equation}
\label{fact1}
 F^{B\to T}\, (\phi_{T}^{(i)})= \frac{\epsilon_{\bullet}}{\epsilon_T}\, F^{B\to V}\, (\phi_{V}^{(i)})=
 \frac{2m_B m_T}{m_{B}^2 - q^2}\, F^{B\to V}\, (\phi_{V}^{(i)})
\end{equation}
While extracting the form factors in a non-perturbative way by using QCD sum rules, 
the pole structure of the form factors are constrained in an analytic way whereas, in pQCD platform which uses the perturbative 
properties of the form factors such as the factorization, construct the parametrization form in a phenomenological way.
In general, the polar form of $B\to V$ form factors include pole form, dipole form, exponential form and the BK parametrization~\cite{Becirevic:1999kt}. 
By adopting this approach one defines the dipole form of $B\to V$ form factors in a pQCD as,
\begin{equation}
 F(q^2)=\frac{F(0)}{1-a(q^2/m_B^2)+b(q^2/m_B^2)^2}
\end{equation}
The only difference for the case of $B\to T$ form factors is that it receives an additional $q^2$ dependency.
This can be seen in the factorization formula of Eq.~\ref{fact1} and the formula for $F(q^2)$ in the Eq.~\ref{ffdef}. 
Hence, this modification is appropriate for the $q^2$ distribution of $B\to T$ form factors.
We refer to Ref.~\cite{Wang:2010ni} for all the omitted details.

\subsection{Angular distribution and the transversity amplitudes 
for $B_s \to f_{2}^{\prime}(1525)(\to K^+\,K^-)\,\mu^+ \,\mu^-$ decays}
The decay amplitude for $B_s \to f_{2}^{\prime}(1525)\,l^+ \,l^-$ can be obtained from the effective Hamiltonian 
of Eq~\ref{eff_ham}. Using the helicity techniques of Ref~\cite{Li:2010ra}, the differential decay width of the four-body 
decay of $B_s \to f_{2}^{\prime}(1525)(\to K^+\,K^-)\,\mu^+ \,\mu^-$ can be written in terms of several angular coefficients as
\begin{eqnarray}
\label{ddrate}
 \frac{d^4\Gamma}{dq^2 d\cos\theta_K d\cos\theta_l d\phi} &=& \frac{3}{8}\,\Bigg[\,
 I_{1}^{c}\,C^2\,+\,2\,I_{1}^{s}\,S^2\,+\,(I_{2}^{c}\,C^2\,+\,2\,I_{2}^{s}\,S^2)\,cos\,2\theta_l\,+\,
 2\,I_{3}\,S^2\,\sin^2 \theta_l\,\cos\,2\phi\,+\,
  \nonumber \\ &&
 2\,\sqrt{2}\,I_{4}\,C\,S\,\sin 2\theta_l\,\cos\phi\,+\,
 2\,\sqrt{2}\,I_{5}\,C\,S\,\sin \theta_l\,\cos\phi\,+\,
 2\,I_{6}\,S^2\,\cos \theta_l\,+\,
 \nonumber \\ &&
 2\,\sqrt{2}\,I_{7}\,C\,S\,\sin \theta_l\,\sin\phi\,+\,
 2\,\sqrt{2}\,I_{8}\,C\,S\,\sin2 \theta_l\,\sin\phi\,+\,
 2\,I_{9}\,S^2\,\sin^2 \theta_l\,\sin2\phi\, \Bigg]\,,
\end{eqnarray}
where $C=C(f_{2}^{\prime}) \equiv \sqrt{\frac{5}{16\pi}}(3\cos^2\theta_K -1)$ and $S=S(f_{2}^{\prime}) \equiv \sqrt{\frac{15}{32\pi}}
\sin(2\theta_K)$. The direction of $f_2^{\prime}$ is chosen along the $z$ direction in the $B_s$ meson rest frame. The polar angle
$\theta_K~(\theta_l)$ 
is defined as the angle between the direction of $K^-~(\mu^-)$ and the $z$ axis in the rest frame of the lepton pair.
Similarly, $\phi$ is the angle between the decay planes of $f_{2}^{\prime}$
and the lepton pair. Moreover, the angular coefficients $I_i (q^2)$ are defined as 

\begin{eqnarray}
 I_{1}^{c} &=& \bigg(|A_{L0}|^2 + |A_{R0}|^2\bigg) + 8\frac{m_{l}^2}{q^2} Re\bigg[A_{L0}A_{R0}^{*}\bigg] + 4\frac{m_{l}^2}{q^2}|A_{t}|^2, \nonumber \\
 I_{2}^{c} &=& -\beta_{l}^2 \bigg(|A_{L0}|^2 + |A_{R0}|^2\bigg), \nonumber \\
 I_{1}^{s} &=& \frac{3}{4} \bigg[|A_{L\perp}|^2 + |A_{L\parallel}|^2 + |A_{R\perp}|^2 + |A_{R\parallel}|^2\bigg] \bigg(1-\frac{4m_{l}^2}{3q^2}\bigg) +
 \frac{4m_{l}^2}{q^2} Re\bigg[A_{L\perp} A_{R\perp}^{*} + A_{L\parallel} A_{R\parallel}^{*}\bigg], \nonumber \\
 I_{2}^{s} &=& \frac{1}{4} \beta_{l}^2 \bigg[|A_{L\perp}|^2 + |A_{L\parallel}|^2 + |A_{R\perp}|^2 + |A_{R\parallel}|^2\bigg],   \nonumber \\
 I_{3} &=& \frac{1}{2} \beta_{l}^2 \bigg[|A_{L\perp}|^2 - |A_{L\parallel}|^2 + |A_{R\perp}|^2 - |A_{R\parallel}|^2\bigg], \nonumber \\
 I_{4} &=& \frac{1}{\sqrt{2}} \beta_{l}^2 \bigg[Re\bigg(A_{L0}A_{L\parallel}^{*}\bigg) + Re\bigg(A_{R0}A_{R\parallel}^{*}\bigg)\bigg],  \nonumber \\
 I_{5} &=& \sqrt{2} \beta_l \bigg[Re\bigg(A_{L0}A_{L\perp}^{*}\bigg) - Re\bigg(A_{R0}A_{R\perp}^{*}\bigg)\bigg], \nonumber \\
 I_{6} &=& 2 \beta_l \bigg[Re\bigg(A_{L\parallel}A_{L\perp}^{*}\bigg) - Re\bigg(A_{R\parallel}A_{R\perp}^{*}\bigg)\bigg], \nonumber \\
 I_{7} &=& \sqrt{2} \beta_l \bigg[Im\bigg(A_{L0}A_{L\parallel}^{*}\bigg) - Im\bigg(A_{R0}A_{R\parallel}^{*}\bigg)\bigg], \nonumber \\
 I_{8} &=& \frac{1}{\sqrt{2}} \beta_{l}^2 \bigg[Im\bigg(A_{L0}A_{L\perp}^{*}\bigg) + Im\bigg(A_{R0}A_{R\perp}^{*}\bigg)\bigg], \nonumber \\
 I_{9} &=& \beta_{l}^2 \bigg[Im\bigg(A_{L\parallel}A_{L\perp}^{*}\bigg) + Im\bigg(A_{R\parallel}A_{R\perp}^{*}\bigg)\bigg]\,,
\end{eqnarray}

where $\beta_l = \sqrt{1-4\,m_l^2/q^2}$ is the mass correction factor.
In our analysis, we assume all the angular coefficients to be real and $CP$ conserving. 
For convenience, we introduce here the transversity amplitudes $A_{L0}$, $A_{R0}$, $A_{L\perp}$, $A_{R\perp}$, $A_{L\parallel}$ and
$A_{R\parallel}$.
However, they are nothing but linear combinations of the helicity amplitudes as mentioned in the Ref.~\cite{Li:2010ra}. 
The subscripts $L$ and $R$ represent the chiralities of the lepton current where the right chiral amplitudes differ by left
chiral amplitudes as $A_{Ri}={A_{Li}}|_{C_{10}\to -C_{10}}$.
The amplitudes $A_i$ are obtained from the hadronic $B \to f_{2}^{\prime} V$ amplitudes $\mathcal{H}_i$ through 
$A_i=\sqrt{\frac{\sqrt{\lambda}q^2 \beta_l\, \mathcal{B}(f_{2}^{\prime}(1525)\to K^+\,K^-)}{3\cdot 32 m_B^3 \pi^3}}\mathcal{H}_i$.
The details of the helicity amplitudes are discussed in the Appendix B.
The explicit expressions for the transversity
amplitudes for the $B_s \to f_{2}^{\prime}(1525)(\to K^+\,K^-)\,\mu^+ \,\mu^-$ decay are written as follows:
\begin{eqnarray}
 A_{L0} &=& N_{f_{2}^{\prime}} \frac{\sqrt{\lambda}}{\sqrt{6}\,m_{B_s}m_{f_{2}^{\prime}}}\frac{1}{2m_{f_{2}^{\prime}}\sqrt{q^2}}
 \bigg\{(C_{9}^{eff} - C_{10}) \bigg[(m_{B_s}^2 - m_{f_{2}^{\prime}}^2 - q^2)(m_{B_s} + m_{f_{2}^{\prime}})A_1 - \frac{\lambda}{m_{B_s} + m_{f_{2}^{\prime}}}A_2\bigg]+ \nonumber \\  
 && 2\,m_b\, C_{7}^{eff}\, \bigg[(m_{B_s}^2 + 3m_{f_{2}^{\prime}}^2 - q^2)T_2 - \frac{\lambda}{m_{B_s}^2 - m_{f_{2}^{\prime}}^2}T_3 \bigg] \bigg\}\,, \nonumber \\
 A_{L\perp} &=& - N_{f_{2}^{\prime}}\sqrt{2} \frac{\sqrt{\lambda}}{\sqrt{8}\,m_{B_s}m_{f_{2}^{\prime}}} \bigg[(C_{9}^{eff} - C_{10})
 \frac{\sqrt{\lambda}}{m_{B_s} + m_{f_{2}^{\prime}}} V + \frac{\sqrt{\lambda}\,2\,m_b\,C_{7}^{eff}}{q^2} T_1 \bigg]\,, \nonumber \\
 A_{L\parallel} &=& N_{f_{2}^{\prime}}\sqrt{2} \frac{\sqrt{\lambda}}{\sqrt{8}\,m_{B_s}m_{f_{2}^{\prime}}} \bigg[(C_{9}^{eff} - C_{10})
 (m_{B_s} + m_{f_{2}^{\prime}}) A_1 + \frac{2\,m_b\,C_{7}^{eff}(m_{B_s}^2 - m_{f_{2}^{\prime}}^2)}{q^2} T_2 \bigg]\,, \nonumber \\
 A_{Lt} &=& N_{f_{2}^{\prime}} \frac{\sqrt{\lambda}}{\sqrt{6}\,m_{B_s}m_{f_{2}^{\prime}}} (C_{9}^{eff} - C_{10}) \frac{\sqrt{\lambda}}{\sqrt{q^2}}A_0\,,
\end{eqnarray}
where $\lambda=m_{B_s}^4\,+m_{f_{2}^{\prime}}^4\,+q^4\,-\,2\,(m_{B_s}^2m_{f_{2}^{\prime}}^2+m_{f_{2}^{\prime}}^2 q^2+q^2 m_{B_s}^2)$ and 
$N_{f_{2}^{\prime}}$ is the normalization constant defined as 
\begin{equation}
 N_{f_{2}^{\prime}} = \bigg[\frac{G_{F}^2\alpha_{em}^2}{3\cdot 2^{10}\pi^5\,m_{B_s}^3}|V_{tb}V_{ts}^{*}|^2 q^2 \sqrt{\lambda}\bigg(1-\frac{4m_{l}^2}{q^2}
 \bigg)^{1/2} \mathcal{B}(f_{2}^{\prime}\, \to\, K^+\, K^-)\bigg]^{1/2}\,.
\end{equation}

\subsection{Decay distribution and LFU observables}
By integrating Eq.~\ref{ddrate} with respect to $\theta_K$, $\theta_l$ and $\phi$, we obtain the differential decay rate. That is
\begin{equation}
 \frac{d\,\Gamma}{d\,q^2}=\frac{1}{4}\bigg[3I_{1}^c + 6I_{1}^s - I_{2}^c -2I_{2}^s \bigg]\,.
\end{equation}
We define several other $q^2$ dependent observables such as the differential branching ratio, the longitudinal polarization 
fraction and the forward-backward asymmetry for the $B_s \to f_{2}^{\prime}(1525)(\to K^+\,K^-)\,\mu^+ \,\mu^-$ decays. Those are
\begin{equation}
 DBR(q^2)=\frac{d\Gamma/dq^2}{\Gamma_{Total}}, \hspace{0.5cm}
 F_L(q^2)=\frac{3I_{1}^c - I_{2}^c}{3I_{1}^c + 6I_{1}^s - I_{2}^c -2I_{2}^s}, \hspace{0.5cm}
 A_{FB}(q^2)=\frac{3I_{6}}{3I_{1}^c + 6I_{1}^s - I_{2}^c -2I_{2}^s}\,.
\end{equation}
In principle, the angular analysis of $B_s \to f_{2}^{\prime}(1525)(\to K^+\,K^-)\,\mu^+ \,\mu^-$ decay
provides several additional observables in the form of ratios of various angular coefficients. 
These observables are found to be very sensitive to NP.
Here we define some angular observables such as $\langle P_1 \rangle$, $\langle P_2 \rangle$, $\langle P^{\prime}_4 \rangle$
and $\langle P^{\prime}_5 \rangle$ as reported in Refs.~\cite{DescotesGenon:2012zf,Descotes-Genon:2013vna}. The explicit expressions are
as follows: 
\begin{equation}
 \langle P_1 \rangle = \frac{1}{2}\frac{\int_{bin}dq^2 I_{3}}{\int_{bin}dq^2 I_{2}^s}, \hspace{0.5cm}
 \langle P_2 \rangle = \frac{1}{8}\frac{\int_{bin}dq^2 I_{6}}{\int_{bin}dq^2 I_{2}^s}, \hspace{0.5cm}
 \langle P^{\prime}_4 \rangle = \frac{\int_{bin}dq^2 I_{4}}{\sqrt{-\int_{bin}dq^2 I_{2}^c \int_{bin}dq^2 I_{2}^s}},  \hspace{0.5cm}
 \langle P^{\prime}_5 \rangle = \frac{\int_{bin}dq^2 I_{5}}{2 \sqrt{-\int_{bin}dq^2 I_{2}^c \int_{bin}dq^2 I_{2}^s}}\,.
\end{equation}
One can construct several other observables that can be defined in the form of ratios
or in the form of differences between the observables involving two different families of lepton pairs. These observables such as the ratio
of branching ratio $R_{f_{2}^{\prime}}$ and $\langle Q_{F_L} \rangle$, $\langle Q_{A_{FB}} \rangle$, $\langle Q_{i}^{(\prime)} \rangle$
($i\in1,2,4,5$) are quite sensitive to NP. In the SM, we expect the value of $R_{f_{2}^{\prime}}$ to be very close to $1$. 
Similarly, since the observables $Q^{(\prime)}$~\cite{Capdevila:2016ivx}
are defined to be the differences between the $e$ and $\mu$ modes, one would expect these quantities to be almost zero in the SM.  
 Hence any deviation from zero would be a clear signal of NP in $b \to s\,l^+\,l^-$ quark level transition decays.
Measurement of these observables in future may provide crucial information regarding LFUV observed in various $B$ meson decays.
The explicit expressions for these observables are as follows:
\begin{equation}
 R_{f_{2}^{\prime}}(q^2)=\frac{\mathcal{B}(B_s \to f_{2}'\,\mu^+ \,\mu^-)}{\mathcal{B}(B_s \to f_{2}'\,e^+ \,e^-)}\,
\end{equation}
and
\begin{equation}
 \langle Q_{F_L} \rangle = \langle {F_L}^{\mu} \rangle - \langle {F_L}^{e} \rangle, \hspace{0.5cm}
 \langle Q_{A_{FB}} \rangle = \langle {A_{FB}}^{\mu} \rangle - \langle {A_{FB}}^{e} \rangle, \hspace{0.5cm}
 \langle Q_{i}^{(\prime)} \rangle = \langle P_{i}^{(\prime)\mu} \rangle - \langle P_{i}^{(\prime)e} \rangle.
\end{equation}

\section{Results and Discussions}
\label{results}
\subsection{Input Parameters}

We report here all the relevant input parameters that are used in our numerical analysis. 
Masses of the mesons, leptons and quarks are in GeV, the Fermi coupling constant is in GeV$^{-2}$ and the life time of $B_s$ meson 
is in seconds. We consider the masses of $b$ quark and $c$ quark evaluated at the $\overline{MS}$ scheme. 
The uncertainties associated with the CKM matrix element and $\mathcal{B}(f_{2}^{\prime} \to K^+\,K^-)$ are reported within parentheses.
We do not report the uncertainties associated with other input parameters as they are not important for our analysis.
In Table~\ref{tab_wc}, we report the values of Wilson 
coefficients $C_{i}(m_b)$ that are evaluated in the leading logarithmic approximation.
The form factor input parameters evaluated in the pQCD approach are reported in Table~\ref{tab_ff}
where, $F(0)$ denote the form factors at $q^2=0$
i.e., at the maximum recoil point and $a$ and $b$ are the two fitted parameters. There are two kinds of errors associated with $F(0)$, $a$ 
and $b$.
The first error is coming from the decay constant of the $B_s$ meson and the shape parameter $\omega_b$ and the second error is coming 
from the $\Lambda_{QCD}$, the scales $ts$ and the threshold resummation parameter $c$. We refer to Ref.~\cite{Wang:2010ni} for all the omitted
details. 

\begin{table}[htbp]
\centering
\setlength{\tabcolsep}{2pt} % Default value: 6pt
\renewcommand{\arraystretch}{1.5} % Default value:
\begin{tabular}{cccccccccc}
\hline
Parameter & Value & Parameter & Value & Parameter & Value & Parameter & Value & Parameter & Value \\
\hline
$m_{B_s}$ & 5.36689 & $m_{f_{2}^{\prime}}$ & 1.525 & $m_{b}^{\overline{MS}}$ & 4.20 & $m_{c}^{\overline{MS}}$ & 1.28 & $m_{b}^{pole}$ & 4.80\\
$\tau_{B_s}$ & $1.509 \times 10^{-12}$ & $G_F$ & $1.1663787 \times 10^{-5}$ & $\alpha_e$ & 1/133.28 & $|V_{tb} V_{ts}^*|$ & 0.04088(55) & $\mathcal{B}(f_{2}^{\prime} \to K^+{K^-})$ & 0.4435(11)\\
\hline 
\hline
\end{tabular}
\caption{Theory input parameters~\cite{Tanabashi:2018oca}}
\label{tab_input}
\end{table}

\begin{table}[htbp]
\centering
\setlength{\tabcolsep}{8pt} % Default value: 6pt
\renewcommand{\arraystretch}{1.5} % Default value:
\begin{tabular}{ccccccccc}
\hline
$C_1$ & $C_2$ & $C_3$ & $C_4$ & $C_5$ & $C_6$ & $C_{7}^{eff}$ & $C_9$ & $C_{10}$ \\
\hline
-0.248 & 1.107 & 0.011 & -0.026 & 0.007 & -0.031 & -0.313 & 4.344 & -4.669 \\
\hline 
\hline
\end{tabular}
\caption{Wilson coefficients $C_{i}(m_b)$ in the leading logarithmic approximation~\cite{Ali:1999mm}}
\label{tab_wc}
\end{table}

\begin{table}[htbp]
\centering
\setlength{\tabcolsep}{5pt} % Default value: 6pt
\renewcommand{\arraystretch}{1.5} % Default value: 
\begin{tabular}{ccccccc}
\hline
       & $V$ & $A_0$ & $A_1$ & $T_1$ & $T_2$ & $T_3$ \\
\hline
$F(0)$ & $0.20^{+0.04+0.05}_{-0.03-0.03}$ & $0.16^{0.03+0.03}_{-0.02-0.02}$ & $0.12^{+0.02+0.03}_{-0.02-0.02}$ & $0.16^{+0.03+0.04}_{-0.03-0.02}$ 
& $0.16^{+0.03+0.04}_{-0.03-0.02}$ & $0.13^{+0.03+0.03}_{-0.02-0.02}$ \\ 
$a$ & $1.75^{+0.02+0.05}_{-0.00-0.03}$ & $1.69^{+0.00+0.04}_{-0.01-0.03}$ & $0.80^{+0.02+0.07}_{-0.00-0.03}$ & $1.75^{+0.01+0.05}_{-0.00-0.05}$
& $0.82^{+0.00+0.04}_{-0.04-0.06}$ & $1.64^{+0.02+0.06}_{-0.00-0.06}$ \\
$b$ & $0.69^{+0.05+0.08}_{-0.01-0.01}$ & $0.64^{+0.00+0.01}_{-0.04-0.02}$ & $-0.11^{+0.05+0.06}_{-0.00-0.00}$ & $0.71^{+0.03+0.06}_{-0.01-0.08}$
& $-0.08^{+0.00+0.03}_{-0.09-0.08}$ & $0.57^{+0.04+0.05}_{-0.01-0.09}$ \\
\hline 
\hline
\end{tabular}
\caption{Form factor input parameters~\cite{Wang:2010ni}}
\label{tab_ff}
\end{table}

\subsection{Standard Model predictions}
We now proceed to discuss our results in the SM. 
We report in Table~\ref{tab_sm1} and~\ref{tab_sm2}, the central values and the corresponding $1\sigma$ uncertainties for each of the 
observables such as
the differential branching ratio, the normalized longitudinal polarization fraction $\langle F_L \rangle$, the normalized forward-backward 
asymmetry 
$\langle A_{FB} \rangle$, $\langle P_1 \rangle$, $\langle P_2 \rangle$, $\langle P^{\prime}_4 \rangle$, $\langle P^{\prime}_5 \rangle$ 
and also LFUV sensitive observables such as the ratio of branching ratio $R_{f_{2}^{\prime}}$, $\langle Q_{F_L} \rangle$,
$\langle Q_{A_{FB}} \rangle$, $\langle Q_{i}^{(\prime)} \rangle$ in different $q^2$ bins for both $e$ and the $\mu$ mode.
Here, we restrict our analysis to the low dilepton invariant mass region ranging from $q^2\in[0.045,6.0]$ GeV$^2$ that excludes the charmonium
contributions. We have considered several $q^2$ bins with similar bin sizes such as [0.10, 0.98], [1.1, 2.5], [2.5, 4.0] and [4.0, 6.0] 
as reported by LHCb in the measurements of $B_s\, \to\, \phi\, \mu^+\,\mu^-$ decays~\cite{Aaij:2013aln,Aaij:2015esa}. 
In addition, we include [1.1, 6.0] and [0.045, 6.0] bins as well.
The central values for each observables are obtained by considering the central values of each input parameters. 
The corresponding $1\sigma$ uncertainties are obtained by using the uncertainties associated with input parameters such as the form factors, 
the CKM matrix elements $|V_{tb}V^*_{ts}|$ and the branching ratio $\mathcal{B}(f_{2}^{\prime} \to K^+\,K^-)$.
We notice that the branching ratio for 
$B_s \to f_{2}^{\prime}(1525)(\to K^+\,K^-)\,\{\mu^+/e^+\} \,\{\mu^-/e^-\}$ decays is of the order of $\mathcal{O}(10^{-7})$ in the SM.
As expected, in the SM, both the $e$  and $\mu$ modes
show similar behavior for all the observables. Obviously, this is a clear confirmation of the LFU in the SM.
To account for the LFU, we expect $\langle Q_{F_L} \rangle$, $\langle Q_{A_{FB}} \rangle$, $\langle Q_{i}^{(\prime)} \rangle$s 
($i\in1,2,4,5$) 
to be almost zero, although a slight non-zero contribution may occur due to the difference in 
the masses of $e$ and $\mu$. In addition, we expect the ratio of branching ratio $R_{f_{2}^{\prime}}$ to be almost equal to unity. 
These are observed to be true from the entries reported in Table~\ref{tab_sm3}. In addition to the bins reported for the branching
ratio in Table~\ref{tab_sm3}, for completeness we also report the branching ratios for $\mu$ and $e$ modes in the full $q^2$ range 
to be $2.13\pm0.43 \times 10^{-7}$ and $2.49\pm0.44 \times 10^{-7}$ respectively (exluding the branching ratio of $f_{2}^{\prime}$ 
decay into $K^+\,K^-$ explicitly) and these values are found to agree with~\cite{Wang:2010ni}.

\begin{table}[htbp]
\centering
\begin{tabular}{|c|c|c|c|c|c|c|c|c|c|c|}
\hline
\hline
    \multirow{2}{*}{$q^{2}$ bins (GeV$^2$)}
     &\multicolumn{2}{c|}{BR$\times 10^{-7}$} &\multicolumn{2}{c|}{$\langle F_{L} \rangle$} &\multicolumn{2}{c|}{$\langle A_{FB} \rangle$} \\
    \cline{2-7}
    &$e$ mode&$\mu$ mode&$e$ mode&$\mu$ mode&$e$ mode&$\mu$ mode \\
    \hline
    \hline
    [0.10, 0.98]      & $0.116 \pm  0.021$ & $0.114 \pm  0.021$ & $0.502 \pm  0.108$ & $0.503 \pm  0.108$ & $0.096 \pm  0.017$ & $0.086 \pm  0.016$ \\
     \hline
    [1.1, 2.5]        & $0.105 \pm  0.025$ & $0.105 \pm  0.025$ & $0.854 \pm  0.043$ & $0.855 \pm  0.047$ & $0.082 \pm  0.034$ & $0.082 \pm  0.036$ \\
    \hline
    [2.5, 4.0]        & $0.111 \pm  0.026$ & $0.110 \pm  0.026$ & $0.841 \pm  0.045$ & $0.843 \pm  0.045$ & $-0.014 \pm  0.040$ & $-0.014 \pm  0.039$ \\
    \hline
    [4.0, 6.0]        & $0.154 \pm  0.035$ & $0.153 \pm  0.035$ & $0.760 \pm  0.062$ & $0.762 \pm  0.062$ & $-0.116 \pm  0.050$ & $-0.116 \pm  0.049$ \\
    \hline
%   [6.0, 8.0]        & $0.157 \pm  0.035$ & $0.156 \pm  0.035$ & $0.672 \pm  0.069$ & $0.673 \pm  0.069$ & $-0.205 \pm  0.056$ & $-0.204 \pm  0.056$ \\
%   \hline
    [1.1, 6.0]        & $0.370 \pm  0.085$ & $0.368 \pm  0.085$ & $0.810 \pm  0.050$ & $ 0.812 \pm  0.050$ & $-0.029 \pm  0.040$ & $-0.030 \pm  0.040$ \\
    \hline
    [0.045, 6.0]      & $0.524 \pm  0.103$ & $0.512 \pm  0.103$ & $0.700 \pm  0.071$ & $0.712 \pm  0.069$ & $0.004 \pm  0.030$ & $-0.000 \pm  0.030$ \\
     \hline 
\hline
\end{tabular}
\caption{The central values and the corresponding $1\sigma$ uncertainties for each of the observables such as
the branching ratio, the normalized longitudinal polarization fraction $\langle F_L \rangle$, the normalized forward-backward 
asymmetry 
$\langle A_{FB} \rangle$ for both $e$ mode and $\mu$ mode of $B_s \to f_{2}'(1525)(\to K^+\,K^-)\,l^+ \,l^-$ decays.}
\label{tab_sm1}
\end{table}

%%%%%%%%%%%%%%%%%%%%%%%%%%%%%%%%%
\begin{table}[htbp]
\centering
\resizebox{\columnwidth}{!}{
\begin{tabular}{|c|c|c|c|c|c|c|c|c|c|c|c|}
\hline
\hline
\multirow{2}{*}{$q^{2}$ bins (GeV$^2$)}
&\multicolumn{2}{c|}{$\langle P_1 \rangle$} &\multicolumn{2}{c|}{$\langle P_2 \rangle$} &\multicolumn{2}{c|}{$\langle P^{\prime}_4 \rangle$} &\multicolumn{2}{c|}{$\langle P^{\prime}_5 \rangle$}\\
\cline{2-9}
&$e$ mode&$\mu$ mode&$e$ mode&$\mu$ mode&$e$ mode&$\mu$ mode&$e$ mode&$\mu$ mode \\
    \hline
    \hline
    [0.10, 0.98]   & $-0.008 \pm  0.265$ & $-0.008 \pm  0.267$ & $0.132 \pm  0.024$ & $0.158 \pm  0.029$ & $-0.468 \pm  0.086$ & $-0.457 \pm  0.089$ & $0.554 \pm  0.103$ & $0.593 \pm  0.114$ \\
    \hline
    [1.1, 2.5]     & $-0.043 \pm  0.197$ & $-0.043 \pm  0.197$ & $0.373 \pm  0.080$ & $0.378 \pm  0.081$ & $0.248 \pm  0.234$ & $0.251 \pm  0.234$ & $-0.076 \pm  0.263$ & $-0.079 \pm  0.267$ \\
    \hline
    [2.5, 4.0]     & $-0.112 \pm  0.240$ & $-0.112 \pm  0.240$ & $-0.046 \pm  0.157$ & $-0.046 \pm  0.158$ & $0.810 \pm  0.197$ & $0.811 \pm  0.197$ & $-0.616 \pm  0.235$ & $-0.620 \pm  0.236$ \\
    \hline
    [4.0, 6.0]     & $-0.159 \pm  0.282$ & $-0.159 \pm  0.282$ & $-0.314 \pm  0.081$ & $-0.315 \pm  0.082$ & $0.995 \pm  0.156$ & $0.995 \pm  0.156$ & $-0.794 \pm  0.186$ & $-0.797 \pm  0.187$ \\
     \hline
%    [6.0, 8.0]     & $-0.211 \pm  0.279$ & $-0.211 \pm  0.279$ & $-0.412 \pm  0.048$ & $-0.413 \pm  0.048$ & $1.067 \pm  0.138$ & $1.067 \pm  0.138$ & $-0.837 \pm  0.168$ & $-0.840 \pm  0.169$ \\
%     \hline
    [1.1, 6.0]     & $-0.120 \pm  0.221$ & $-0.121 \pm  0.222$ & $-0.095 \pm  0.128$ & $-0.098 \pm  0.128$ & $0.735 \pm  0.188$ & $0.739 \pm  0.188$ & $-0.546 \pm  0.219$ & $-0.552 \pm  0.220$ \\
     \hline
    [0.045, 6.0]   & $-0.060 \pm  0.179$ & $-0.074 \pm  0.175$ & $0.004 \pm  0.067$ & $-0.003 \pm  0.085$ & $0.307 \pm  0.171$ & $0.405 \pm  0.183$ & $-0.167 \pm  0.185$ & $-0.238 \pm  0.204$ \\
     \hline
\hline
\end{tabular}}
\caption{The central values and the corresponding $1\sigma$ uncertainties of various angular observables such as
$\langle P_1 \rangle$, $\langle P_2 \rangle$, $\langle P^{\prime}_4 \rangle$, $\langle P^{\prime}_5 \rangle$
for both $e$ mode and $\mu$ mode of $B_s \to f_{2}'(1525)(\to K^+\,K^-)\,l^+ \,l^-$ decays.}
\label{tab_sm2}
\end{table}

%%%%%%%%%%%%%%%%%%%%%%%%%%%%%%%%%%%%%%%%%%%
\begin{table}[htbp]
\centering
%\resizebox{\columnwidth}{!}{
\begin{tabular}{|c|c|c|c|c|c|c|c|}
\hline
\hline
     $q^{2}$ bins (GeV$^2$) & $\langle R \rangle$ & $\langle Q_1 \rangle$ & $\langle Q_2 \rangle$ & $\langle Q^{\prime}_{4} \rangle$ & $\langle Q^{\prime}_{5} \rangle$ & $\langle Q_{A_{FB}} \rangle$ & $\langle Q_{F_{L}} \rangle$\\
    \hline
    \hline
    [0.10, 0.98]         & $0.979 \pm  0.005$ & $0.000 \pm  0.003$  & $0.026 \pm  0.005$  & $0.011 \pm  0.005$ & $0.039 \pm  0.011$ & $ -0.010 \pm  0.002 $ & $ 0.001 \pm  0.001 $\\
     \hline
    [1.1, 2.5]           & $0.994 \pm  0.005$ & $-0.000 \pm  0.001$ & $0.005 \pm  0.002$  & $0.002 \pm  0.001$ & $-0.002 \pm  0.004$ & $ -0.002 \pm  0.001 $ & $ 0.002 \pm  0.001 $\\
     \hline
    [2.5, 4.0]           & $0.995 \pm  0.005$ & $-0.000 \pm  0.000$ & $-0.001 \pm  0.001$ & $0.000 \pm  0.000$ & $-0.004 \pm  0.002$ & $ 0.000 \pm  0.001 $ & $ 0.002 \pm  0.001 $\\
     \hline
    [4.0, 6.0]           & $0.996 \pm  0.003$ & $-0.000 \pm  0.000$ & $-0.001 \pm  0.001$ & $0.000 \pm  0.000$ & $-0.004 \pm  0.001$ & $ 0.001 \pm  0.000 $ & $ 0.002 \pm  0.000 $ \\
     \hline
%    [6.0, 8.0]           & $0.997 \pm  0.003$ & $-0.000 \pm  0.000$ & $-0.001 \pm  0.000$ & $0.000 \pm  0.000$ & $-0.003 \pm  0.001$ & $ 0.001 \pm  0.000 $ & $ 0.001 \pm  0.000 $ \\
%     \hline
    [1.1, 6.0]           & $0.995 \pm  0.002$ & $-0.001 \pm  0.002$ & $-0.003 \pm  0.001$ & $0.004 \pm  0.001$ & $-0.007 \pm  0.002$ & $ -0.000 \pm  0.001 $ & $ 0.002 \pm  0.001 $\\
    \hline
    [0.045, 6.0]         & $0.976 \pm  0.005$ & $-0.014 \pm  0.040$ & $-0.007 \pm  0.019$ & $0.098 \pm  0.015$ & $-0.071 \pm  0.021$ & $ -0.004 \pm  0.001 $ & $ 0.012 \pm  0.002 $\\
     \hline 
\hline
\end{tabular}
\caption{The central values and the corresponding $1\sigma$ uncertainties of various LFUV sensitive observables such as
the ratio of branching ratio $\langle R_{f_{2}^{\prime}} \rangle$, $\langle Q_{i}^{(\prime)} \rangle$, $\langle Q_{F_L} \rangle$, $\langle Q_{A_{FB}} \rangle$
for $B_s \to f_{2}'(1525)(\to K^+\,K^-)\,l^+ \,l^-$ decays.}
\label{tab_sm3}
\end{table}

\begin{figure}[htbp]
\centering
\includegraphics[width=8.9cm,height=6.0cm]{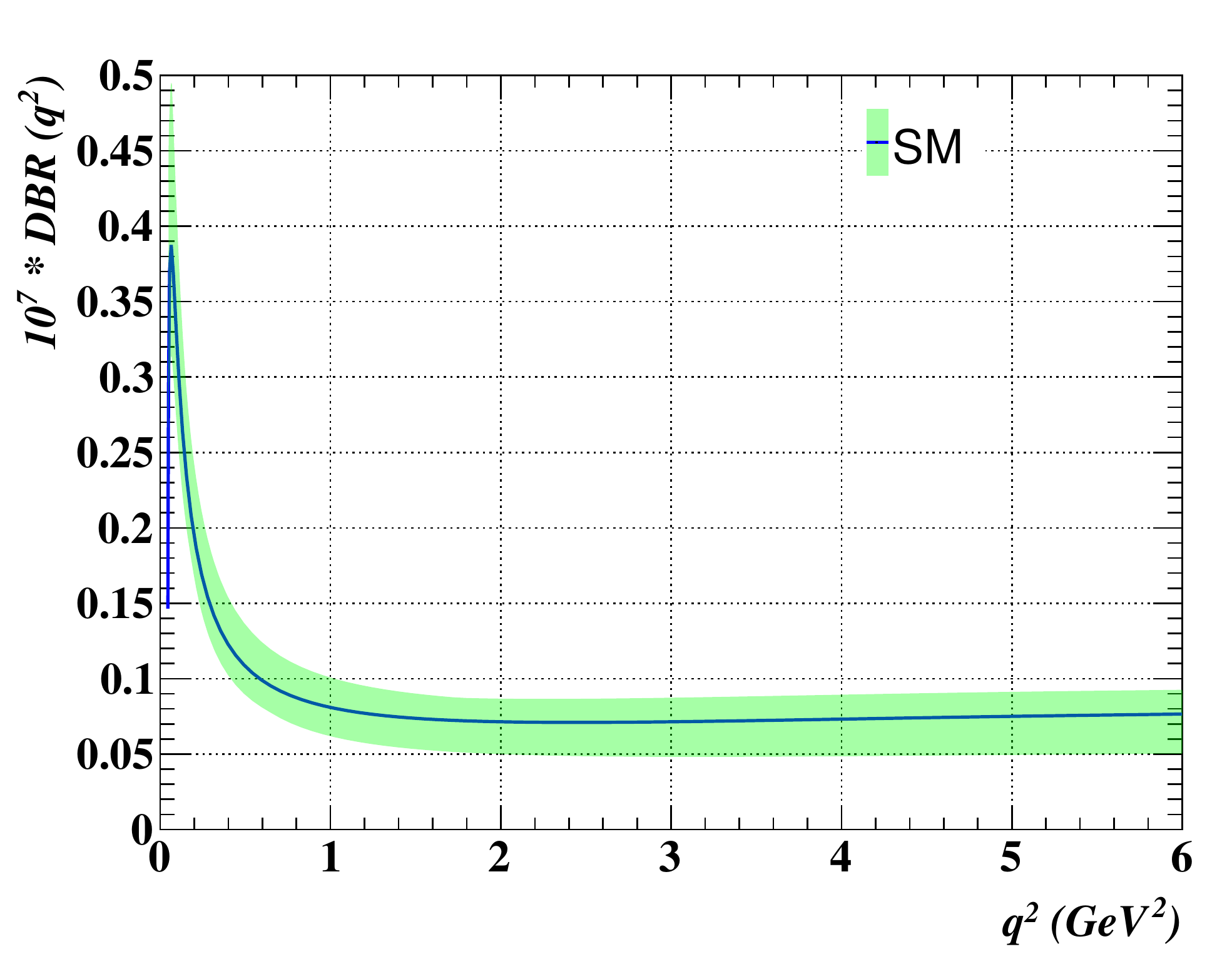}
\includegraphics[width=8.9cm,height=6.0cm]{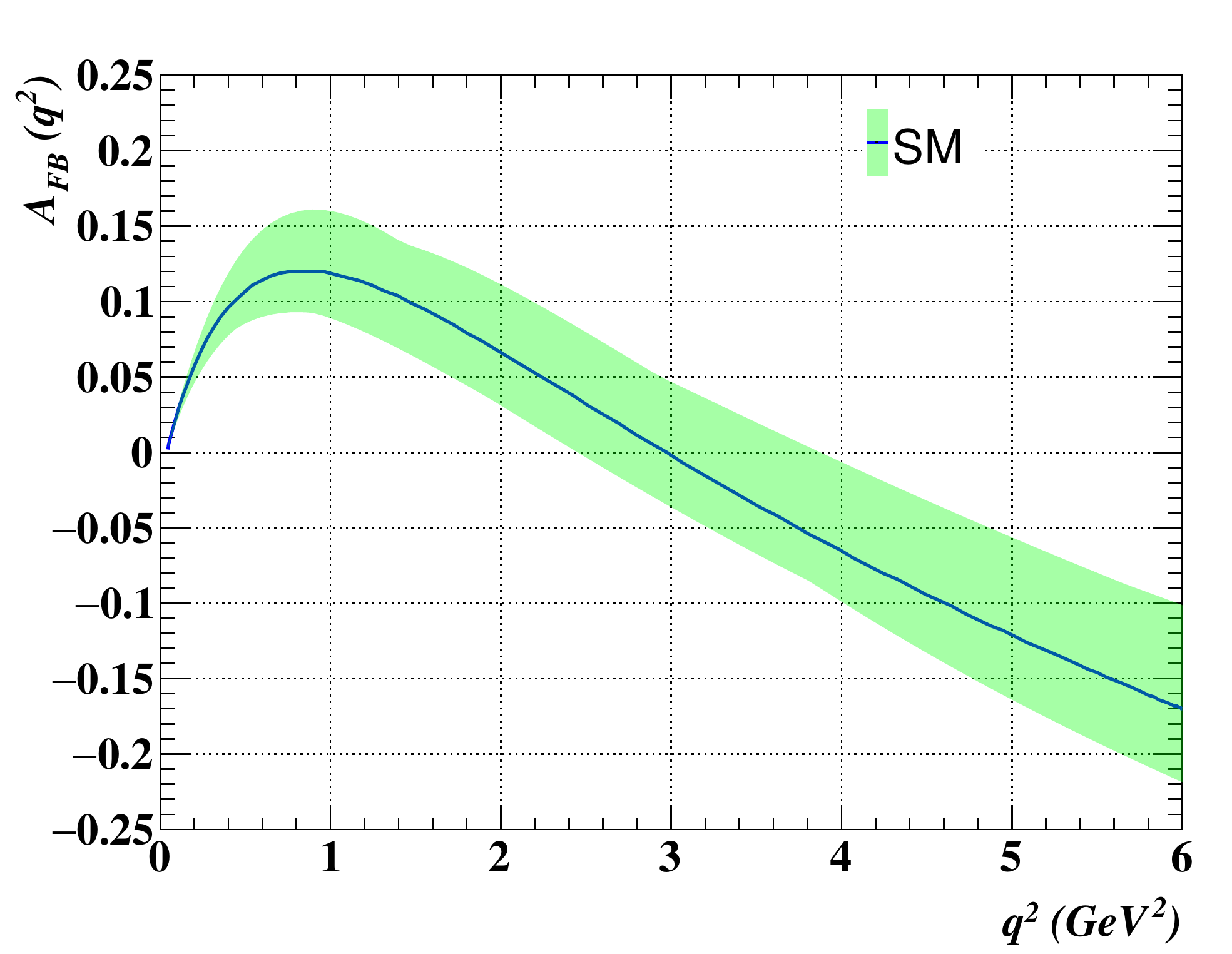}
\includegraphics[width=8.9cm,height=6.0cm]{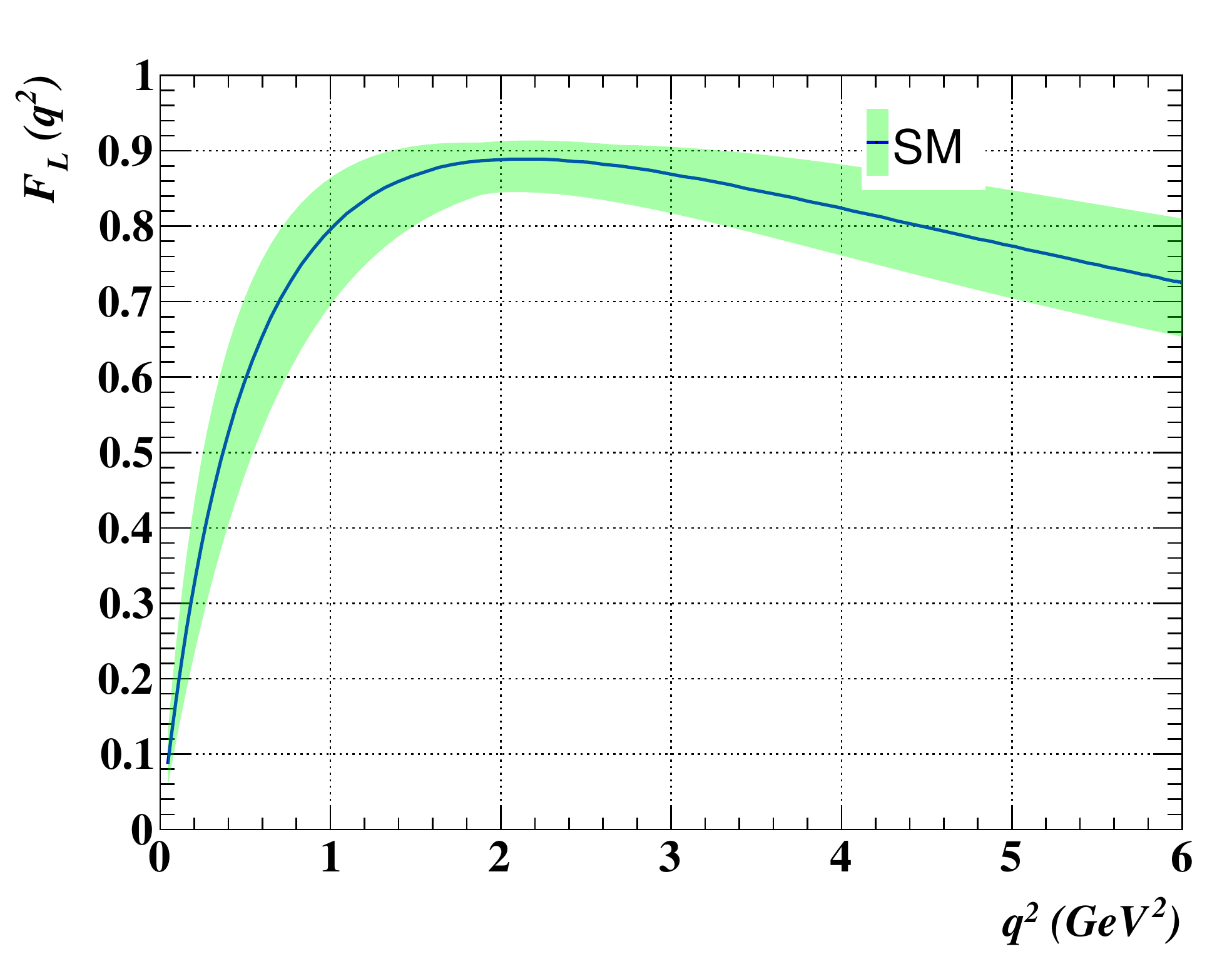}
\includegraphics[width=8.9cm,height=6.0cm]{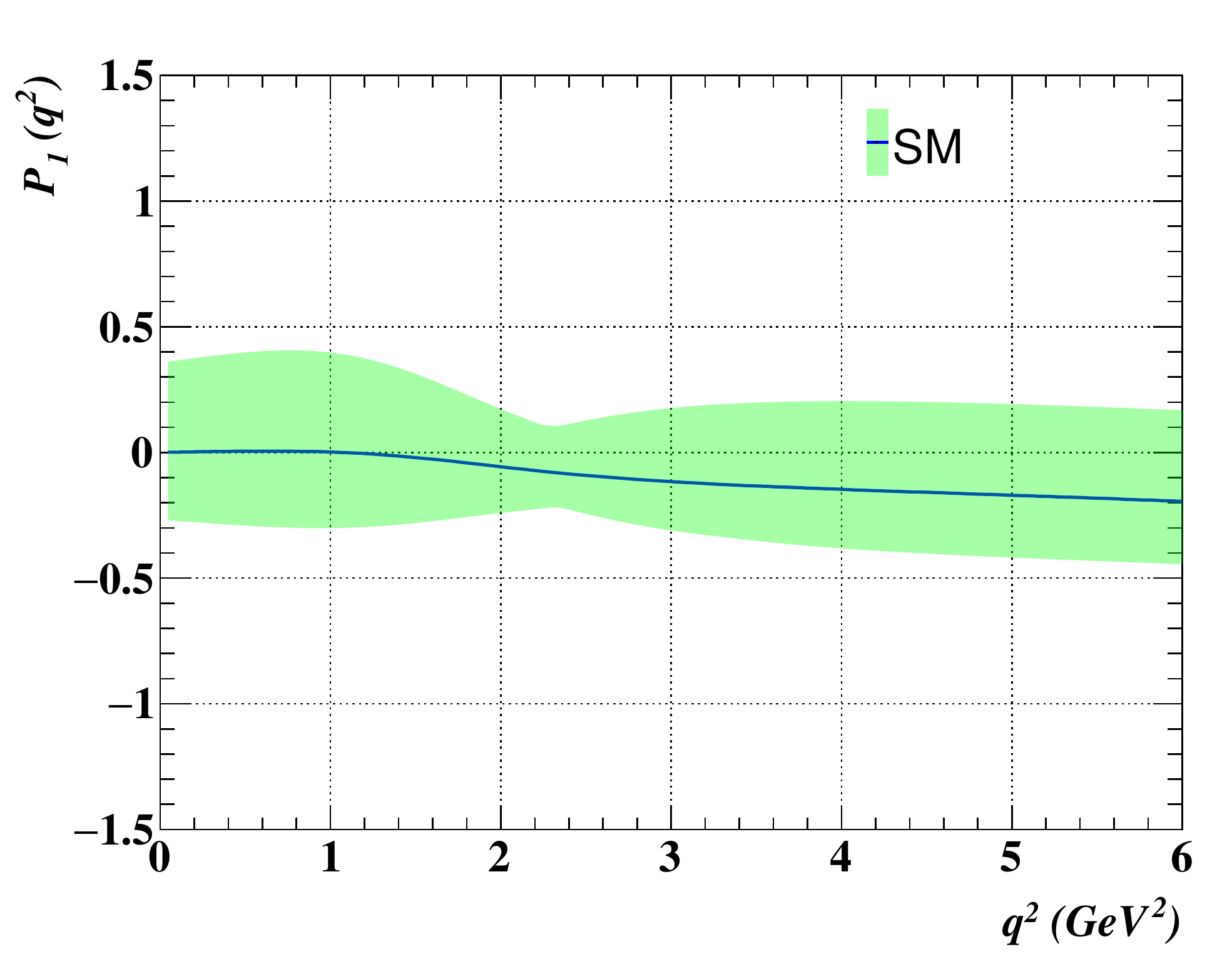}
\includegraphics[width=8.9cm,height=6.0cm]{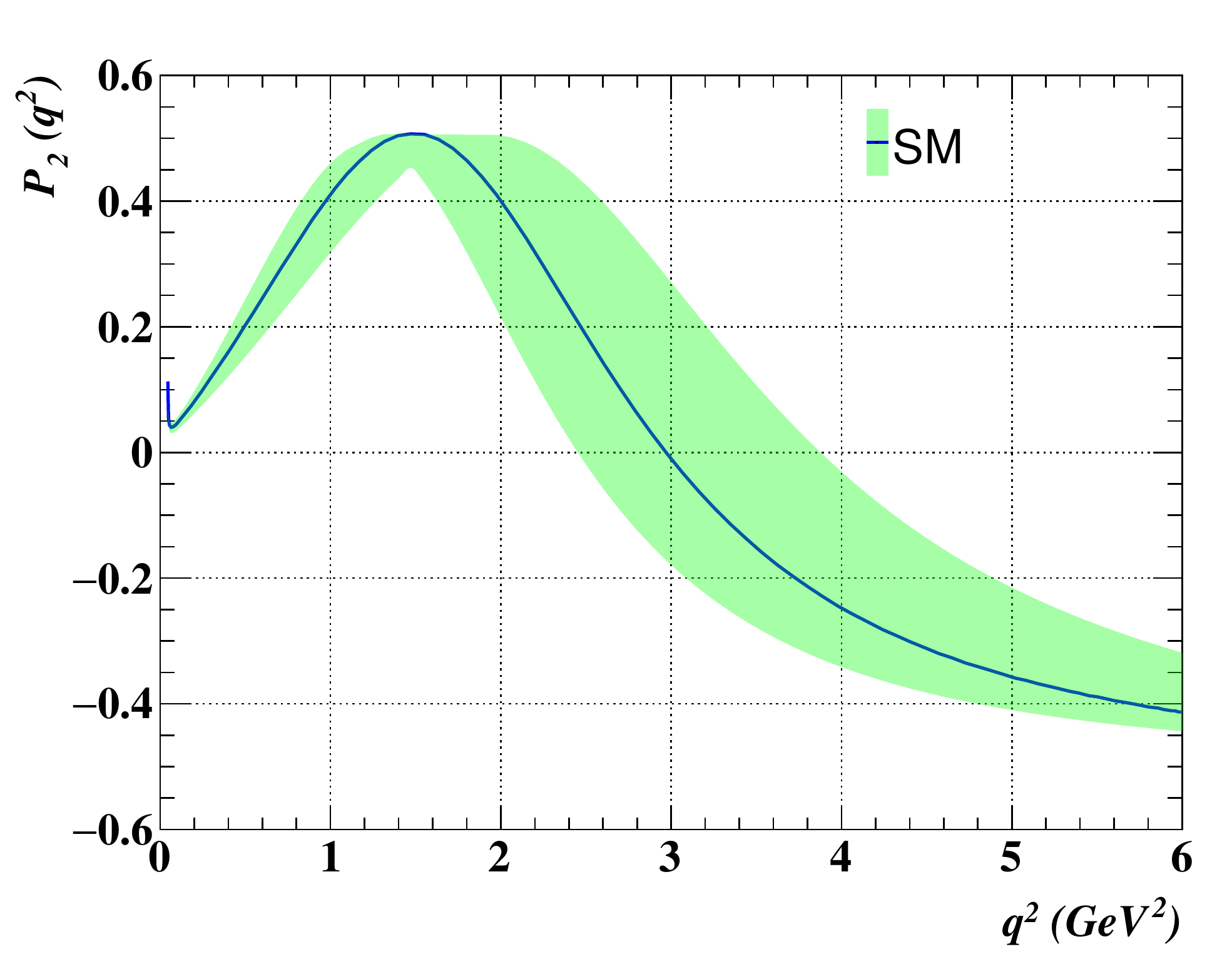}
\includegraphics[width=8.9cm,height=6.0cm]{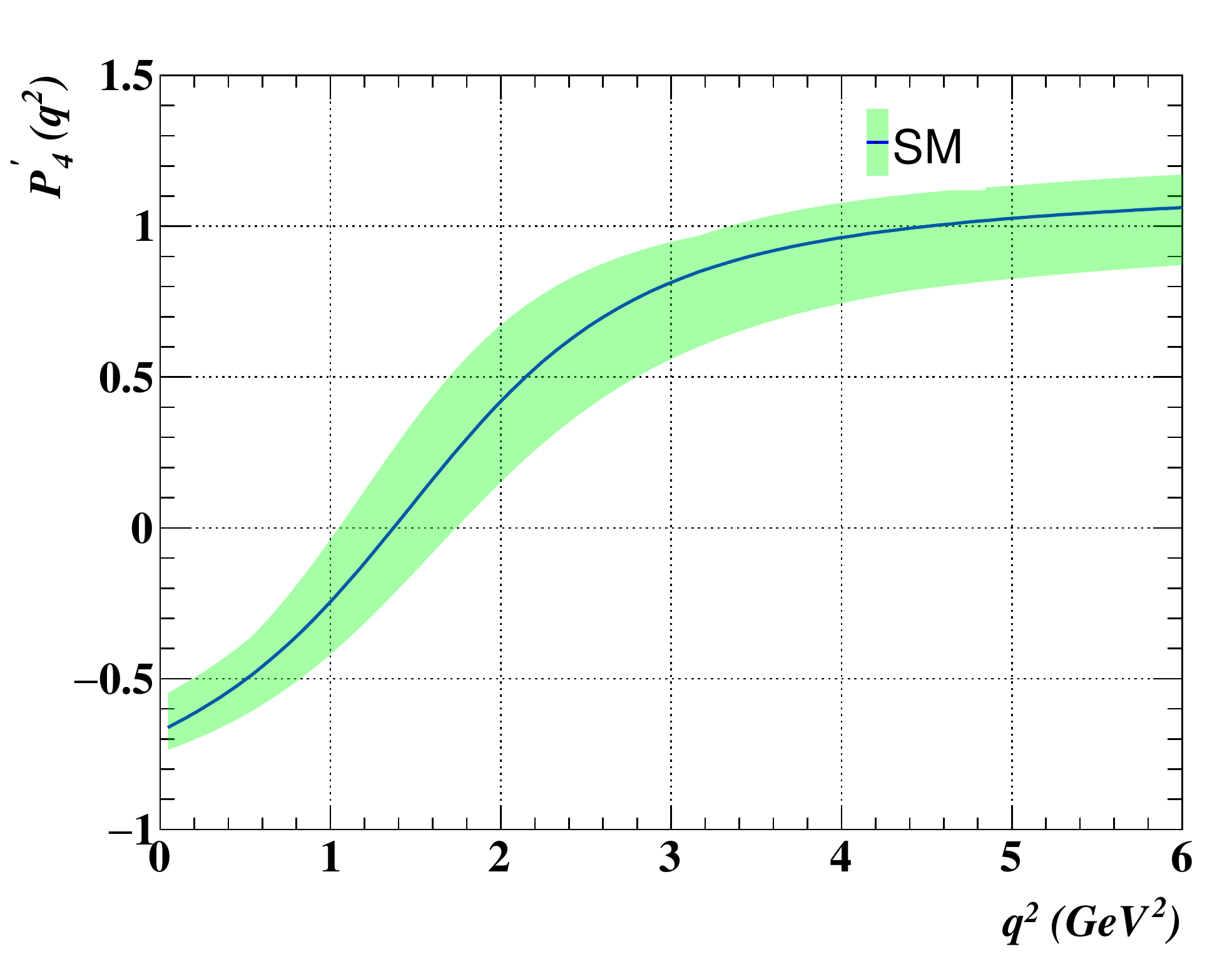}
\includegraphics[width=8.9cm,height=6.0cm]{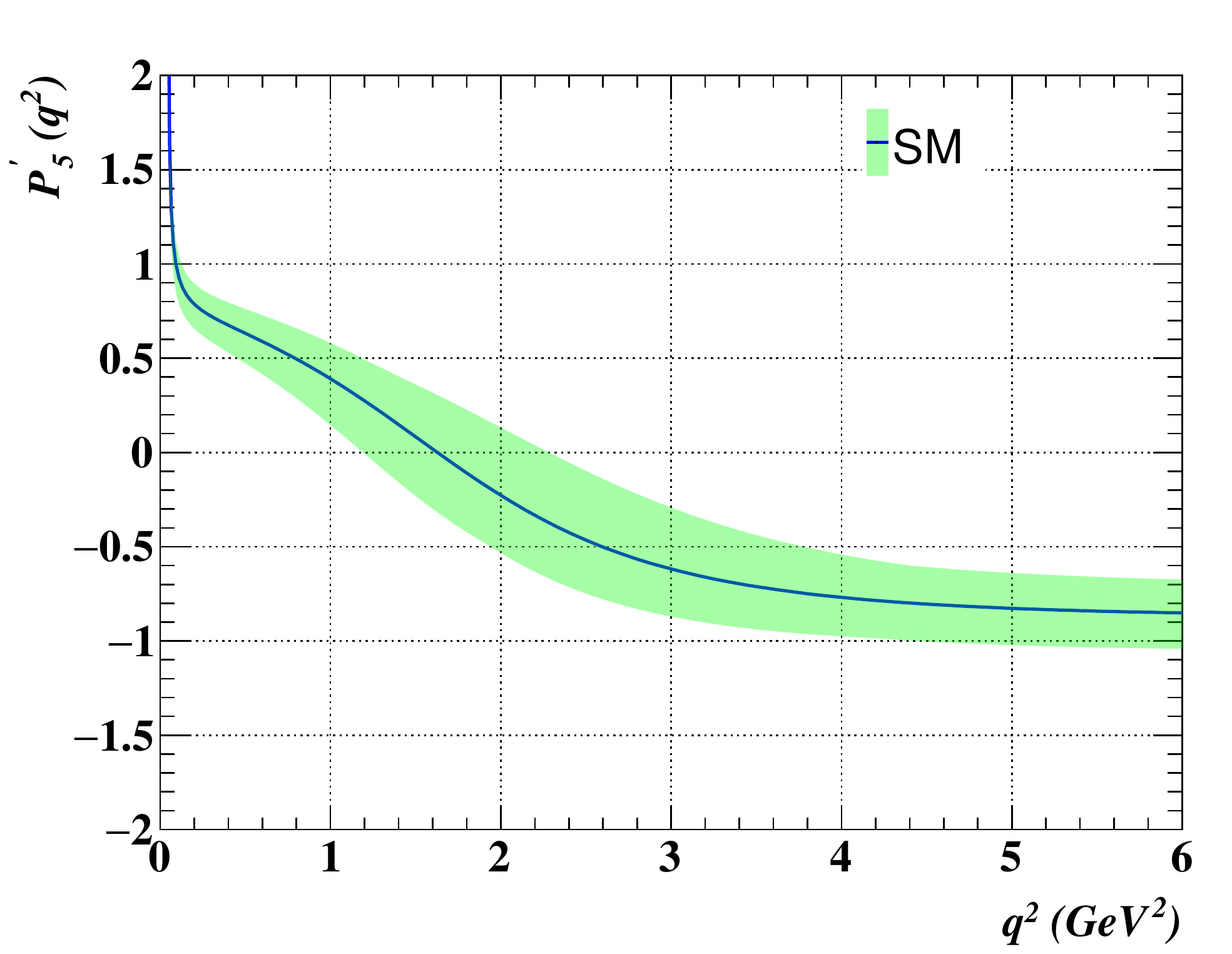}
\caption{The $q^2$ distribution of various observables for the $B_s \to f_{2}'(1525)(\to K^+\,K^-)\,\mu^+ \,\mu^-$ decays in the SM. The band
corresponds to the uncertainties in the input parameters such as the $B_s \to f_2^{\prime}$ transition form factors, CKM matrix element and 
$\mathcal{B}(f_{2}^{\prime} \to K^+\,K^-)$.}
\label{fig_sm}
\end{figure}

We show in Fig.~\ref{fig_sm} the $q^2$ distribution of various observables in the low dilepton invariant mass region $q^2\in[0.045,6.0]{\rm GeV^2}$. 
The central line corresponds to the central values of each input parameters whereas to obtain the uncertainty band, we employ a naive 
$\chi^2$ test on the input parameters. We define $\chi^2$ as
\begin{equation}
 \chi^2=\sum_i \frac{\left(\mathcal{O}_i - \mathcal{O}_i^C\right)^2}{\Delta_{i}^2}\,,
\end{equation}
where $\mathcal{O}_i \in \Big(F(0), a, b, |V_{tb}V^*_{ts}|, \mathcal{B}(f_{2}^{\prime} \to K^+\,K^-)\Big)$ and $\mathcal{O}_i^C$ 
represent the central values of each input parameters. Here $\Delta_{i}$ represent the respective uncertainties
associated with each input parameters. To obtain the uncertainty in each observable, we impose $\chi^2 \le 7.43$ constraint.  
It is important to note that we observe zero crossing in the $q^2$ distribution of $A_{FB}(q^2)$, $P_2(q^2)$, $P^{\prime}_4 (q^2)$ and 
$P^{\prime}_5 (q^2)$.
Interestingly, the $A_{FB}(q^2)$ and $P_2(q^2)$ have same
zero crossing points i.e., at $q^2\sim 3^{+0.8}_{-0.6} \,{\rm GeV^2}$. Similarly, the $P^{\prime}_4 (q^2)$ and $P^{\prime}_5 (q^2)$ 
have the zero crossing points at around $q^2 \sim 1.4 \pm 0.3\,{\rm GeV^2}$ and $q^2 \sim 1.6 \pm 0.4\,{\rm GeV^2}$, respectively. 
Value of $P_1 (q^2)$ is almost zero in the low $q^2$ region and 
becomes negative at higher $q^2$ regions. The uncertainties associated with $P^{(\prime)}_i (q^2)$ observables are more compared to 
$DBR(q^2)$, $F_{L}(q^2)$, and $A_{FB}(q^2)$. 
The ratio of branching ratio $R_{f_{2}^{\prime}}(q^2)$ is almost equal to $\sim 1$ in the whole $q^2$ region and the 
uncertainty associated with $R_{f_{2}^{\prime}}(q^2)$ is quite negligible in comparison to the uncertainties present in other observables.

\subsection{New Physics}
In order to explain the anomalies present in $b\,\to\,s\,l^+\,l^-$ transition decays, various global fits have been performed by several 
groups~\cite{Capdevila:2017bsm,Altmannshofer:2017yso,DAmico:2017mtc,Hiller:2017bzc,Geng:2017svp,Ciuchini:2017mik,Celis:2017doq,Alok:2017sui,
Alok:2017jgr,Ciuchini:2019usw}. In principle,
the NP can enter the effective Hamiltonian through several NP Lorentz structures such as vector, axial vector, scalar, pseudoscalar 
and tensor operators. 
But few measurements particularly, $B_s\, \to\,\mu^+\,\mu^-$ and $b\,\to\,s\,\gamma$ put severe constraint on the scalar, pseudoscalar and 
tensor NP Lorentz structures~\cite{Alok:2010zd,Alok:2011gv,Bardhan:2017xcc} and hence they are omitted from our analysis.
We refer to Ref.~\cite{Alok:2019ufo} for the global fit results that are performed on the new Wilson coefficients by considering 
${C}_{9,10}^{NP}$ and ${C}_{9,10}^{\prime}$. In particular, these NP operators have V-A structure. 
The authors perform a global fit to these Wilson coefficients by using the constraints coming from 
observables such as $R_K$, $R_{K^*}$, $P_{5}^{\prime}$ and $\mathcal{B}({B_{s} \to \phi\, \mu^+\, \mu^-})$.  
In addition, the fits also include the constraints coming from the branching ratio of $B_s\, \to\, \mu^+\,\mu^-$,
the differential branching ratio of $B^0\,\to K^{0*}\,\mu^+\,\mu^-$, $B^+\,\to K^{+*}\,\mu^+\,\mu^-$, $B^0\,\to K^{0}\,\mu^+\,\mu^-$,
$B^+\,\to K^{+}\,\mu^+\,\mu^-$ and $B\,\to X_s\,\mu^+\,\mu^-$ in several $q^2$ bins and also the constraints from the 
angular observables in $B^0\,\to K^{0*}\,\mu^+\,\mu^-$ and $B_{s}^{0}\,\to \phi\,\mu^+\,\mu^-$ decays in the several $q^2$ bins.
All the omitted details can be found in Ref.~\cite{Alok:2019ufo}. Out of various $1D$ and $2D$ scenarios, we
consider total seven NP scenarios that are having high $\Delta \chi^2$ values: four from $1D$ scenarios and three from $2D$ scenarios. 
We give bin wise predictions as well as the $q^2$ distributions of various observables and make a comparative study
among different NP scenarios and the SM for the $B_s \to f_{2}'(1525)(\to K^+\,K^-)\,l^+ \,l^-$ decay mode. 
The best fit values of the NP Wilson coefficients pertinent for our analysis taken from Ref.~\cite{Alok:2019ufo} are reported in 
Table~\ref{tab_newwc}.

\begin{table}[htbp]
\centering
\setlength{\tabcolsep}{5pt} % Default value: 6pt
\renewcommand{\arraystretch}{1.5} % Default value:
\begin{tabular}{|c|c|c|c|c|c|c|c|}
\hline
Wilson coefficients & ${C}_{9}^{NP}$ & ${C}_{10}^{NP}$ & ${C}_{9}^{NP}=-{C}_{10}^{NP}$ & ${C}_{9}^{NP}=-{C}_{9}^{\prime}$ & 
(${C}_{9}^{NP},{C}_{10}^{NP}$) & (${C}_{9}^{NP}=-{C}_{9}^{\prime}$) & (${C}_{9}^{NP}=-{C}_{10}^{\prime}$) \\
\hline
Best fit values & $-1.07$ & $+0.78$ & $-0.52$ & $-1.11$ & ($-0.94, +0.23$) & ($-1.27, +0.68$) & ($-1.36, -0.46$) \\
\hline 
\hline
\end{tabular}
\caption{Best fit values of NP Wilson coefficients~\cite{Alok:2019ufo}}
\label{tab_newwc}
\end{table}

\subsubsection{New Physics: 1D scenario}
Let us now discuss the four 1D NP scenarios that arises due to contributions coming from ${C}_{9}^{NP}$, ${C}_{10}^{NP}$, 
${C}_{9}^{NP}=-{C}_{10}^{NP}$ and ${C}_{9}^{NP}=-{C}_{9}^{\prime}$. The ${C}_{9,10}^{NP}$ new Wilson coefficients are associated with 
similar interactions as that of ${C}_{9,10}$ SM Wilson coefficients whereas, ${C}_{9,10}^{\prime}$ new Wilson coefficients 
arises due to the right chiral currents which are basically absent in the SM. 
We report in the Appendix in Tables~\ref{tab_npdbr},~\ref{tab_npfl},~\ref{tab_npafb},~\ref{tab_npp1},~\ref{tab_npp2},~\ref{tab_npp4},
~\ref{tab_npp5} the average values of various observables such as
the $BR$, $\langle F_L \rangle$, $\langle A_{FB} \rangle$, $\langle P_1 \rangle$, $\langle P_2 \rangle$, 
$\langle P^{\prime}_4 \rangle$, $\langle P^{\prime}_5 \rangle$ 
for the $\mu$ mode in several $q^2$ bins. The corresponding bin wise plots have been displayed in Fig.~\ref{fig_np1dbin}. 
Our observations are as follows:

\begin{figure}[htbp]
\centering
 \includegraphics[width=8.9cm,height=6.0cm]{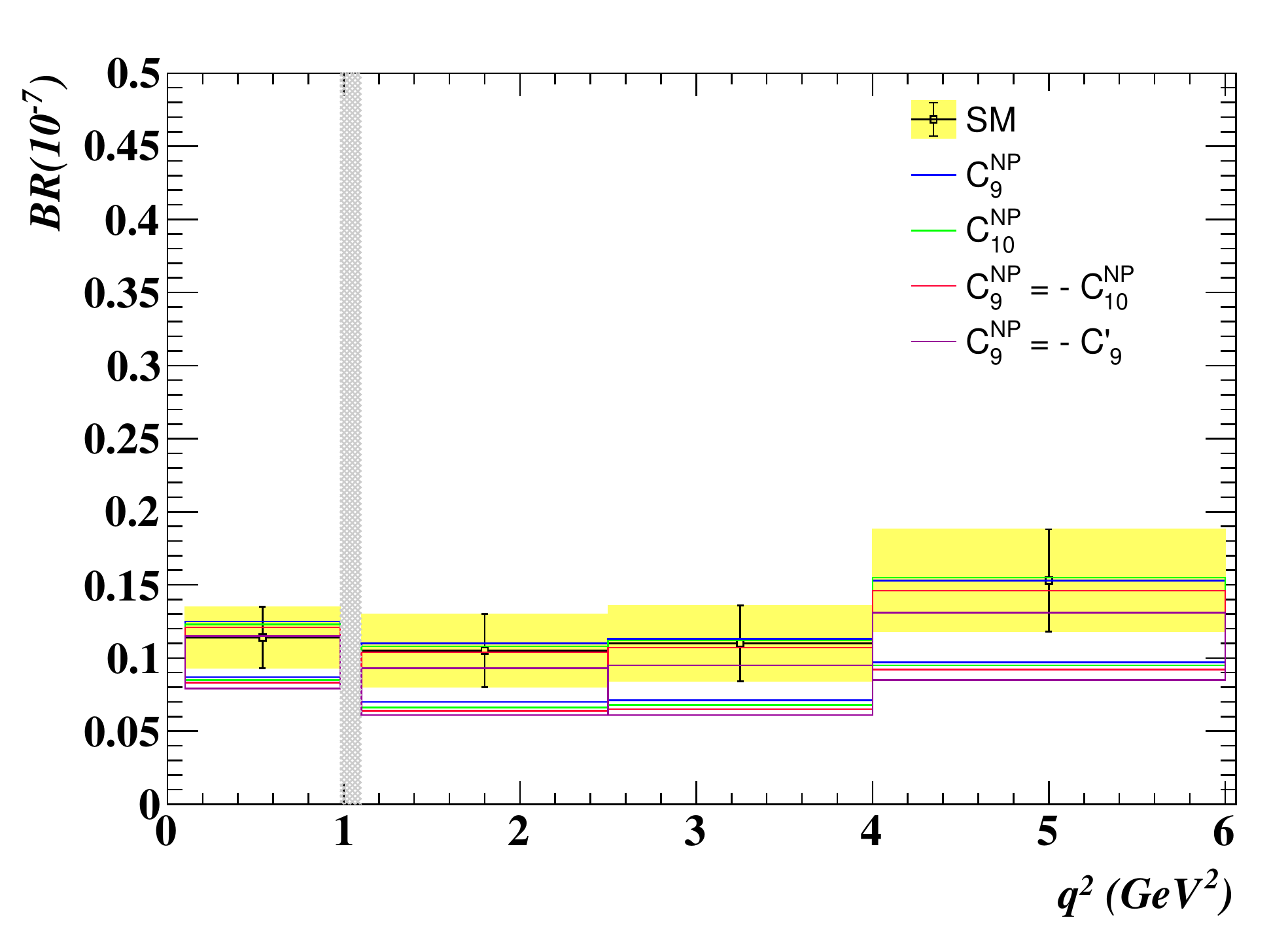}
 \includegraphics[width=8.9cm,height=6.0cm]{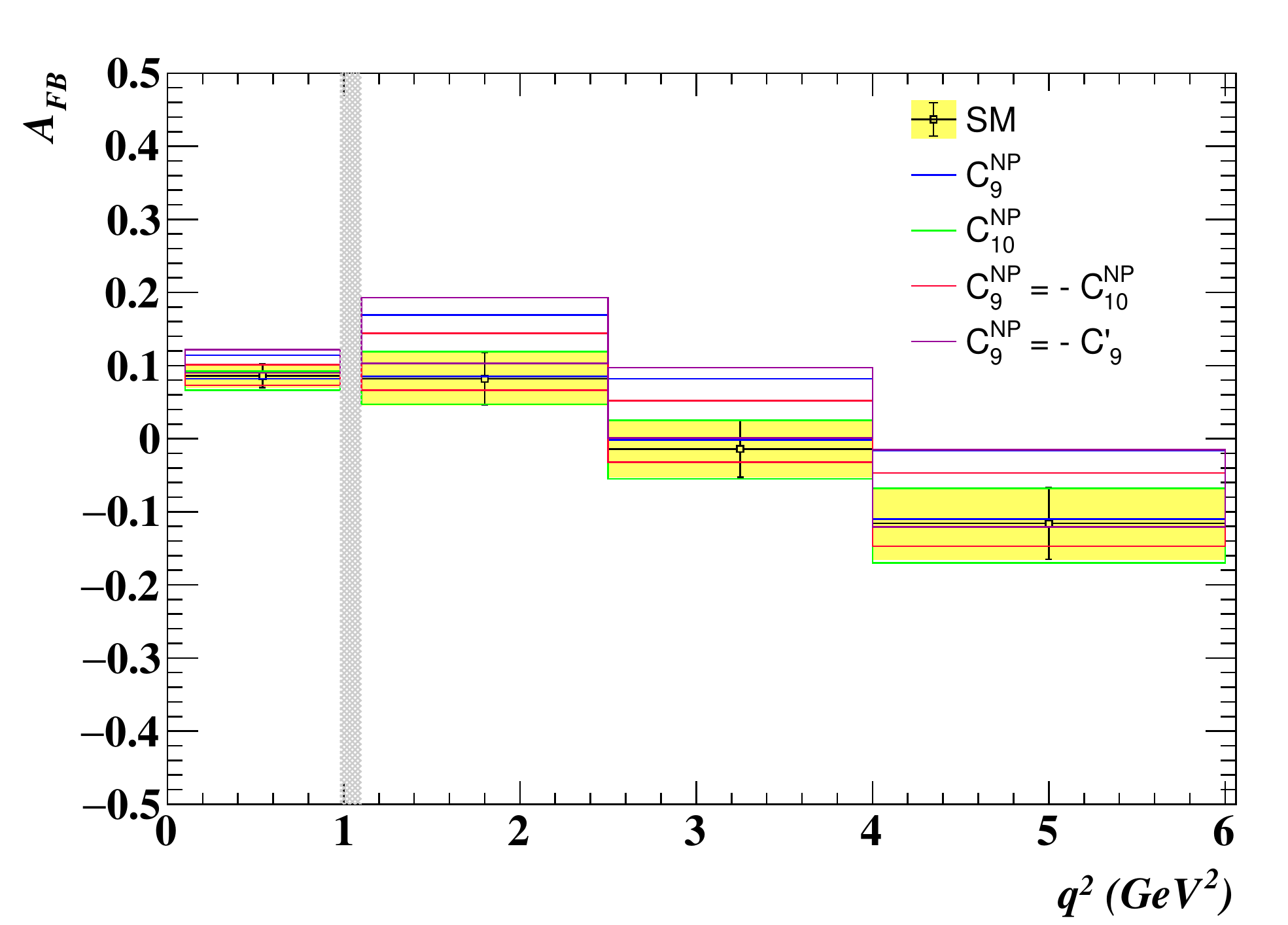}
 \includegraphics[width=8.9cm,height=6.0cm]{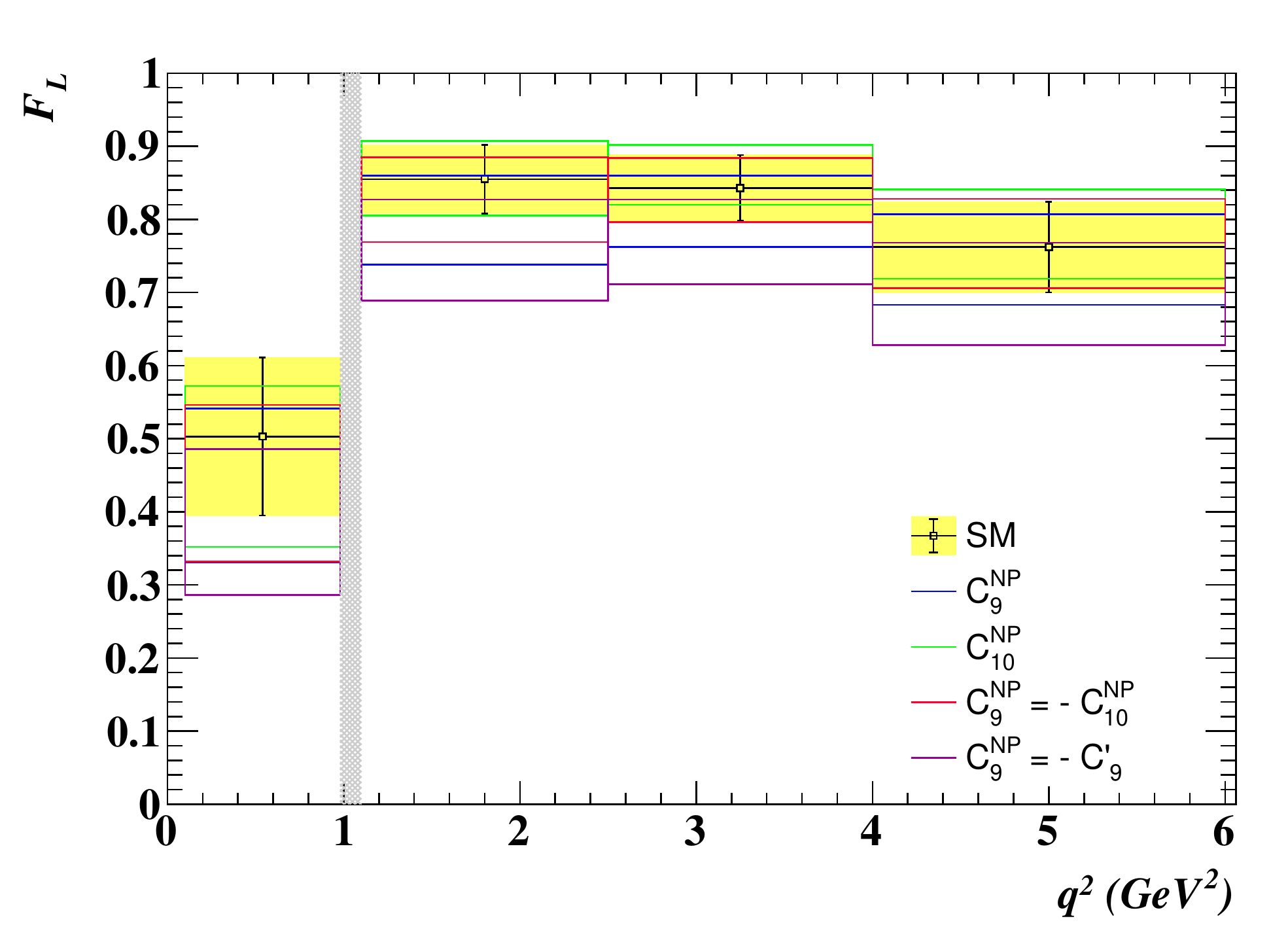}
 \includegraphics[width=8.9cm,height=6.0cm]{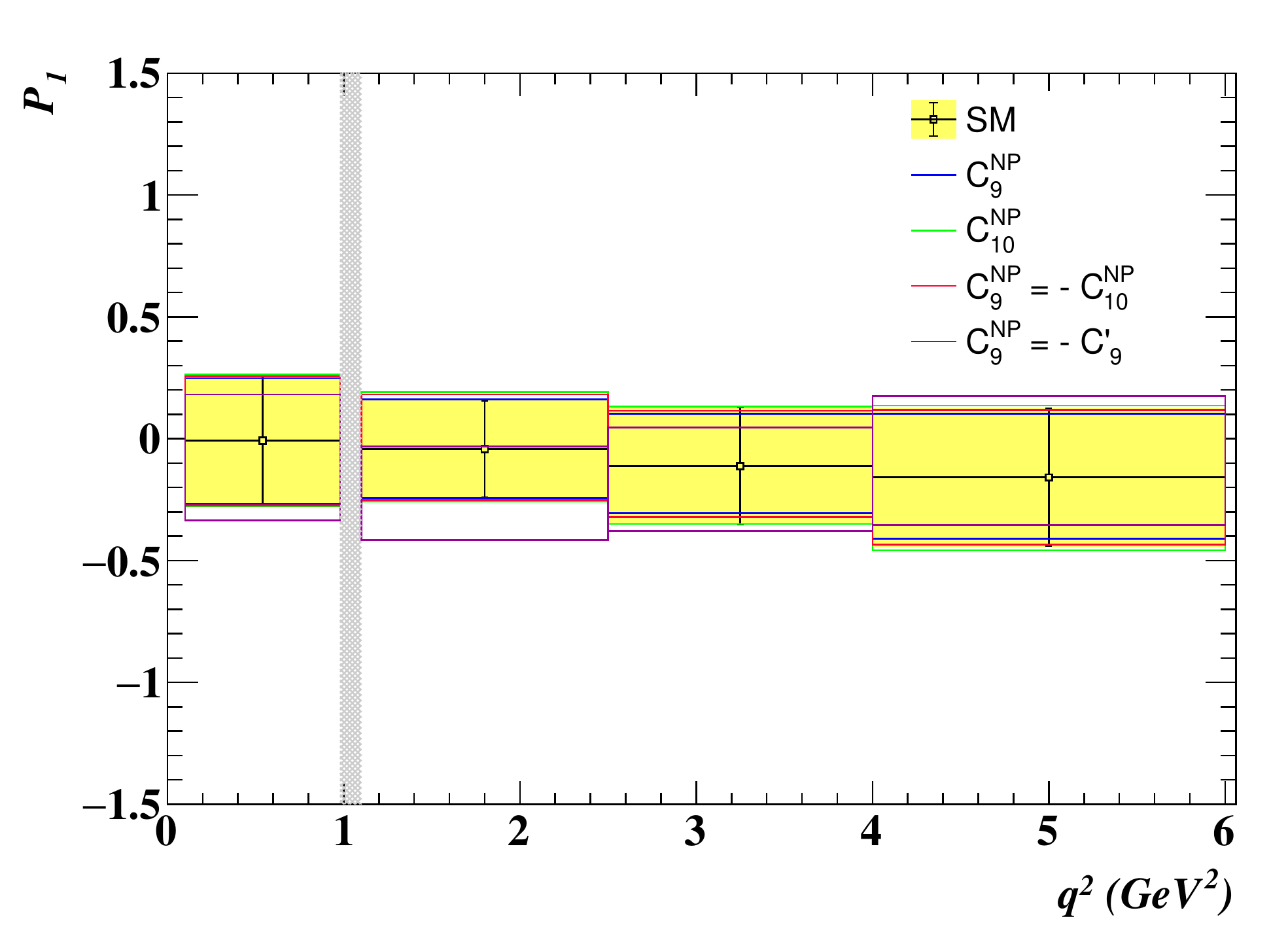}
 \includegraphics[width=8.9cm,height=6.0cm]{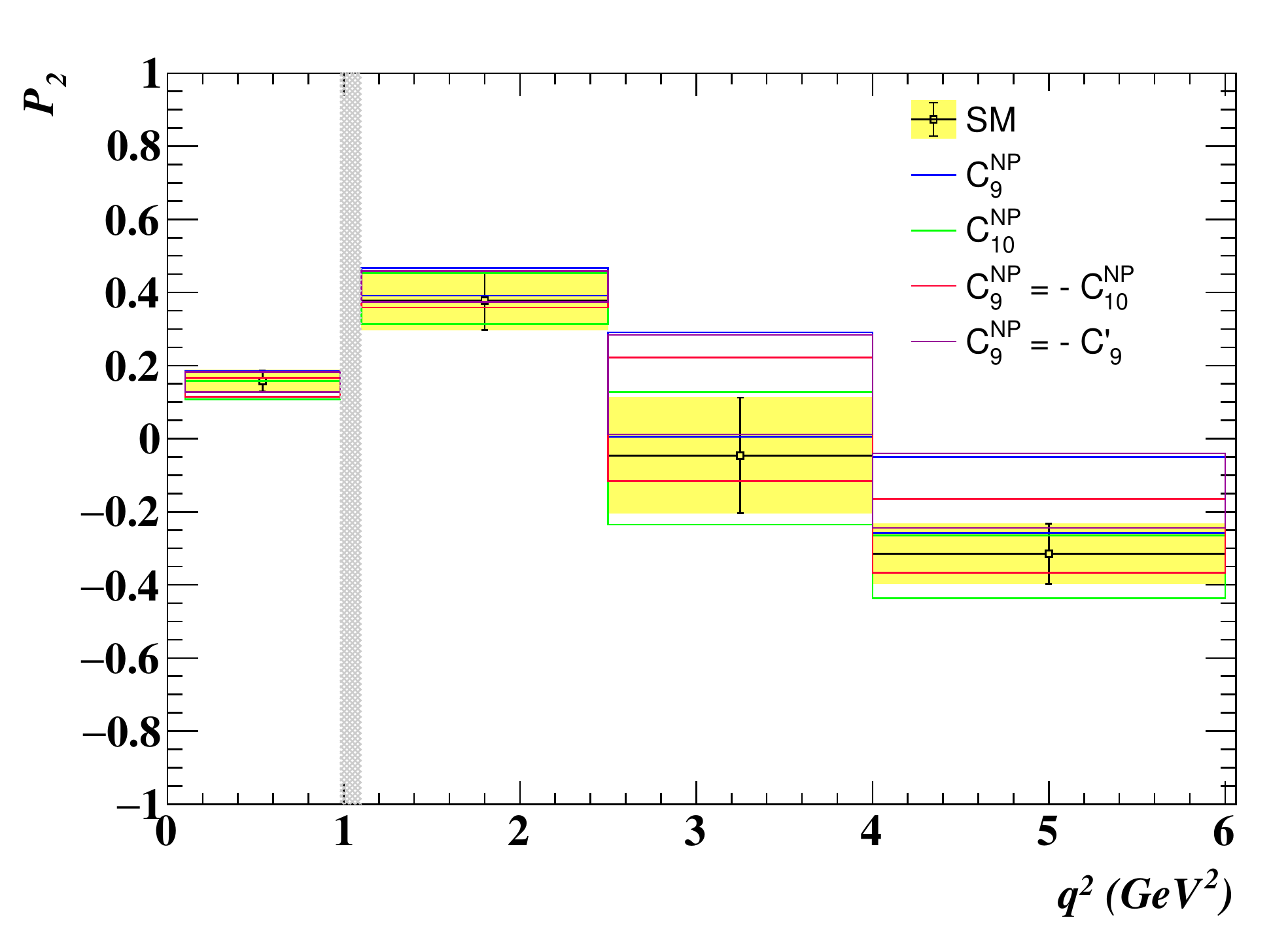}
 \includegraphics[width=8.9cm,height=6.0cm]{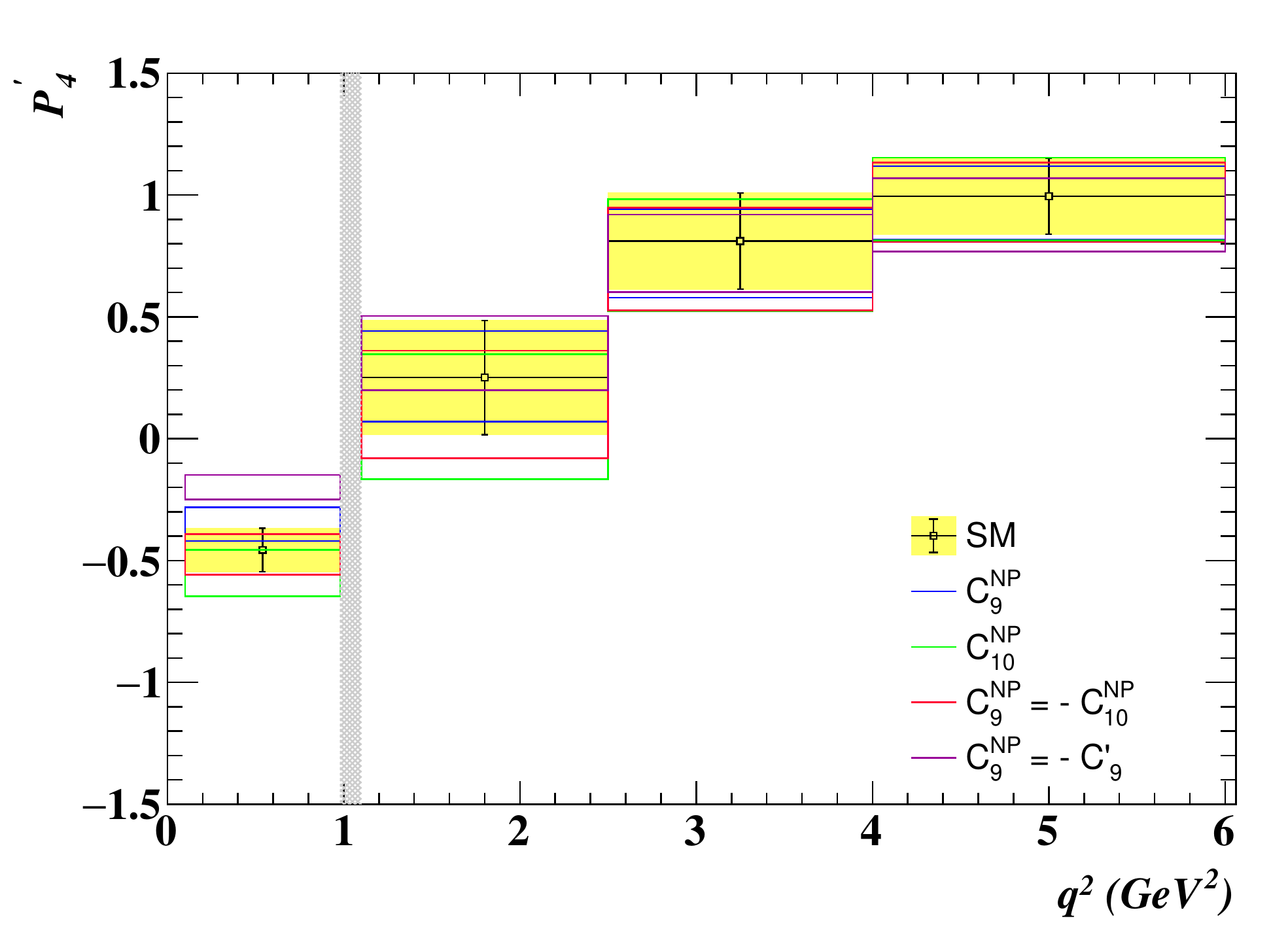}
 \includegraphics[width=8.9cm,height=6.0cm]{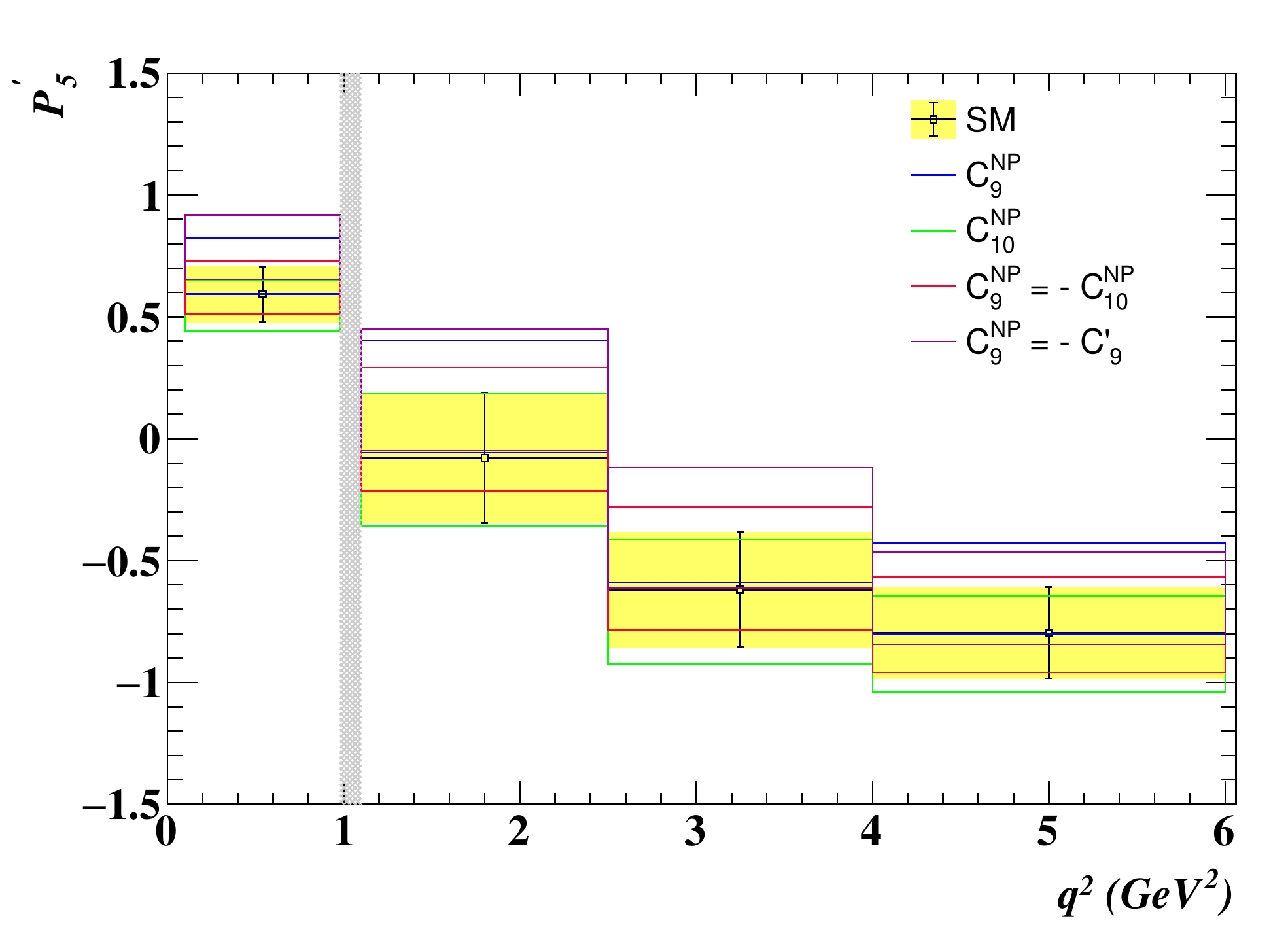}
\caption{The central values and the corresponding $1\sigma$ error bands of various observables such as the branching ratio, 
the longitudinal polarization fraction $F_L$, the forward-backward asymmetry $A_{FB}$, and $P_1$, $P_2$, $P^{\prime}_4$, $P^{\prime}_5$ 
for the $B_s \to f_{2}'(1525)(\to K^+\,K^-)\,\mu^+ \,\mu^-$ decays in 
several $q^2$ bins in the SM and in the presence of four 1D NP scenarios.}
\label{fig_np1dbin}
\end{figure}

\begin{itemize}
 \item $BR$: In the first bin [0.045, 0.98], although the central values of all the NP scenarios
 differ slightly from the SM, they all lie within the SM $1\sigma$ error band.
 In the bins [1.1, 2.5], [2.5, 4.0] and [4.0, 6.0],
 although the central values differ from the SM prediction, no significant deviations are observed, whereas, 
 the central value obtained in case of ${C}_{9}^{NP}=-{C}_{9}^{\prime}$ NP scenario deviates by $1-1.3\sigma$ from the SM expectations. This 
is true for the larger bin [1.1, 6.0] as well. 
 
 \item $F_L$: In the bin [0.045, 0.98], a deviation of around $1\sigma$ from the SM prediction is observed for the 
${C}_{9}^{NP}=-{C}_{9}^{\prime}$ NP scenario. For the rest of the NP scenarios, the deviation, however, is quite negligible. 
 In the bin [1.1, 2.5], a deviation of around $1.3\sigma$ and $2.2\sigma$ from the SM prediction is observed in case of ${C}_{9}^{NP}$ 
and ${C}_{9}^{NP}=-{C}_{9}^{\prime}$ NP scenarios, respectively.
Similarly, in the bin [2.5, 4.0], the ${C}_{9}^{NP}=-{C}_{9}^{\prime}$ NP scenario shows a deviation of around $1.5\sigma$ from the SM 
prediction. Moreover, in the bin [1.0, 6.0], a deviation of around $1.5\sigma$ from the SM prediction is observed in case of 
${C}_{9}^{NP}=-{C}_{9}^{\prime}$ NP scenario.
 
 \item $A_{FB}$: In the bin [0.045, 0.98], the value of $A_{FB}$ obtained in case of ${C}_{9}^{NP}=-{C}_{9}^{\prime}$ NP scenario lies 
outside the SM $1\sigma$ error band, whereas, for rest of the NP scenarios, it seems to lie within the SM $1\sigma$ error band.
 In the bin [1.1, 2.5], the ${C}_{10}^{NP}$ is exactly like the SM, whereas, ${C}_{9}^{NP}$ and ${C}_{9}^{NP}=-{C}_{9}^{\prime}$ show 
around $1.5\sigma$ and $2\sigma$ deviation from the SM prediction.
 In the bin [2.5, 4.0], a deviation of around $1.4\sigma$ and $1.6\sigma$ is observed in case of ${C}_{9}^{NP}$ and 
${C}_{9}^{NP}=-{C}_{9}^{\prime}$  NP scenarios, whereas, in case of ${C}_{10}^{NP}$, it is exactly like the SM.
 
 \item $P_1$: Although the central values of $P_1$ obtained in each NP scenarios differ from the SM central value, they, however, lie within 
the SM $1\sigma$ error band and hence can not be distinguished from the SM predictions.
   
 \item $P_2$: No significant deviations from the SM prediction are observed in the first two bins i.e., in [0.045, 0.98] and 
[1.1, 2.5]. However, in the bins [2.5, 4.0] and [4.0, 6.0], the deviations observed in case of 
${C}_{9}^{NP}$ and ${C}_{9}^{NP}=-{C}_{9}^{\prime}$ NP scenarios are distinguishable from the SM prediction at the level of $1.3\sigma$ and 
$2\sigma$ significance.
 
 \item $P_{4}^{\prime}$: Although there is slight deviation in case of ${C}_{9}^{NP}$ and ${C}_{10}^{NP}$ NP scenarios, they, however, lie
within the SM $1\sigma$ error band in almost all $q^2$ bins. Similarly, with ${C}_{9}^{NP}=-{C}_{10}^{NP}$, it is exactly SM like. 
With ${C}_{9}^{NP}=-{C}_{9}^{\prime}$ NP scenario, we observe a deviation of around $2.5\sigma$ from the SM expectations in $[0.045, 0.98]$ 
bin which is clearly distinguishable from the SM prediction.
 
 \item $P_{5}^{\prime}$: No significant deviation from the SM prediction is observed. The only exception is ${C}_{9}^{NP}=-{C}_{9}^{\prime}$
NP scenario in which a deviation of around $1\sigma$ from the SM prediction is observed in the $q^2 \in [0.045, 0.98]$ bin. It should be
noted that the value of $P_{5}^{\prime}$ obtained with rest of the NP couplings lies within the SM error band.

\end{itemize}

\begin{figure}[htbp]
\centering
\includegraphics[width=8.9cm,height=6.0cm]{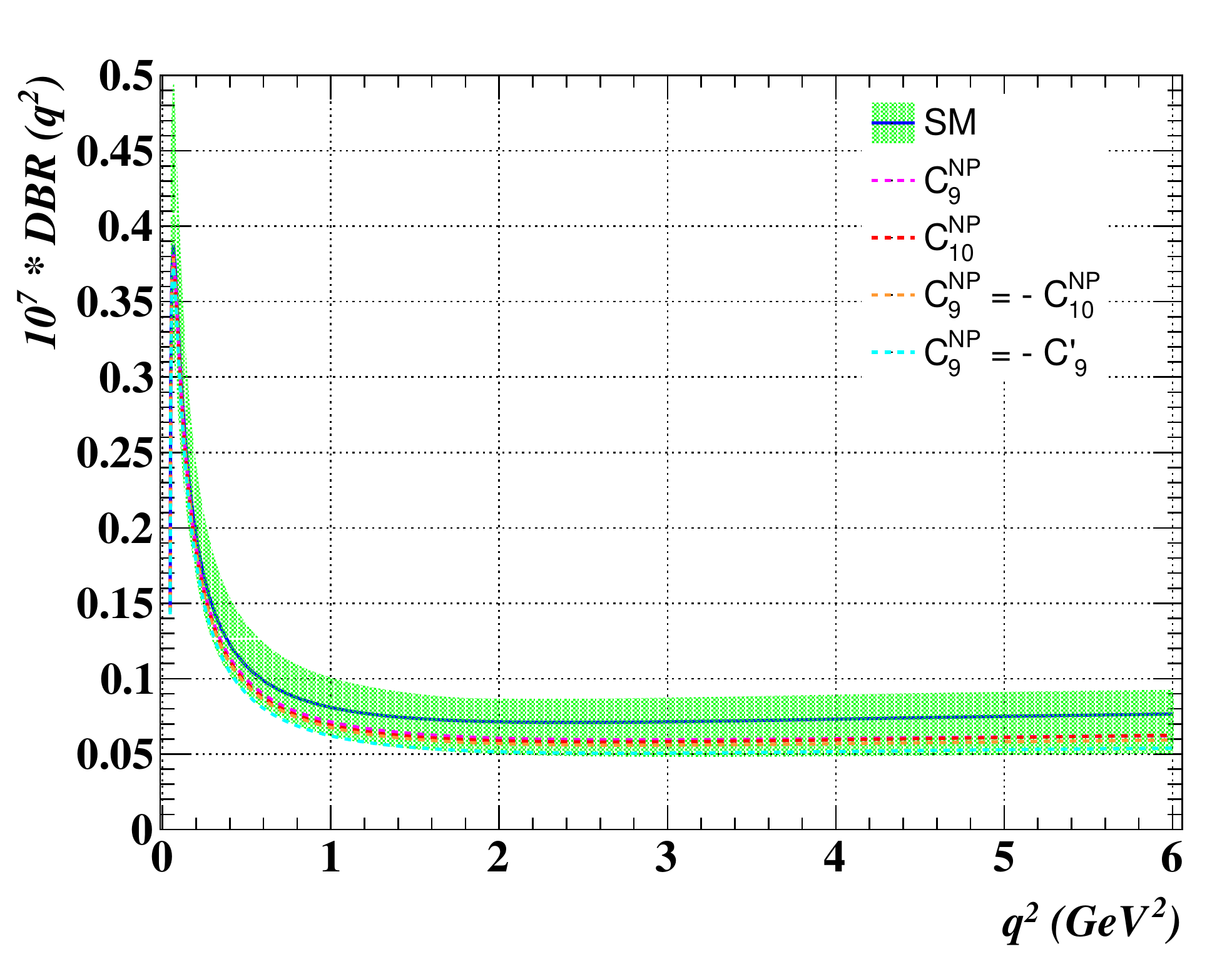}
\includegraphics[width=8.9cm,height=6.0cm]{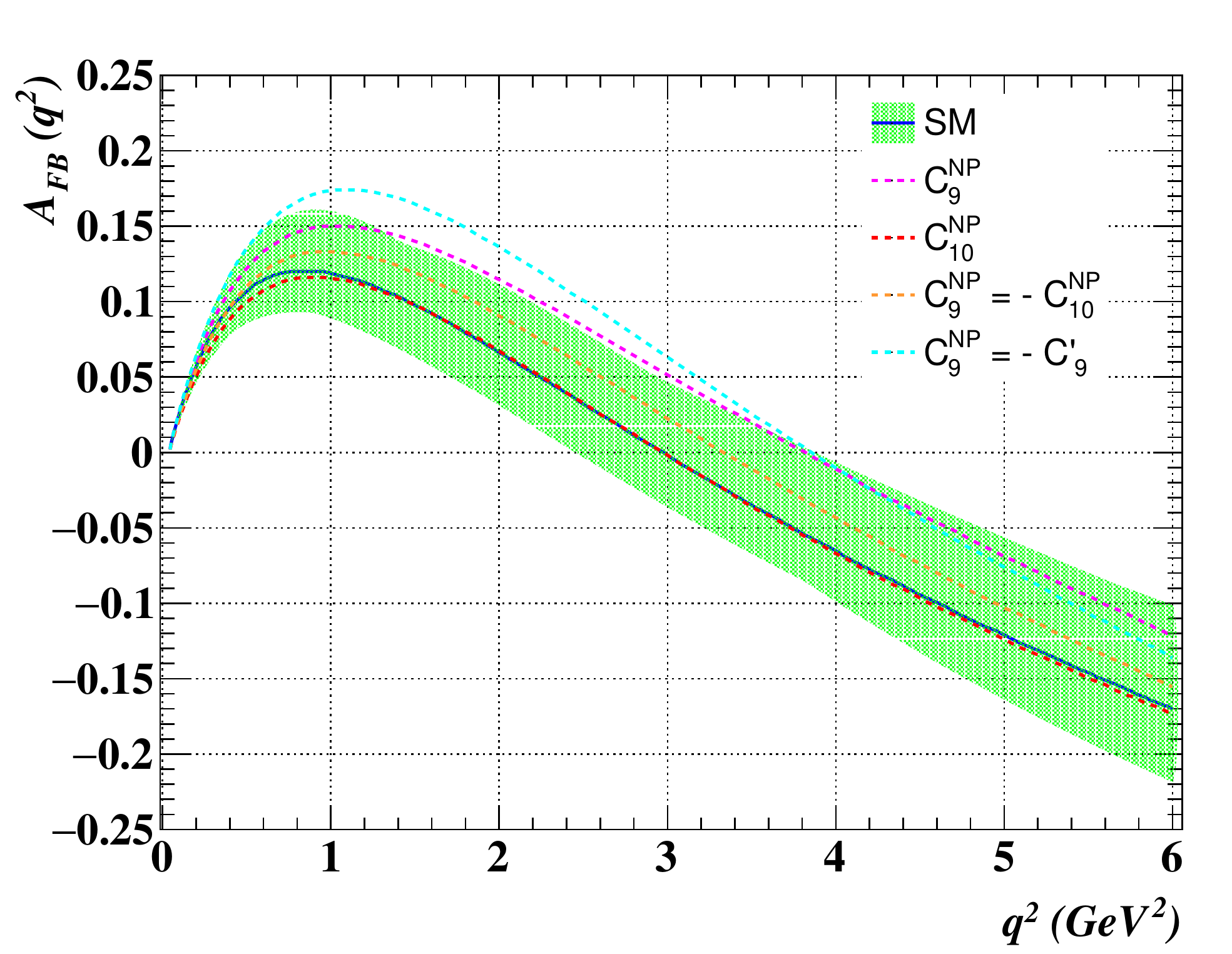}
\includegraphics[width=8.9cm,height=6.0cm]{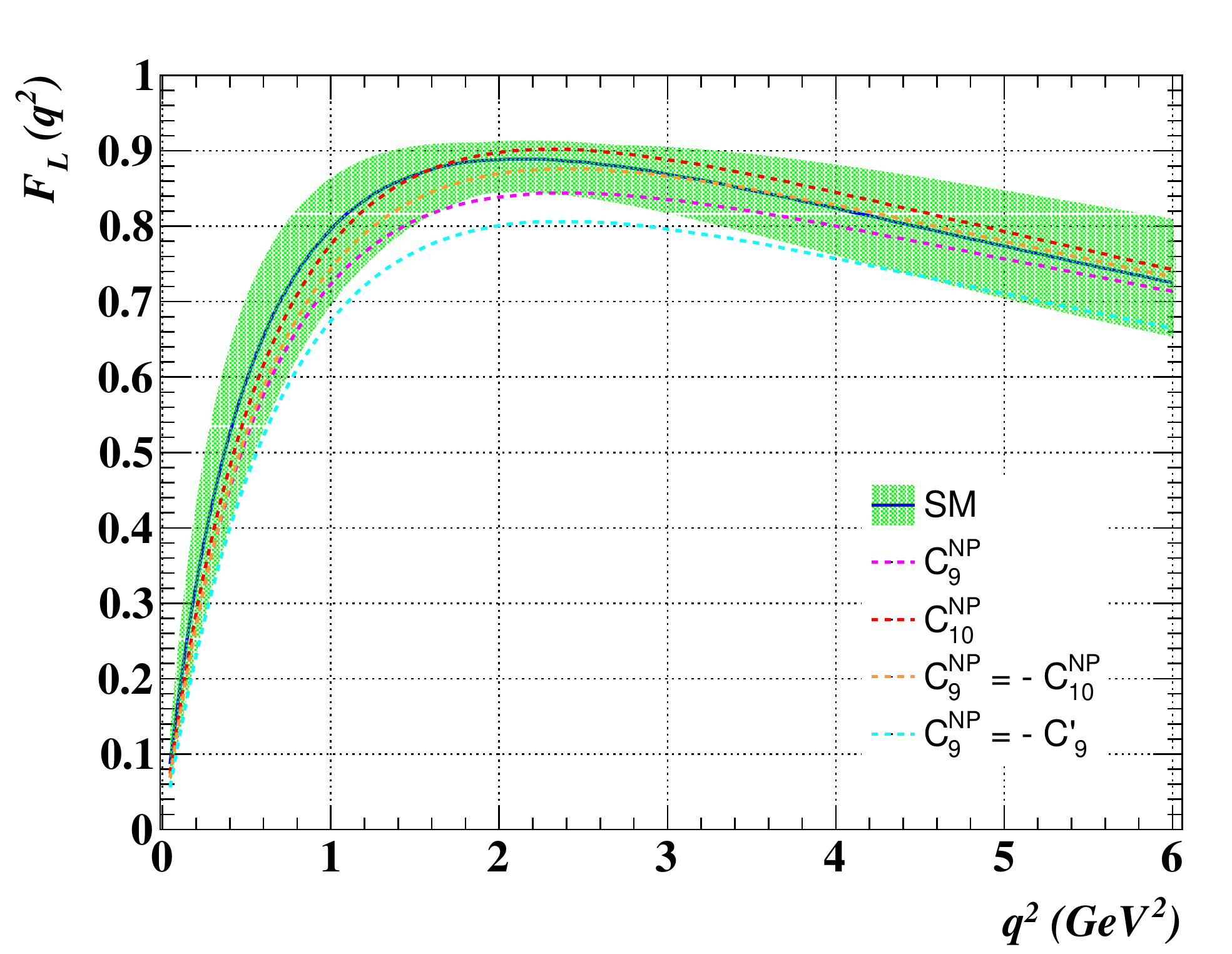}
\includegraphics[width=8.9cm,height=6.0cm]{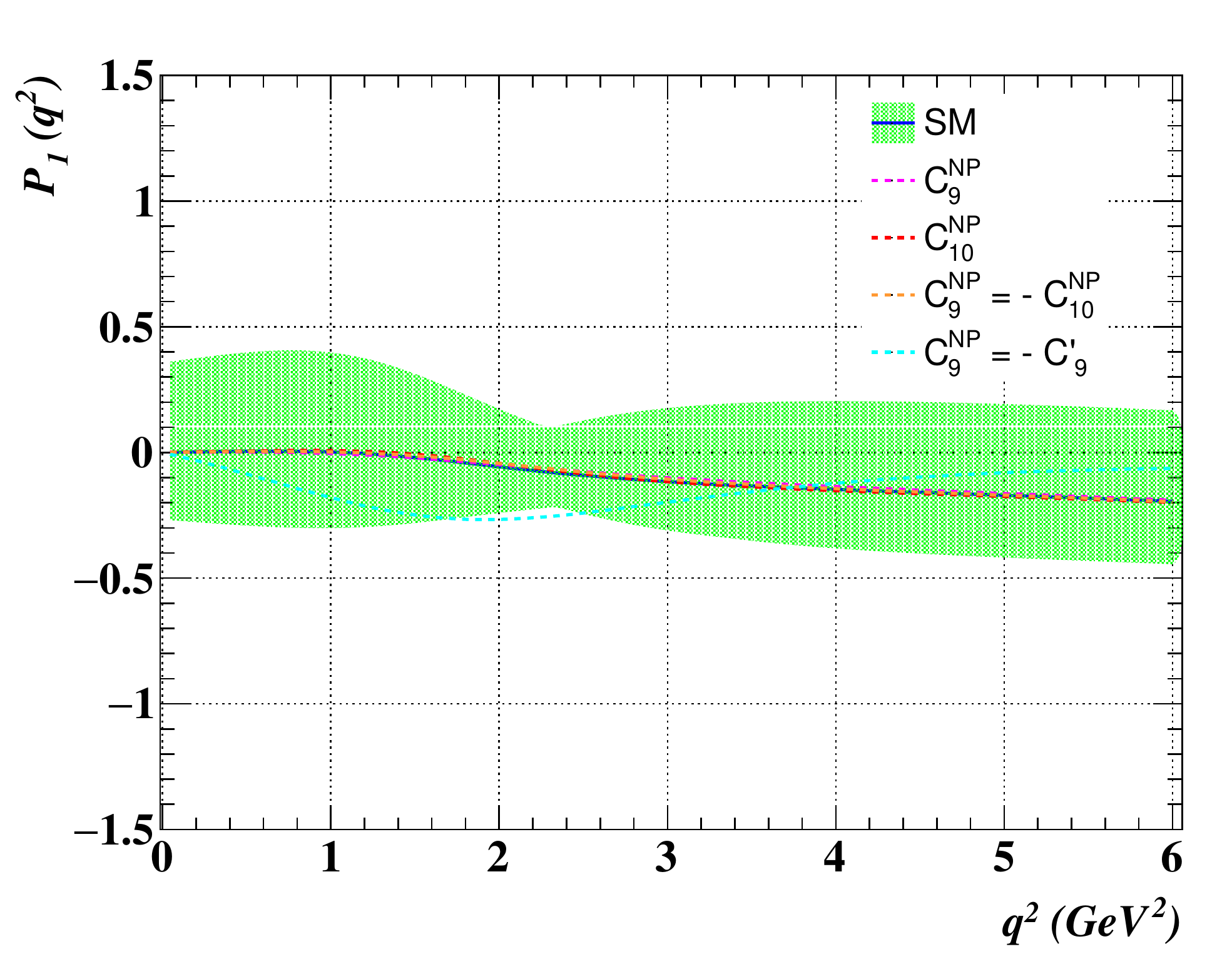}
\includegraphics[width=8.9cm,height=6.0cm]{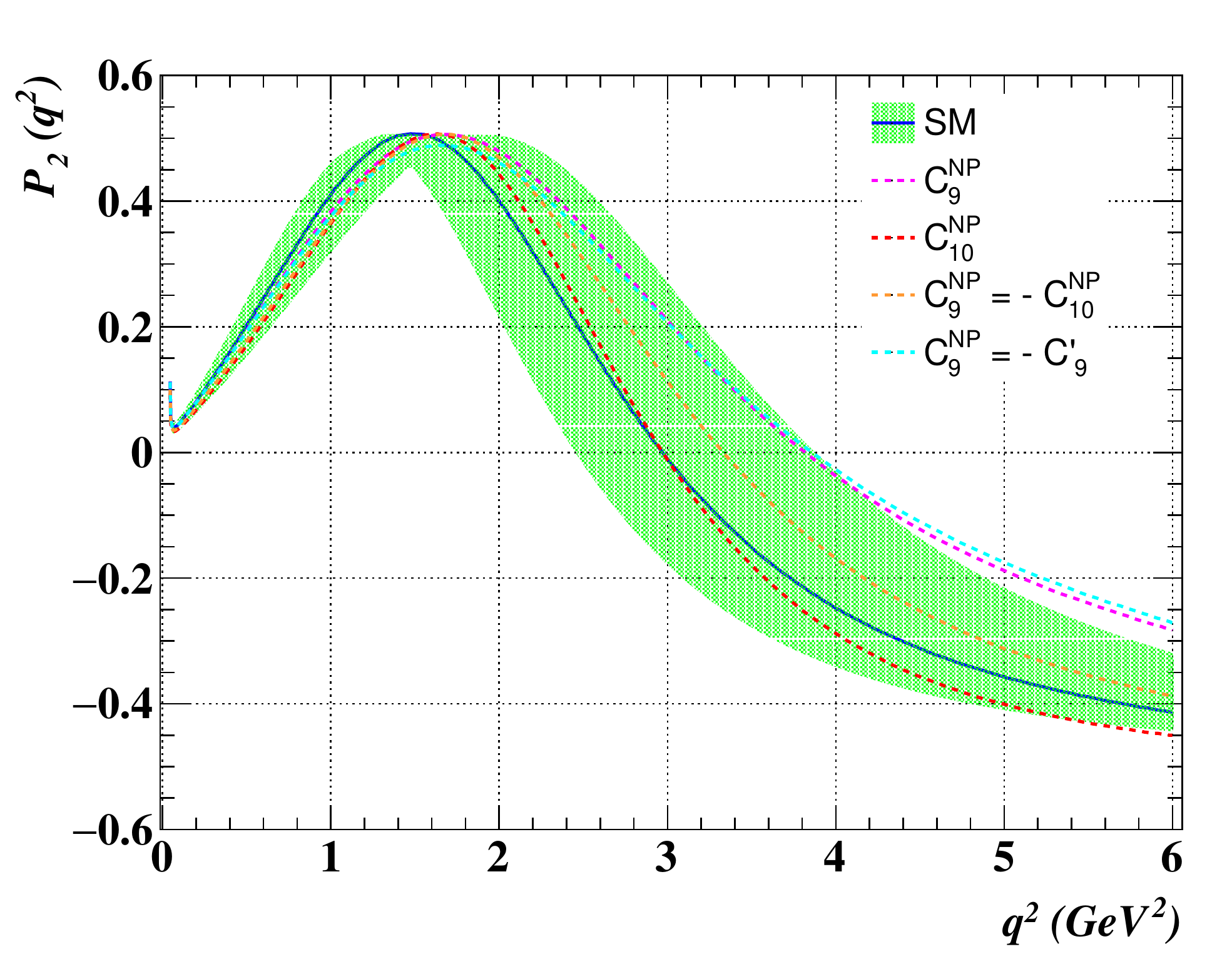}
\includegraphics[width=8.9cm,height=6.0cm]{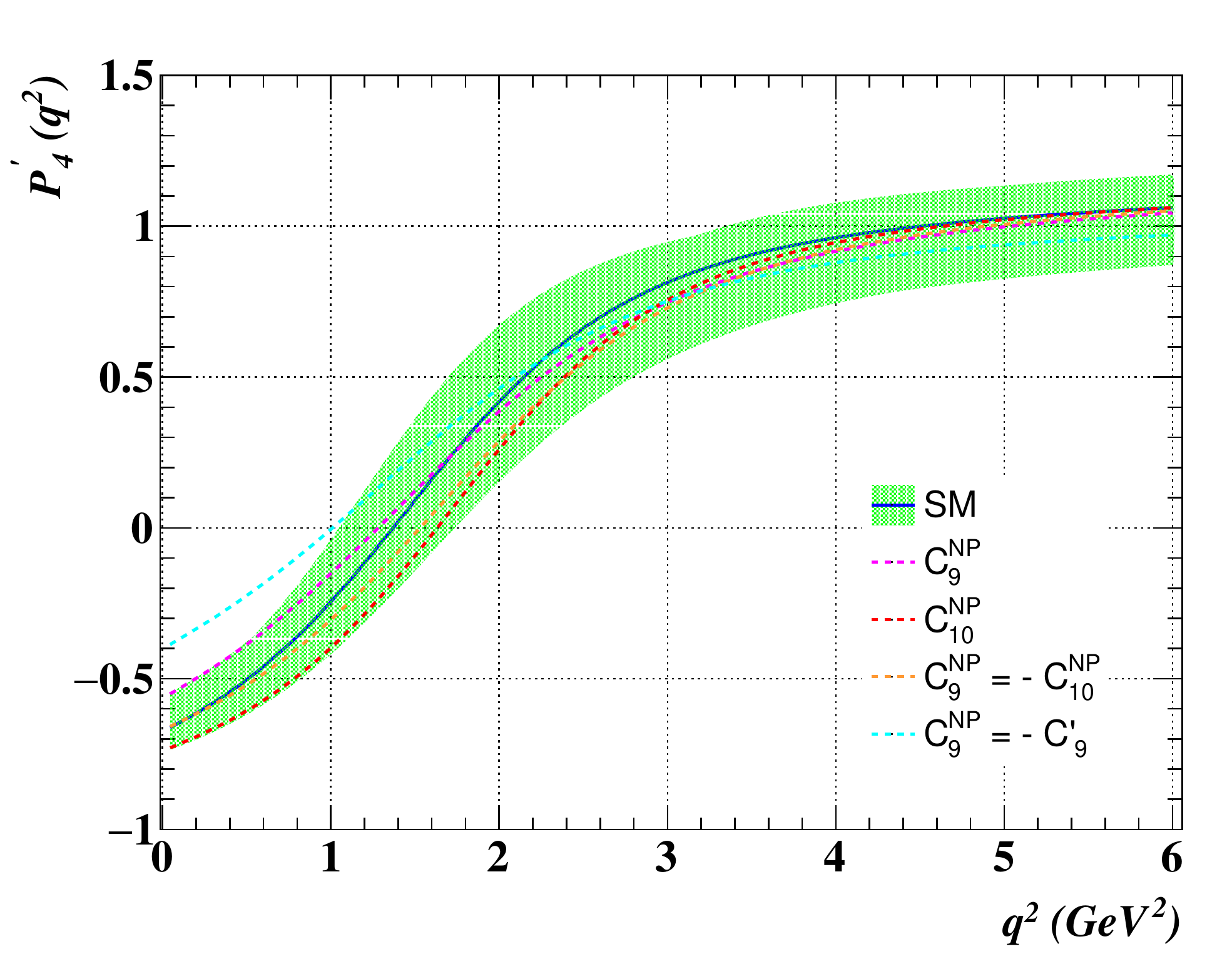}
\includegraphics[width=8.9cm,height=6.0cm]{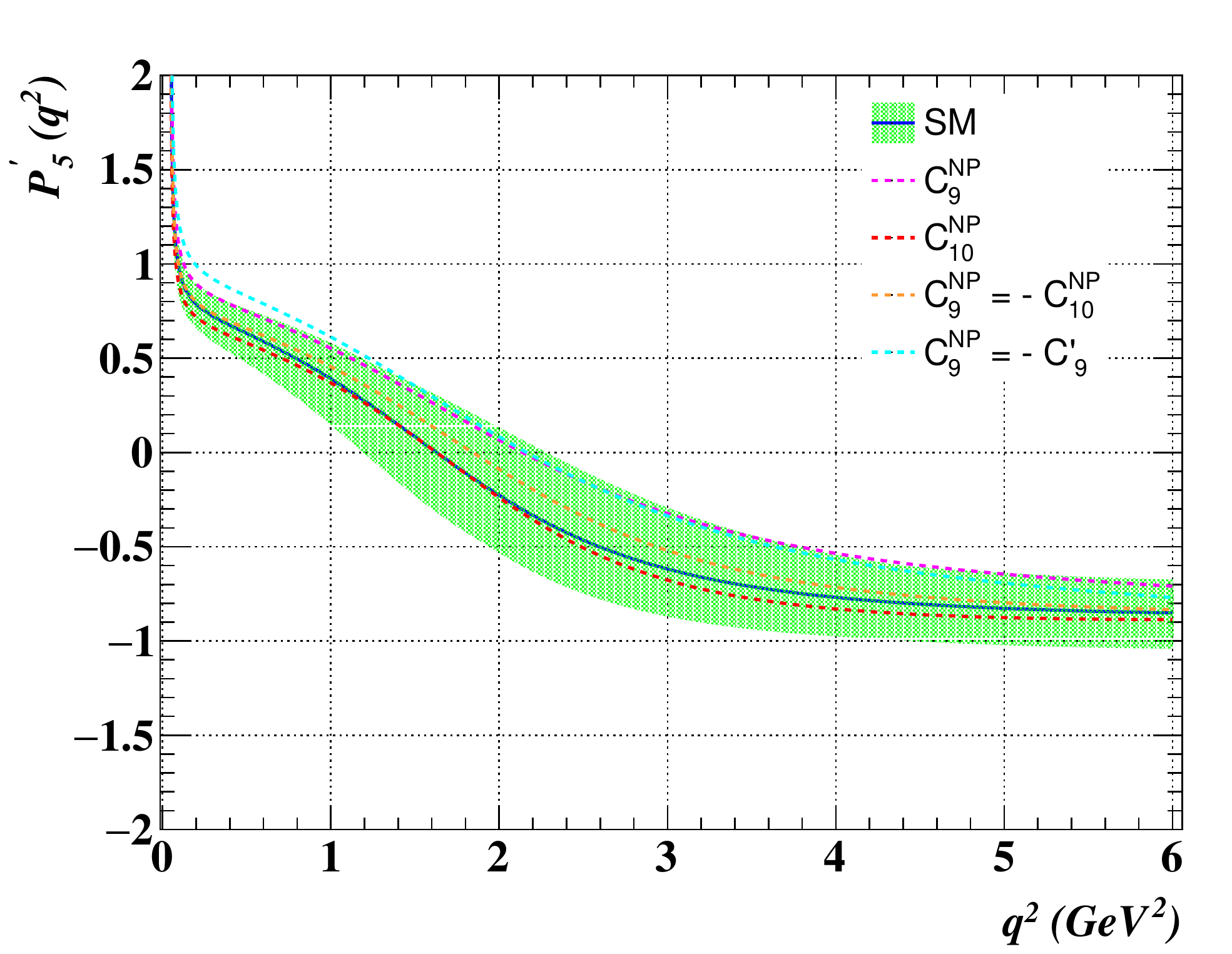}
\caption{The $q^2$ distributions of various observables such as the differential branching ratio $DBR(q^2)$, 
the longitudinal polarization fraction $F_L(q^2)$, the forward-backward asymmetry $A_{FB}(q^2)$, and $P_1(q^2)$, $P_2(q^2)$, 
$P^{\prime}_4(q^2)$, $P^{\prime}_5(q^2)$ for the $B_s \to f_{2}'(1525)(\to K^+\,K^-)\,\mu^+ \,\mu^-$ decays in the SM and in the presence of
${C}_{9}^{NP}$, ${C}_{10}^{NP}$, ${C}_{9}^{NP}=-{C}_{10}^{NP}$ and ${C}_{9}^{NP}=-{C}_{9}^{\prime}$ 1D NP scenarios.}
\label{fig_np1dq2}
\end{figure}

We show in Fig~\ref{fig_np1dq2} the $q^2$ dependent observables for the $B_s \to f_2^{\prime}(1525)\,\mu^+\mu^-$ decays in the presence of several
NP WCs in 1D scenario. The SM error band is shown with green. The detailed observations are as follows:

\begin{itemize}
 \item The differential branching ratio $DBR(q^2)$ is slightly reduced at all $q^2$ for each NP scenarios and it lies within the SM $1\sigma$ 
error band. 

\item It is interesting to note that the zero crossing point of $A_{FB}(q^2)$ is shifted towards the higher $q^2$ regions than in the SM 
for most of the NP
scenarios. It, however, coincides with the SM zero crossing point $q^2\sim 3^{+0.8}_{-0.6} \,{\rm GeV^2}$ for ${C}_{10}^{NP}$ NP coupling. We observe the 
zero crossing of $A_{FB}(q^2)$ at $q^2 \sim 3.3\,{\rm GeV^2}$ for ${C}_{9}^{NP}=-{C}_{10}^{NP}$ scenario. Similarly, the zero crossing is 
observed at around $q^2 \sim 3.8\,{\rm GeV^2}$ for ${C}_{9}^{NP}$ and ${C}_{9}^{NP}=-{C}_{9}^{\prime}$ NP scenarios, respectively. It is
worth mentioning that the zero crossing points for ${C}_{9}^{NP}$ and ${C}_{9}^{NP}=-{C}_{9}^{\prime}$ NP scenarios are distinguishable from
the SM prediction at the level of $1\sigma$ significance.

 \item For the longitudinal polarization fraction $F_L(q^2)$, the $q^2$ distribution obtained for ${C}_{10}^{NP}$ and 
${C}_{9}^{NP}=-{C}_{10}^{NP}$ NP scenarios is quite similar to that of the SM. In case of
${C}_{9}^{NP}$, it lies outside the SM error band in $q^2\in [1.1, 2.5]$ region and becomes very similar to the SM curve in the higher $q^2$ 
regions. The maximum deviation from the SM prediction is observed for ${C}_{9}^{NP}=-{C}_{9}^{\prime}$ NP scenario.

\item For the angular observable $P_1(q^2)$, the $q^2$ distribution obtained for ${C}_{9}^{NP}$, ${C}_{10}^{NP}$ and 
${C}_{9}^{NP}=-{C}_{10}^{NP}$ NP scenarios is quite similar to the SM. The shape, however, is quite different from the SM in case of 
${C}_{9}^{NP}=-{C}_{9}^{\prime}$ NP scenario. The value of $P_1(q^2)$ obtained in this NP scenario is negative in the whole $q^2$ region and 
reaches its minimum of around $-0.25$ at $q^2 = 2\,{\rm GeV^2}$.

 \item In the case of $P_2(q^2)$, similar to $A_{FB}(q^2)$, the zero crossing point is shifted towards the higher $q^2$ regions than in the SM 
for most of the NP scenarios. The maximum deviation in the zero crossing point is observed in case of ${C}_{9}^{NP}$ and 
${C}_{9}^{NP}=-{C}_{9}^{\prime}$ NP scenarios, respectively. 

\item The angular observable $P^{\prime}_4(q^2)$ obtained in each of these $1D$ scenarios lies within the SM error band. There is, however, 
one exception. For ${C}_{9}^{NP}=-{C}_{9}^{\prime}$, it lies outside the SM $1\sigma$ error band in the low $q^2$ region, i.e, for 
$q^2 \le 1\,{\rm GeV^2}$. In addition, the zero crossing points for the 
${C}_{9}^{NP}=-{C}_{10}^{NP}$ and ${C}_{10}^{NP}$ NP scenarios are observed at $q^2 \sim 1.5\,{\rm GeV^2}$ and $q^2 \sim 1.6\,{\rm GeV^2}$, 
whereas, 
the zero crossing points for ${C}_{9}^{NP}$ and ${C}_{9}^{NP}=-{C}_{9}^{\prime}$ are observed at $q^2 \sim 1.3\,{\rm GeV^2}$ and 
$q^2 \sim 1\,{\rm GeV^2}$, respectively. It is worth mentioning that the zero crossing point obtained in case of 
${C}_{9}^{NP}=-{C}_{9}^{\prime}$ NP scenario is distinguishable from the SM zero crossing point $q^2 \sim 1.4 \pm 0.3\,{\rm GeV^2}$
at more than $1\sigma$ significance.

\item For the angular observable $P^{\prime}_5(q^2)$, the zero crossing point obtained in each NP scenarios shifted towards the higher value
of $q^2$ than in the SM except for ${C}_{10}^{NP}$. In case of ${C}_{10}^{NP}$, the zero crossing point coincides with the SM zero crossing point of 
$q^2 \sim 1.6 \pm 0.4\,{\rm GeV^2}$. For ${C}_{9}^{NP}=-{C}_{10}^{NP}$ NP scenario, the zero crossing point is observed at 
$q^2 \sim 1.8\,{\rm GeV^2}$, whereas, for ${C}_{9}^{NP}$ and ${C}_{9}^{NP}=-{C}_{9}^{\prime}$ NP scenarios, we observe
the zero crossing point at $q^2 \sim 2.1\,{\rm GeV^2}$ which deviates from the SM prediction at the level of around $1\sigma$ significance.
\end{itemize}

\subsubsection{New Physics: 2D scenario}

Now we proceed to discuss the impact of several new Wilson coefficients from the $2D$ scenarios.  
We consider three different 2D scenarios: $({C}_{9}^{NP}, {C}_{10}^{NP})$, $({C}_{9}^{NP}, {C}_{9}^{\prime})$ and 
$({C}_{9}^{NP}, {C}_{10}^{\prime})$. We report in the Appendix in the Tables~\ref{tab_npdbr},~\ref{tab_npfl},~\ref{tab_npafb},~\ref{tab_npp1},
~\ref{tab_npp2},~\ref{tab_npp4},~\ref{tab_npp5} the average values of all the observables for the $\mu$ mode. Similarly, 
the bin wise $q^2$ distribution plots are shown in Fig.~\ref{fig_np2dbin}. 
The discussions pertaining to the impact of 2D new WC's on various observables are as follows:

\begin{figure}[htbp]
  \centering
 \includegraphics[width=8.9cm,height=6.0cm]{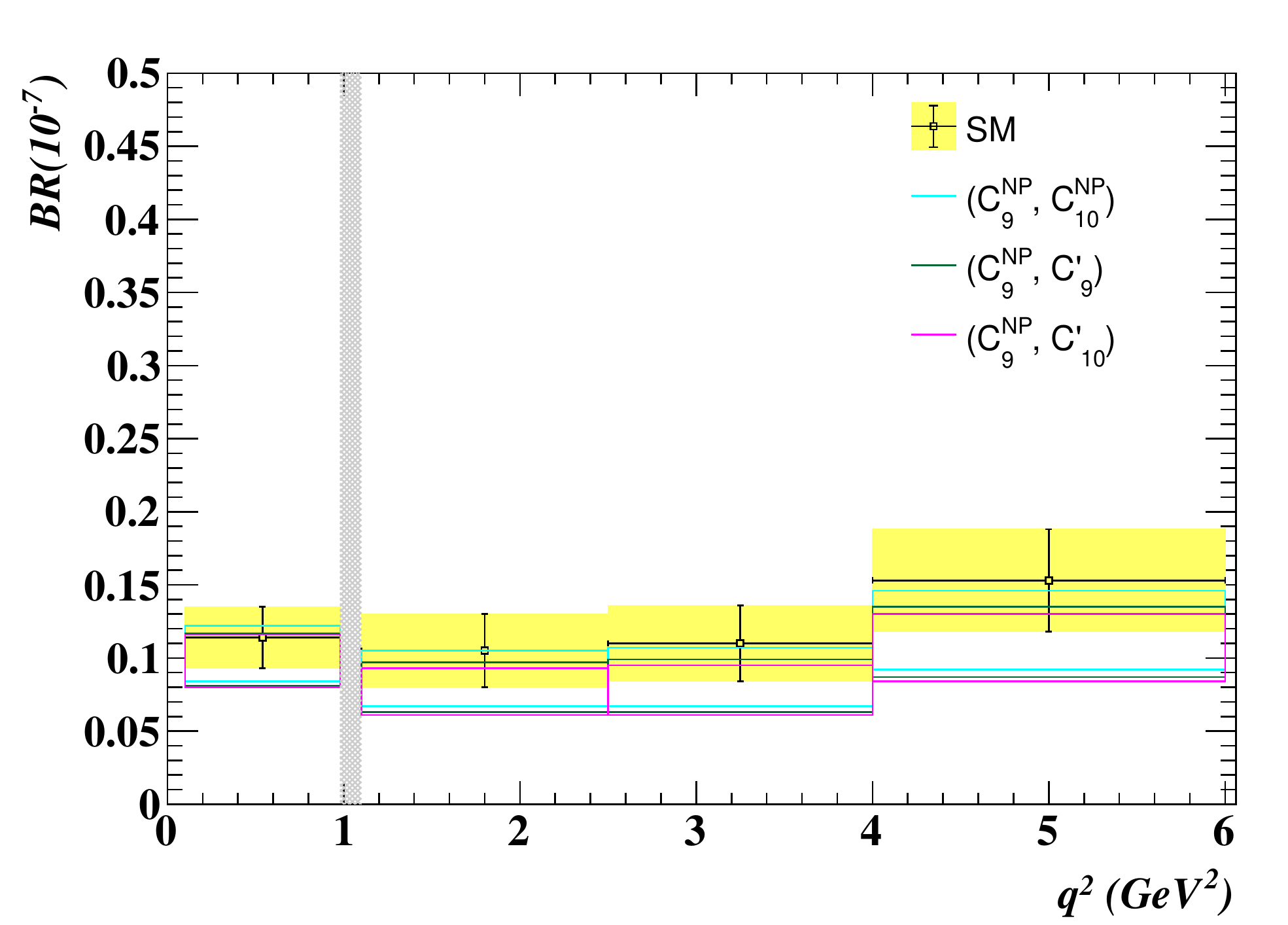}
 \includegraphics[width=8.9cm,height=6.0cm]{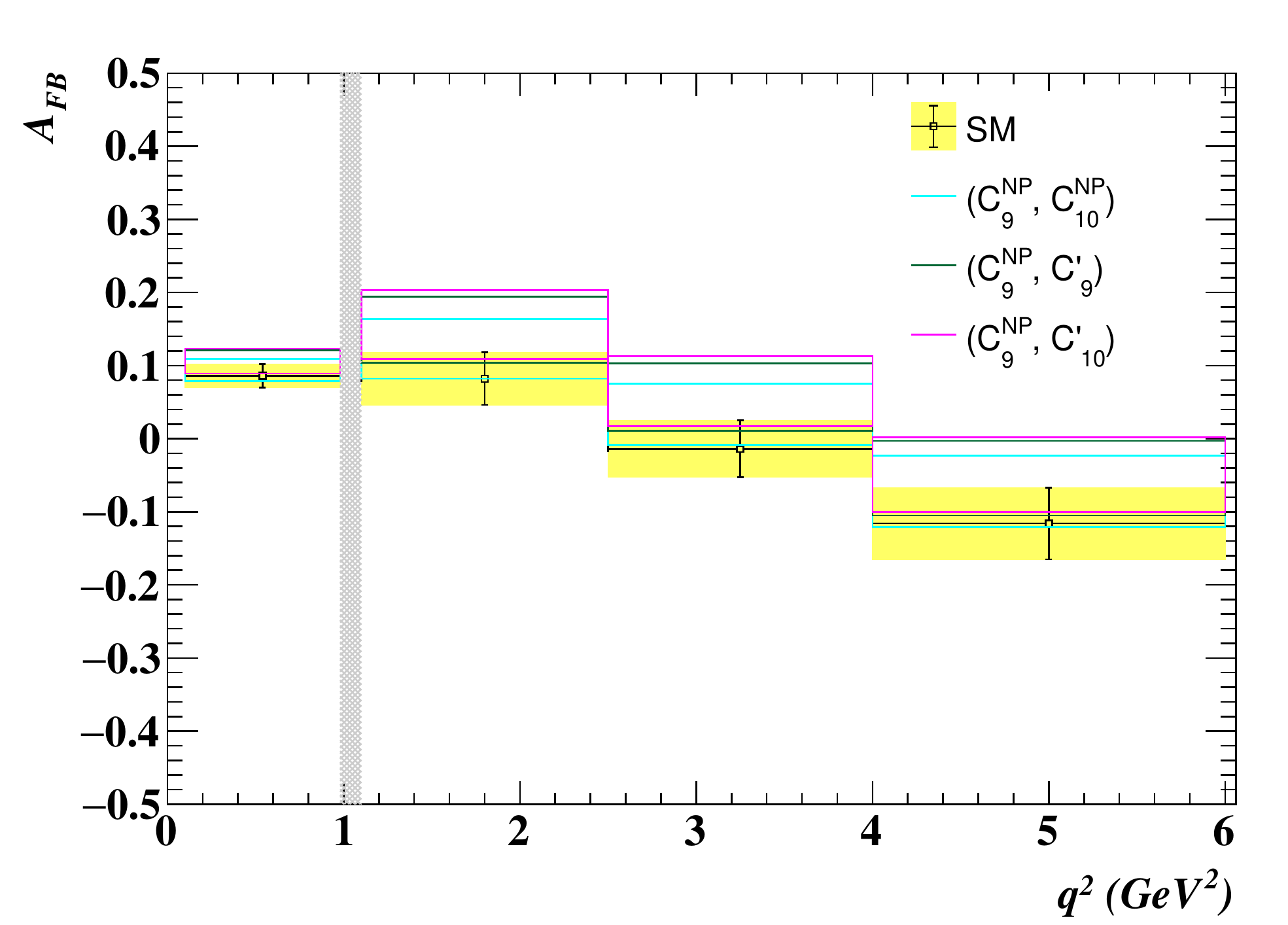}
 \includegraphics[width=8.9cm,height=6.0cm]{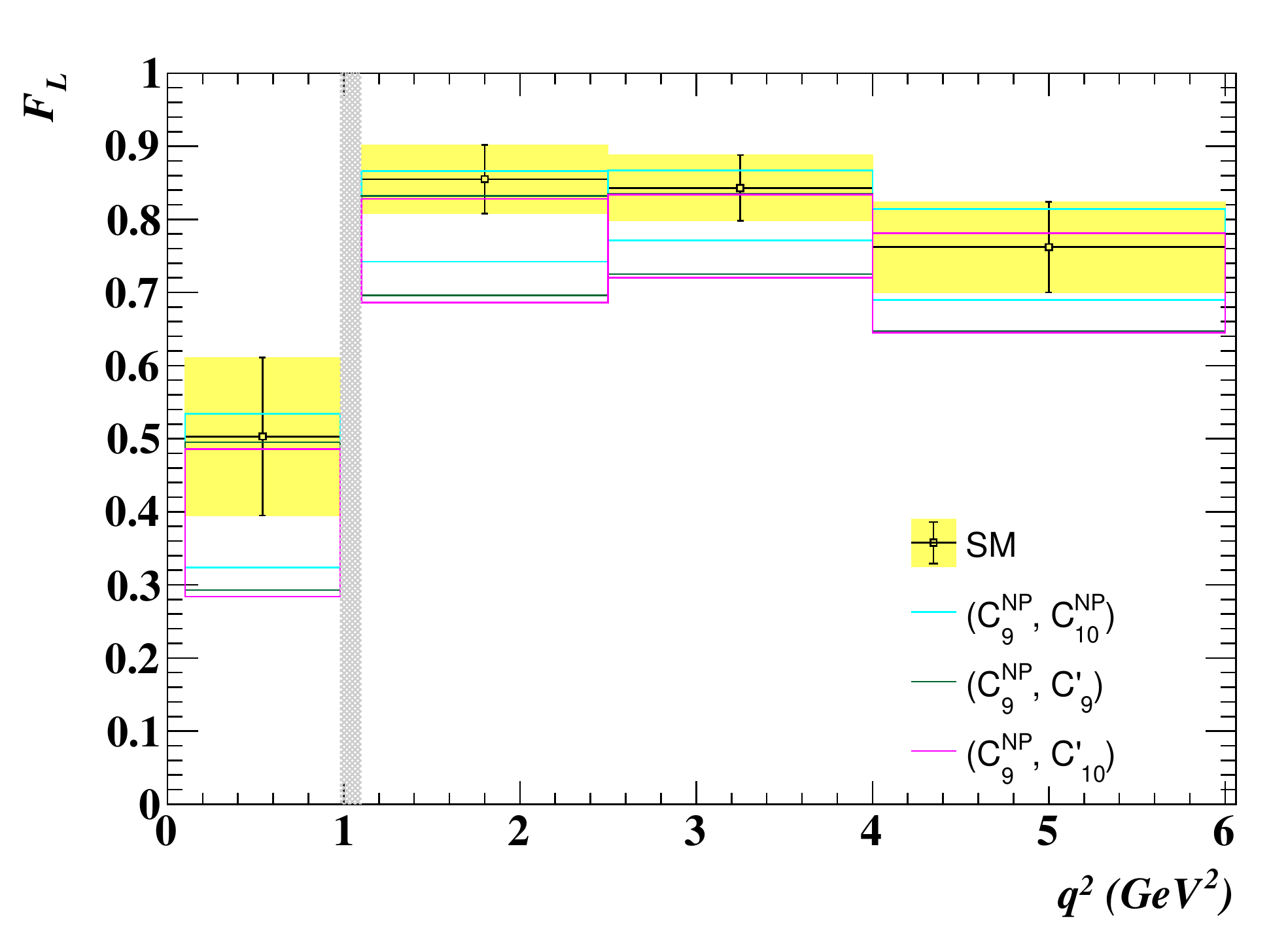}
 \includegraphics[width=8.9cm,height=6.0cm]{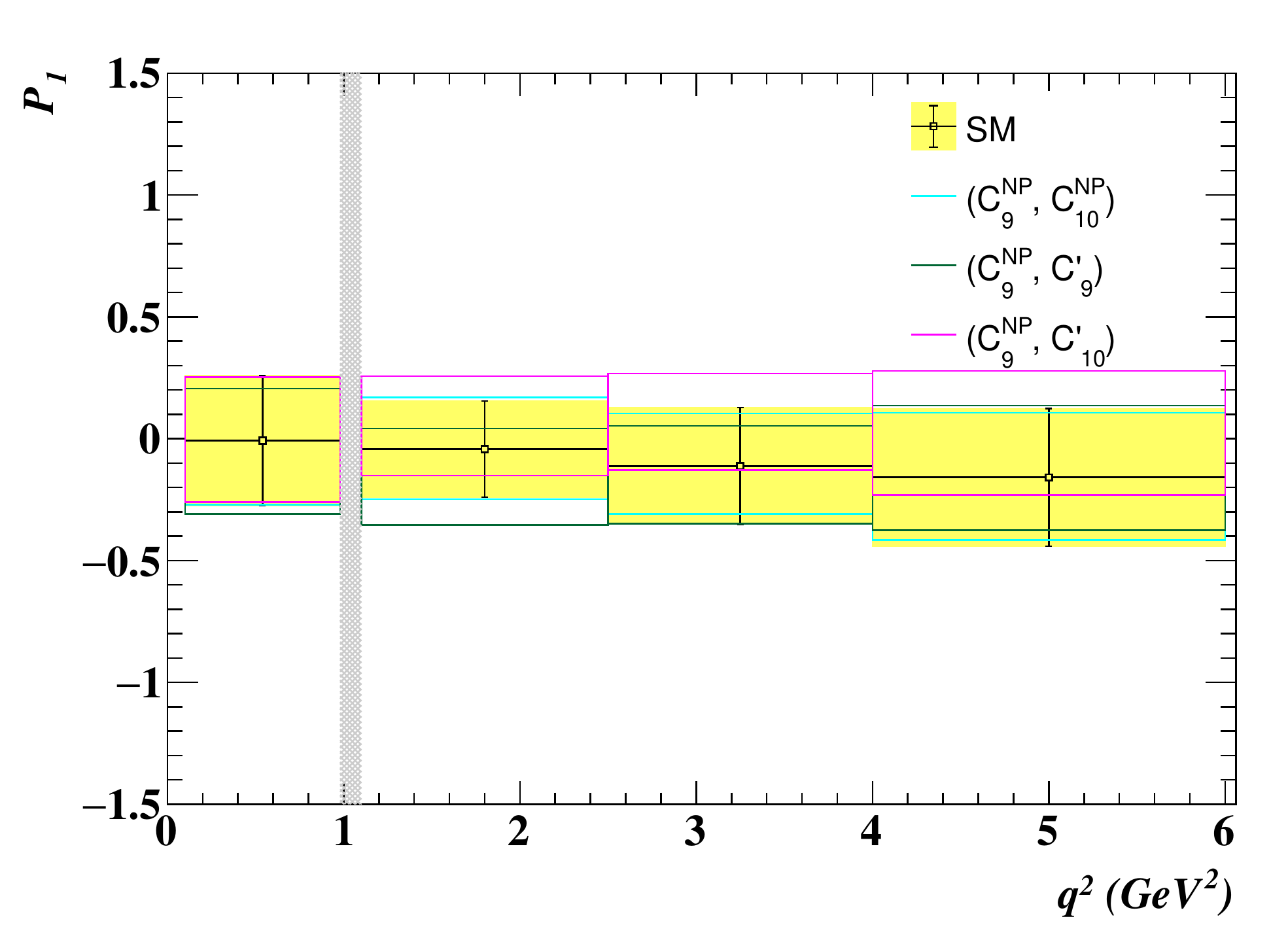}
 \includegraphics[width=8.9cm,height=6.0cm]{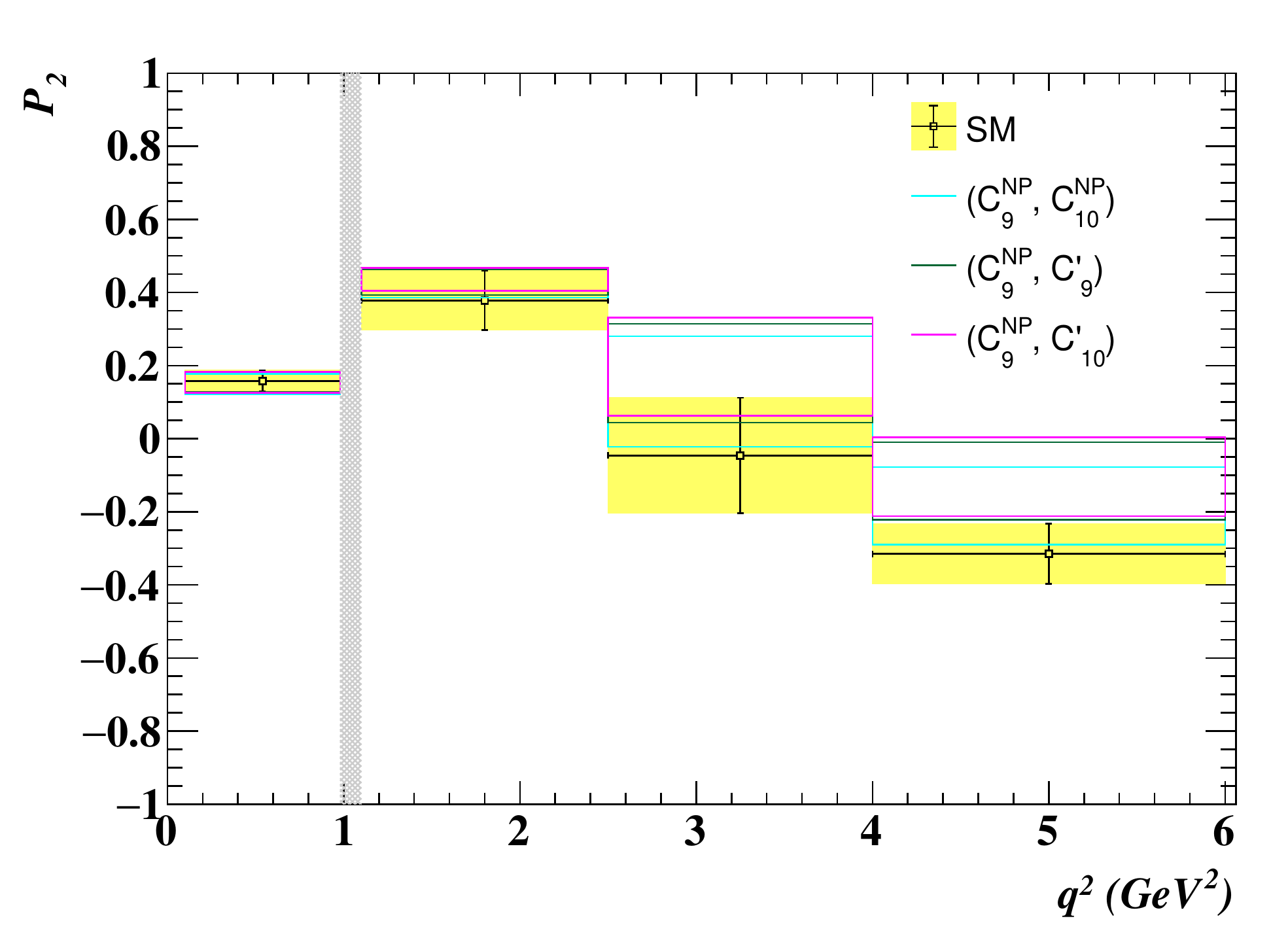}
 \includegraphics[width=8.9cm,height=6.0cm]{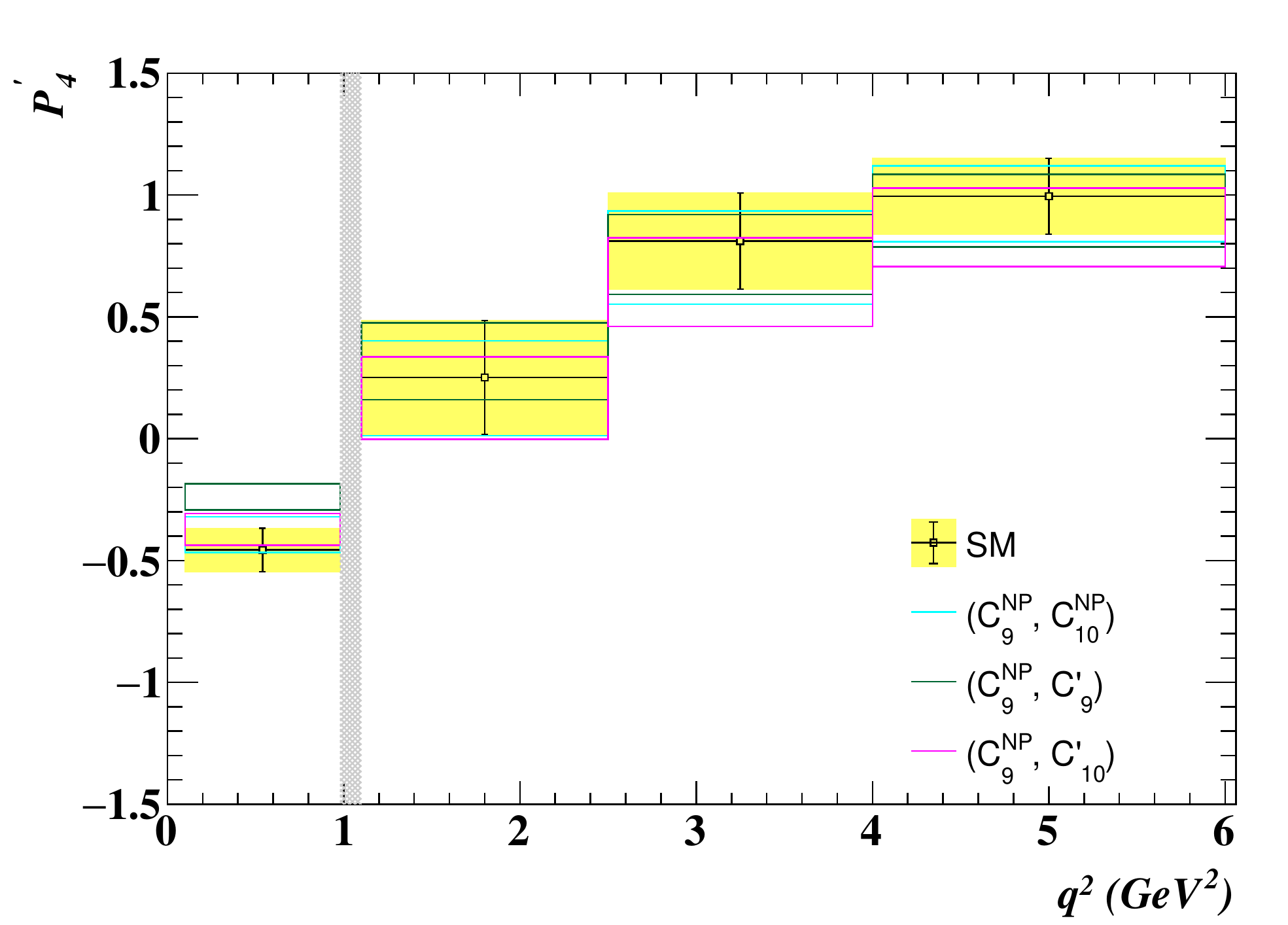}
 \includegraphics[width=8.9cm,height=6.0cm]{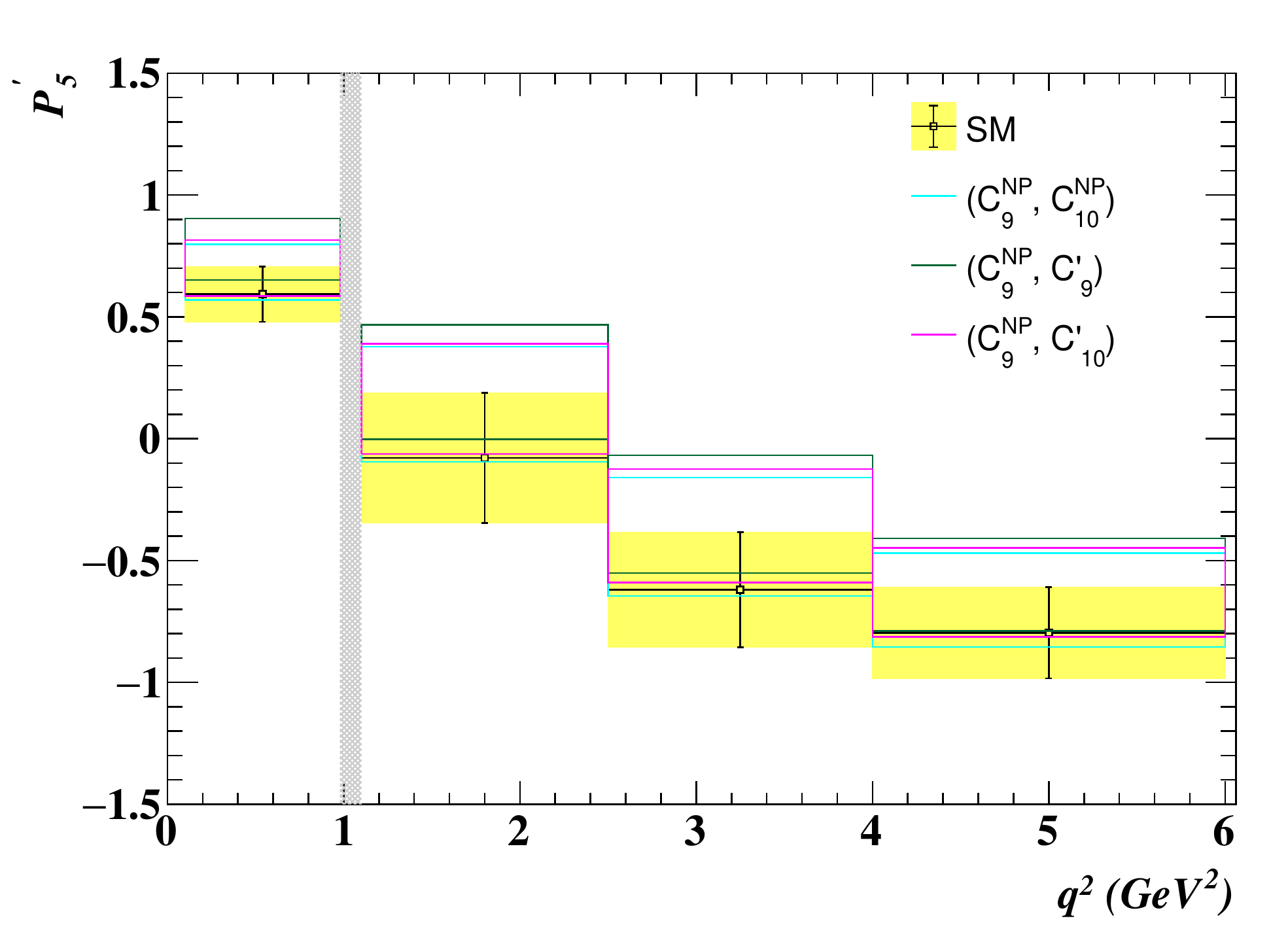}
 \caption{The central values and the corresponding $1\sigma$ error bands of various observables such as the branching ratio, 
 the longitudinal polarization fraction $F_L$, the forward-backward asymmetry $A_{FB}$, and $P_1$, $P_2$, $P^{\prime}_4$, $P^{\prime}_5$ 
 in several $q^2$ bins for the $B_s \to f_{2}'(1525)(\to K^+\,K^-)\,\mu^+ \,\mu^-$ decays in the SM and in the presence of three 2D NP 
 scenarios.}
\label{fig_np2dbin}
\end{figure}

\begin{itemize}
 \item $BR$: Although the central values obtained for each NP scenarios differ from the SM prediction, no significant deviation is observed
 in any $q^2$ bins. The deviation from the SM prediction is observed to be around $1\sigma$ in case of $({C}_{9}^{NP}, {C}_{9}^{\prime})$ and 
$({C}_{9}^{NP}, {C}_{10}^{\prime})$ NP scenarios, whereas, for the $({C}_{9}^{NP}, {C}_{10}^{NP})$ NP scenario, the value of $BR$ lies within 
the SM $1\sigma$ error band.  
 
 \item $F_L$: In the bin $q^2 \in [1.1, 2.5]$, a deviation of around $1.1\sigma$ from the SM prediction is observed in case of
$({C}_{9}^{NP}, {C}_{9}^{\prime})$ and $({C}_{9}^{NP}, {C}_{10}^{\prime})$ NP scenarios. In all other $q^2$ bins, the value of $F_L$, however,
lies within the $1\sigma$ SM error band for each NP scenarios. 
 
 \item $A_{FB}$: In the bin $q^2\in [1.1, 2.5]$ and $q^2 \in [2.5, 4.0]$, the deviation from the SM prediction is observed to be at 
$1.1-1.2\sigma$ level in case of $({C}_{9}^{NP}, {C}_{9}^{\prime})$ and $({C}_{9}^{NP}, {C}_{10}^{\prime})$ NP scenarios.
In all other bins, it however lies within the SM $1\sigma$ error band for each NP scenarios.
 
 \item $P_1$: Although the central values obtained for each NP scenarios differ from the SM central value, no significant deviation is 
observed as they all lie within the SM $1\sigma$ error band. 

 \item $P_2$: A deviation of around $1-1.1\sigma$ from the SM prediction is observed in the bin $q^2\in [2.5, 4.0]$ in case of 
$({C}_{9}^{NP}, {C}_{9}^{\prime})$ and $({C}_{9}^{NP}, {C}_{10}^{\prime})$ NP scenarios. Similarly, in the $q^2\in [4.0, 6.0]$ bin,
a deviation of around $1.5\sigma$ 
is observed in case of $({C}_{9}^{NP}, {C}_{9}^{\prime})$ and $({C}_{9}^{NP}, {C}_{10}^{\prime})$ NP scenarios.

 \item $P_4^{\prime}$: In the bin $q^2 \in [0.045, 0.98]$, the $({C}_{9}^{NP}, {C}_{9}^{\prime})$ NP scenario is distinguishable from the 
SM prediction at the level of $2\sigma$ significance, whereas, in case of $({C}_{9}^{NP}, {C}_{10}^{NP})$ and 
$({C}_{9}^{NP}, {C}_{10}^{\prime})$ NP scenarios, the value of $P_4^{\prime}$ lies within the SM $1\sigma$ error band and hence can not be
distinguished from the SM prediction.  

 \item $P_5^{\prime}$: In the bin $q^2\in[0.045, 0.98]$, the value of $P_5^{\prime}$ obtained in case of $({C}_{9}^{NP}, {C}_{9}^{\prime})$ 
NP scenario shows a deviation around $1\sigma$ from the SM prediction, whereas, with other NP scenarios, it is consistent with the SM 
prediction. Similarly, in the bins $q^2\in[1.1, 2.5]$, $[2.5, 4.0]$ and $[4.0, 6.0]$, no significant deviation from the SM prediction is 
observed and hence indistinguishable from the SM.
 
\end{itemize}

\begin{figure}[htbp]
\centering
\includegraphics[width=8.9cm,height=6.0cm]{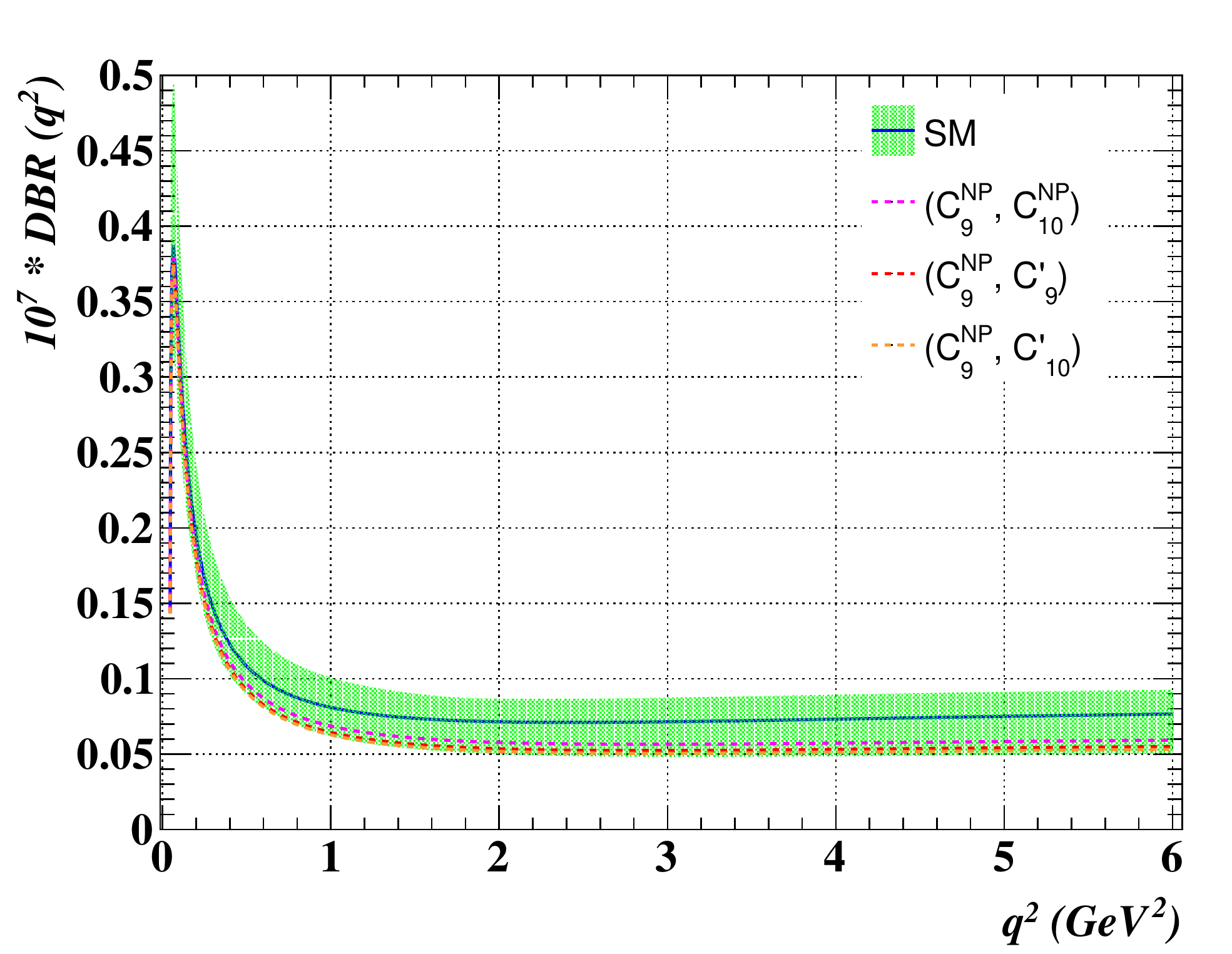}
\includegraphics[width=8.9cm,height=6.0cm]{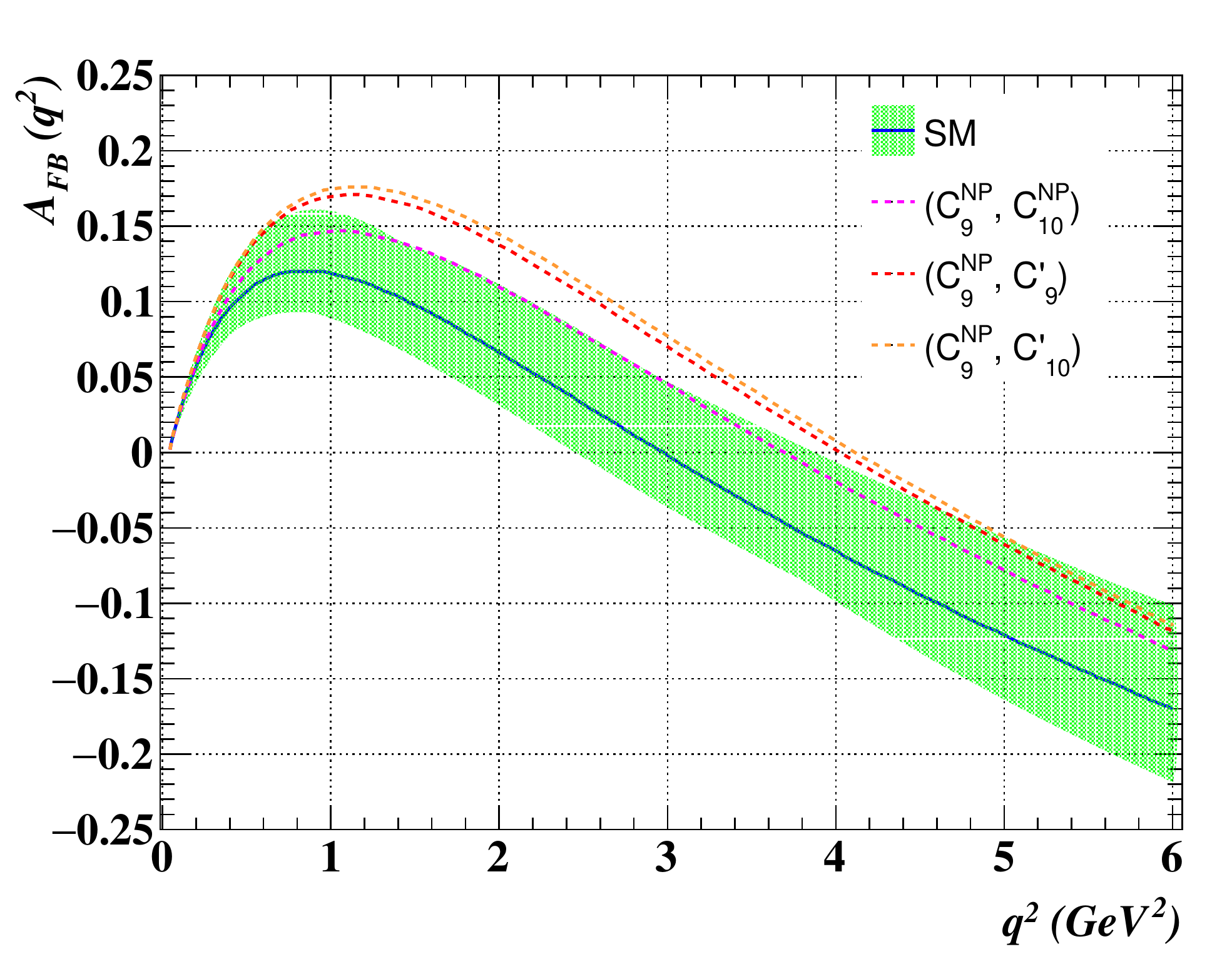}
\includegraphics[width=8.9cm,height=6.0cm]{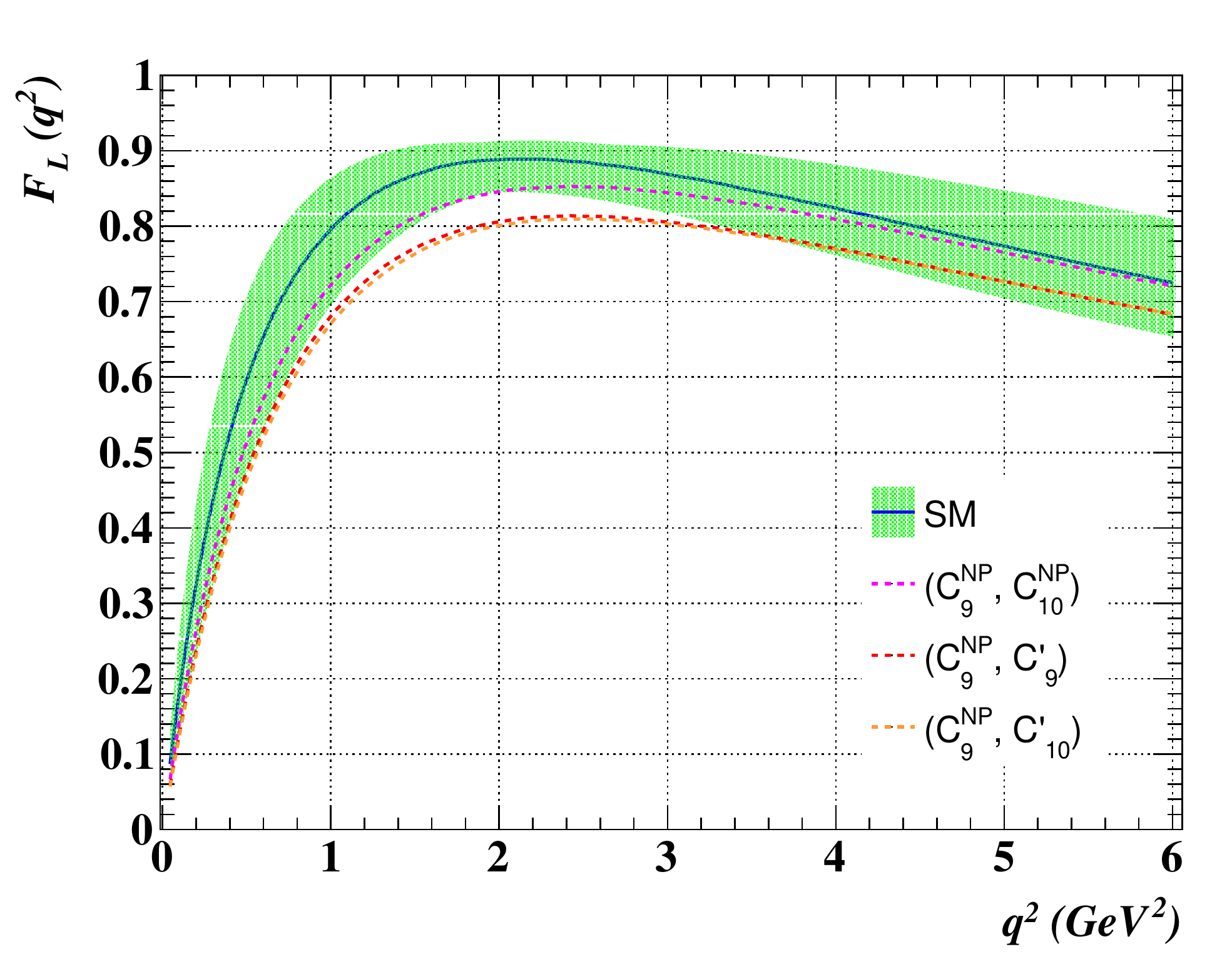}
\includegraphics[width=8.9cm,height=6.0cm]{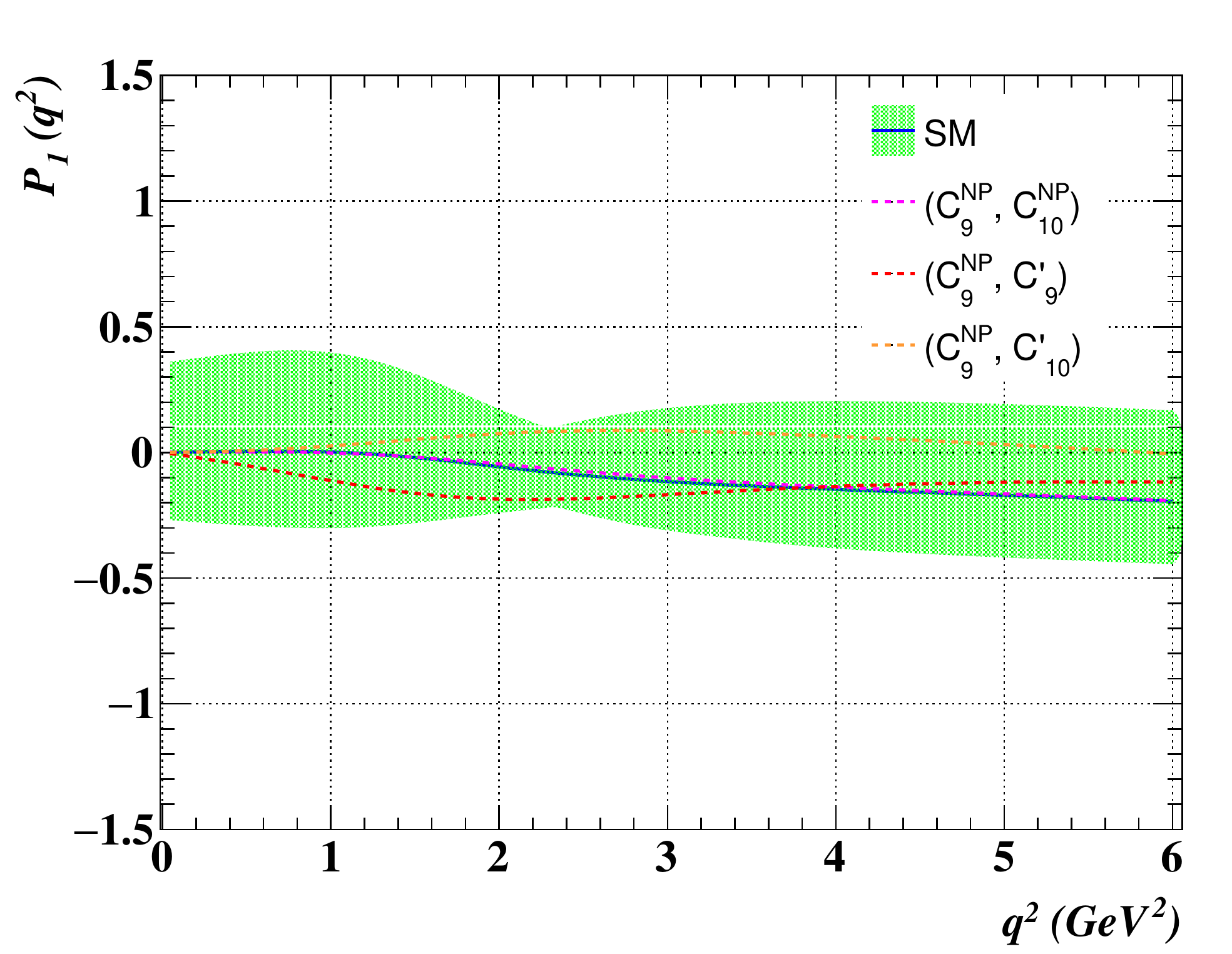}
\includegraphics[width=8.9cm,height=6.0cm]{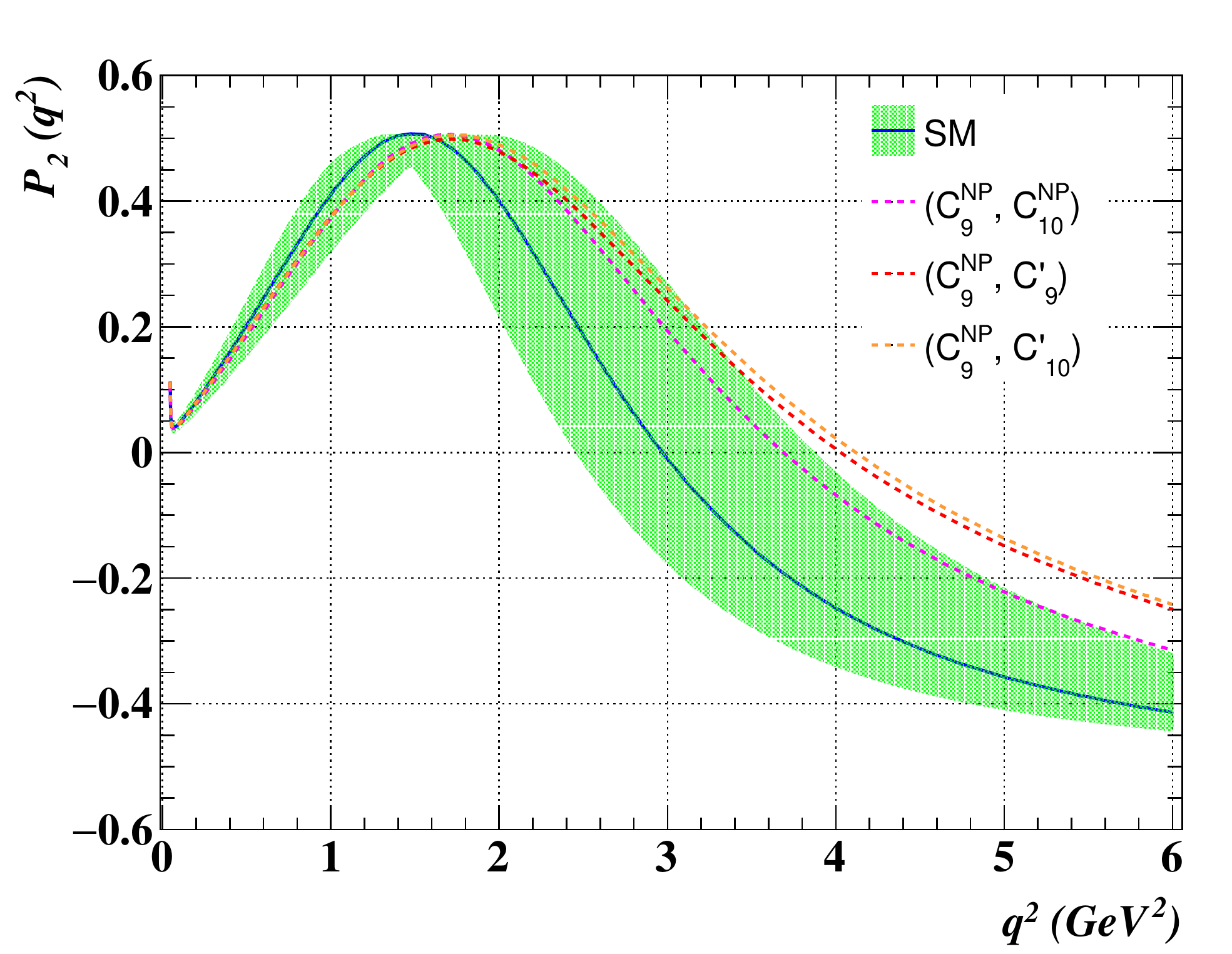}
\includegraphics[width=8.9cm,height=6.0cm]{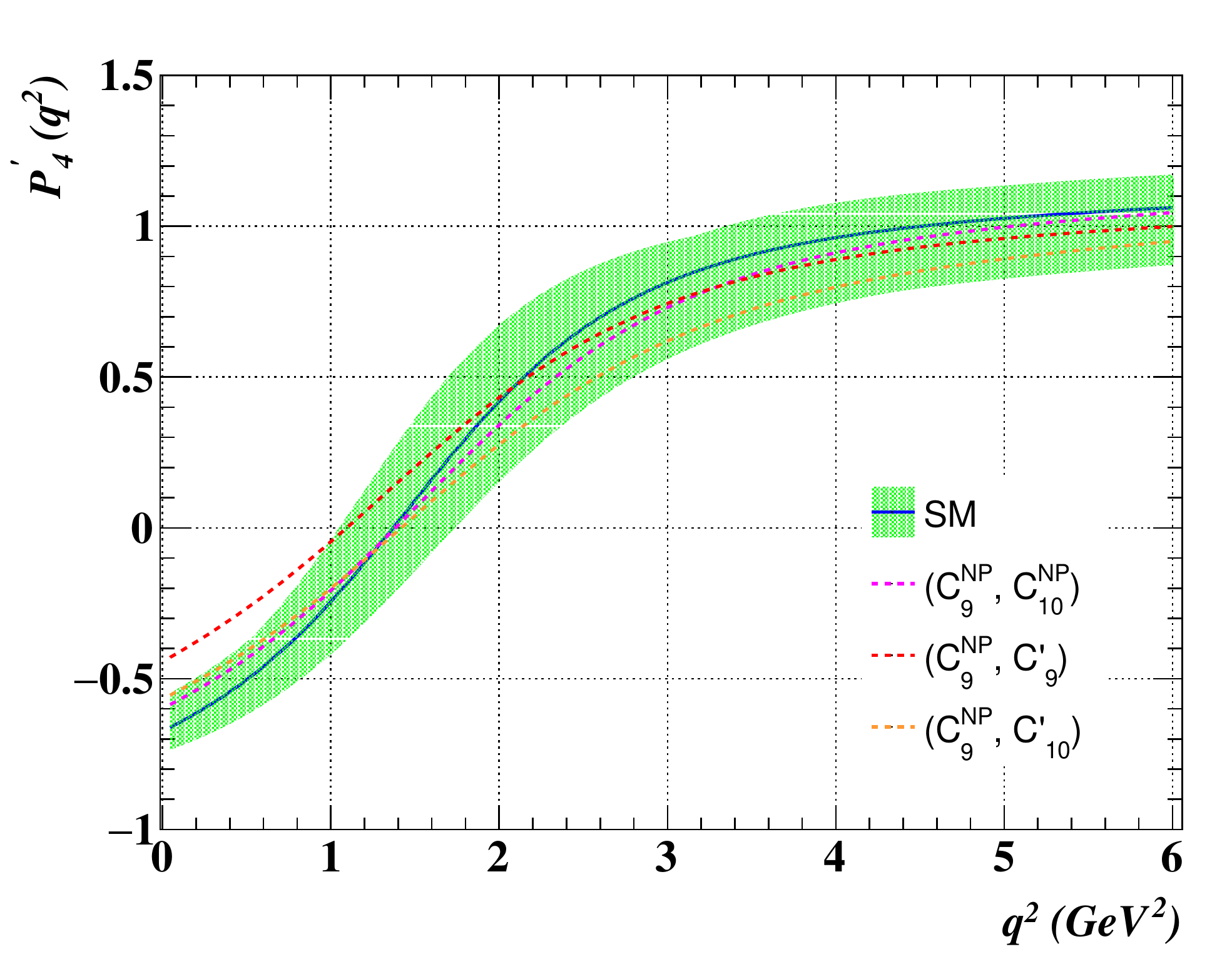}
\includegraphics[width=8.9cm,height=6.0cm]{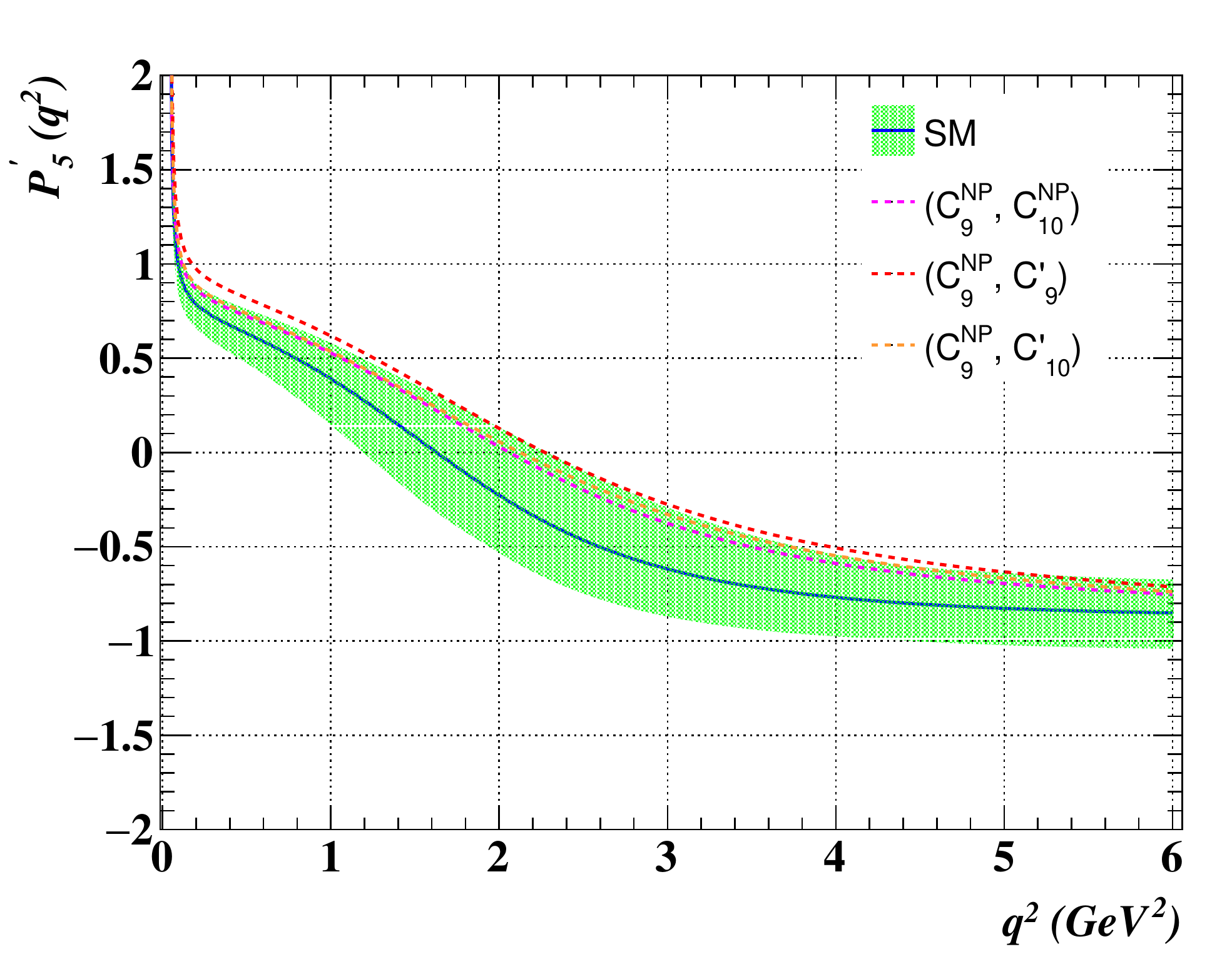}
\caption{The $q^2$ distributions of various observables such as the differential branching ratio $DBR(q^2)$, 
the longitudinal polarization fraction $F_L(q^2)$, the forward-backward asymmetry $A_{FB}(q^2)$, and $P_1(q^2)$, $P_2(q^2)$, 
$P^{\prime}_4(q^2)$, $P^{\prime}_5(q^2)$ for the $B_s \to f_{2}'(1525)(\to K^+\,K^-)\,\mu^+ \,\mu^-$ decays in the SM and in the presence of 
$({C}_{9}^{NP}, {C}_{10}^{NP})$, $({C}_{9}^{NP}, {C}_{9}^{\prime})$ and $({C}_{9}^{NP}, {C}_{10}^{\prime})$ 2D NP scenarios.}
\label{fig_np2dq2}
\end{figure}

We show in Fig.~\ref{fig_np2dq2} the $q^2$ dependence of all the observables for the $B_s \to f_2^{\prime}(1525)\,\mu^+\mu^-$ decays in several 
$2D$ scenarios. The SM $1\sigma$ error band is shown with green. The detailed observations are as follows:

\begin{itemize}
\item Similar to the $1D$ scenario, we observe that the differential branching ratio is slightly reduced at all $q^2$
for each NP scenarios and they all lie within the SM error band. 

\item It is worth mentioning that the zero crossing point for $A_{FB}(q^2)$ is shifted to higher $q^2$ region for all the NP scenarios as
compared to the SM.
The zero crossing points for $A_{FB}(q^2)$ are observed at $q^2 \sim 3.6\,{\rm GeV^2}$, $q^2 \sim 4$ GeV$^2$ and $q^2 \sim 4.1$ GeV$^2$ for 
$({C}_{9}^{NP}, {C}_{10}^{NP})$, $({C}_{9}^{NP}, {C}_{9}^{\prime})$ and for $({C}_{9}^{NP}, {C}_{10}^{\prime})$ NP scenarios, respectively. 
Although all the values are found to be distinct from the SM zero crossing point, it is important to note that the zero crossing point 
obtained in case of $({C}_{9}^{NP}, {C}_{9}^{\prime})$ and 
$({C}_{9}^{NP}, {C}_{10}^{\prime})$ NP scenarios are distinguishable from the SM prediction at the level of more than $1\sigma$ significance.
 
\item The peak of the longitudinal polarization fraction $F_L(q^2)$ may shift towards higher $q^2$ values than in the SM for each NP 
scenarios. It should be
mentioned that the peak of $F_L(q^2)$ obtained in case of $({C}_{9}^{NP}, {C}_{9}^{\prime})$ 
and $({C}_{9}^{NP}, {C}_{10}^{\prime})$ is distinguishable from the SM prediction at the level of more than $1\sigma$ significance.

\item The angular observable $P_1(q^2)$ is zero in SM in the low $q^2$ region, i.e, for $q^2 \le 1.2\,{\rm GeV^2}$ and becomes negative 
as $q^2$ increases. Similar behavior is observed in case of $({C}_{9}^{NP}, {C}_{10}^{NP})$ NP scenario as well. For
$({C}_{9}^{NP}, {C}_{9}^{\prime})$ NP scenario, it deviates slightly away from the SM and reaches minimum value of around $-0.2$ at 
$q^2 = 2\,{\rm GeV^2}$. However, we observe a completely different behavior in case of $({C}_{9}^{NP}, {C}_{10}^{\prime})$ NP scenario.
The value of $P_1(q^2)$ acquires positive values in the whole $q^2$ region and reaches its maximum value of $0.1$ at 
$q^2 \sim 2.2\,{\rm GeV^2}$. Since the SM error band is too large, the $q^2$ distributions of all the NP scenarios lie within the SM
error band.   

\item The peak of $P_2(q^2)$ is slightly reduced and shifted towards the higher $q^2$ values in each NP scenarios as compared to the SM. 
Moreover, the zero crossing point is also shifted to higher values of $q^2$ than in the SM for all the NP scenarios. In case of 
$({C}_{9}^{NP}, {C}_{9}^{\prime})$ and 
$({C}_{9}^{NP}, {C}_{10}^{\prime})$ NP scenarios, the zero crossing points are distinguishable
from the SM zero crossing at the level of more than $1\sigma$ significance. 

\item For the angular observable $P^{\prime}_4(q^2)$, no significant deviation from the SM is observed for each NP scenarios.
However, in the low $q^2$ region, i.e, $q^2 \le 1\,{\rm GeV^2}$, we see significant deviation of $P^{\prime}_4(q^2)$ from the SM prediction
in case of $({C}_{9}^{NP}, {C}_{9}^{\prime})$ NP scenario. 
Similarly, the zero crossing point of $P^{\prime}_4(q^2)$ obtained in case of $({C}_{9}^{NP}, {C}_{10}^{\prime})$ and 
$({C}_{9}^{NP}, {C}_{10}^{NP})$ NP scenarios coincides with the SM zero crossing point of $q^2 \sim 1.4\pm 0.3 \,{\rm GeV^2}$, whereas, 
for $({C}_{9}^{NP}, {C}_{9}^{\prime})$ NP scenario, the zero crossing point is observed at $q^2 \sim 1.1\,{\rm GeV^2}$ and it is 
distinguishable from the SM zero crossing point at the level of $1\sigma$ significance.

\item The $q^2$ distribution of the angular observable $P^{\prime}_5(q^2)$ obtained in each NP scenarios is quite distinct from the SM.
The maximum deviation from the SM prediction is observed for $({C}_{9}^{NP}, {C}_{9}^{\prime})$ NP scenario.
The zero crossing points for all the three NP scenarios lie within $q^2 \sim 2.1-2.3$ GeV$^2$, and interestingly, the zero crossing point for 
$({C}_{9}^{NP}, {C}_{9}^{\prime})$ is distinguishable from the SM at more than $1.5\sigma$ significance.
\end{itemize}

\subsection{Sensitivity of LFUV observables in $B_s \to f_{2}^{\prime}(1525)(\to K^+\,K^-)\,\mu^+ \,\mu^-$ decays}

\begin{figure}[htbp]
  \centering
 \includegraphics[width=8.9cm,height=6.0cm]{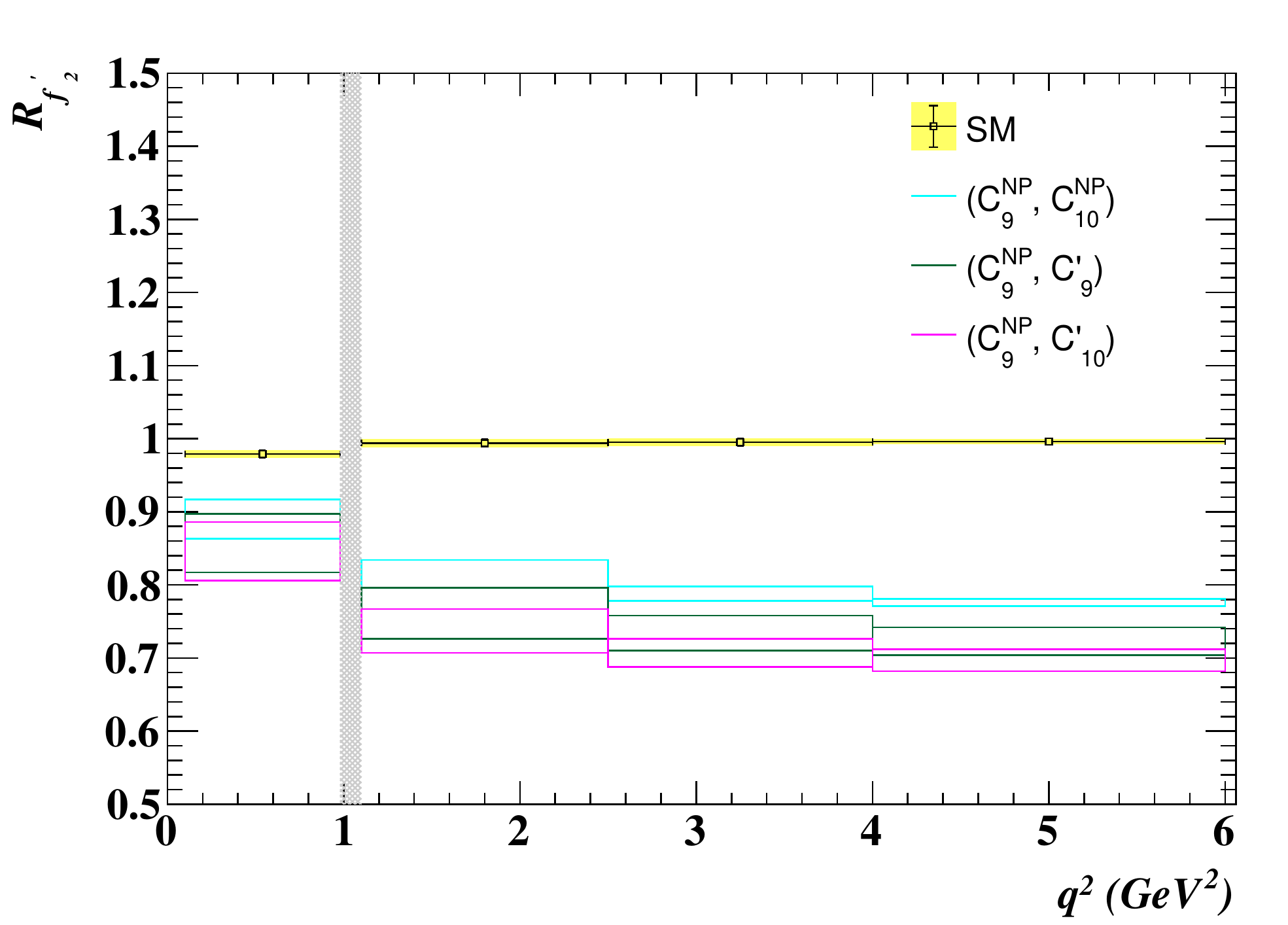}
 \includegraphics[width=8.9cm,height=6.0cm]{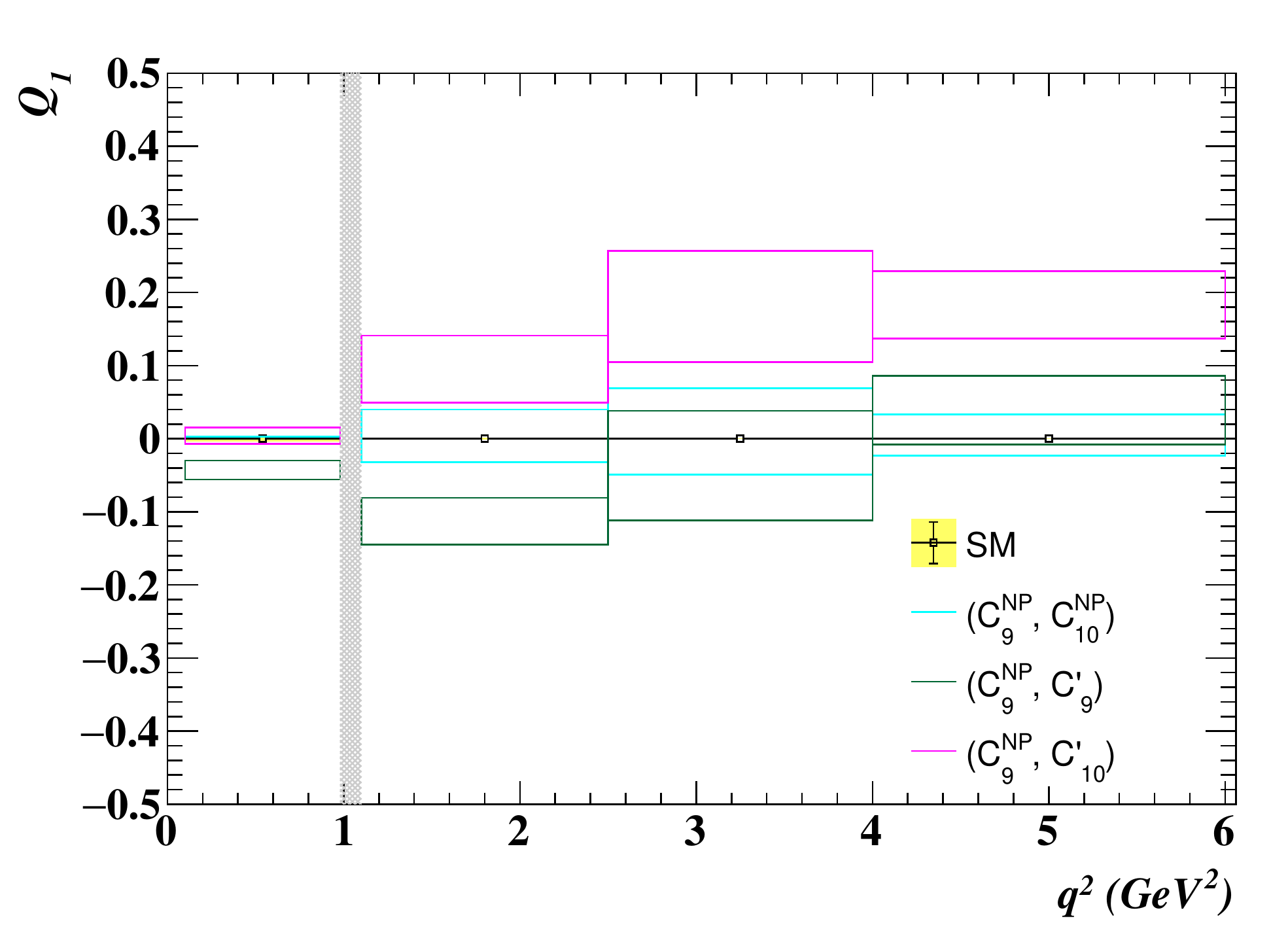}
 \includegraphics[width=8.9cm,height=6.0cm]{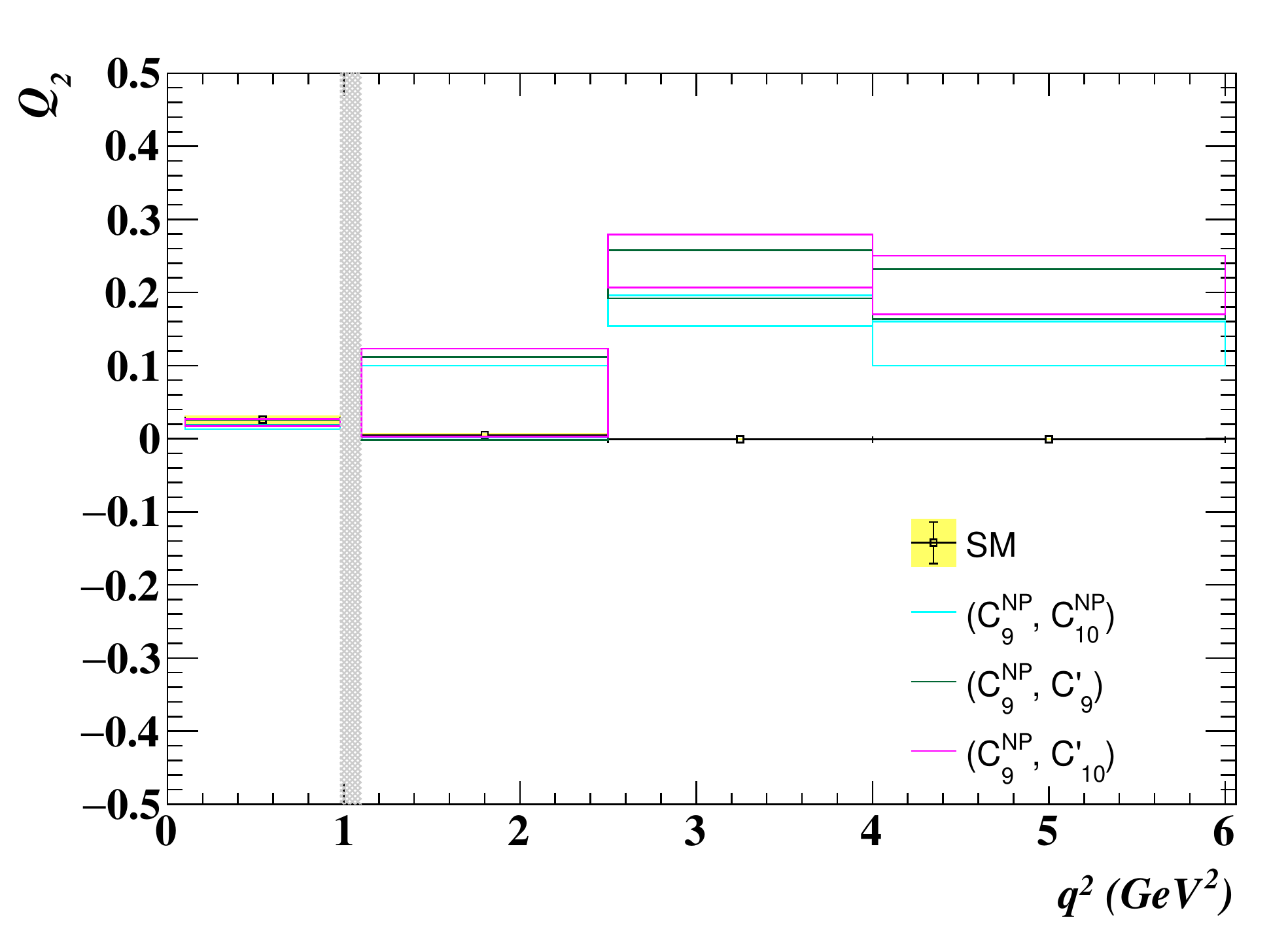}
 \includegraphics[width=8.9cm,height=6.0cm]{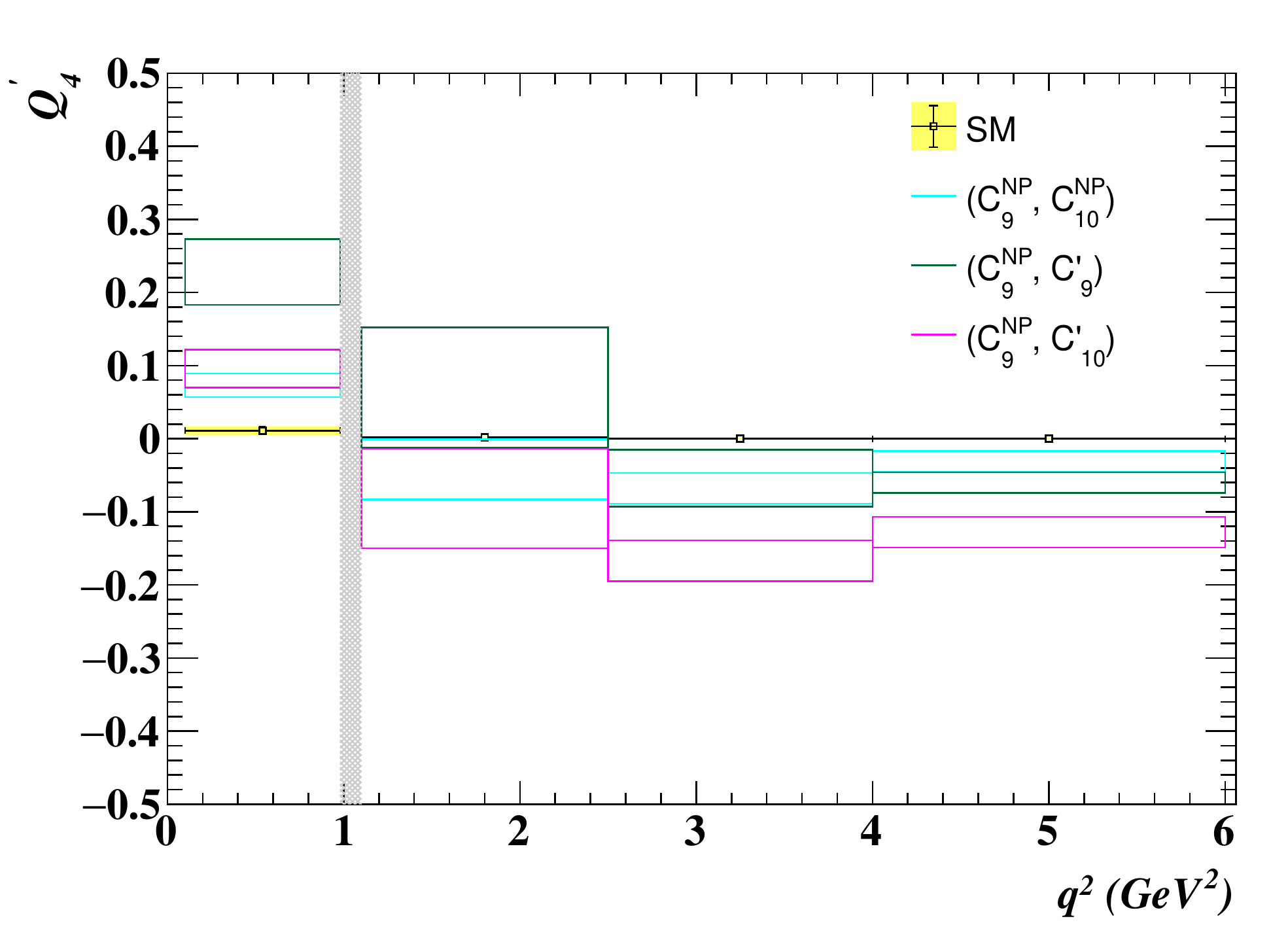}
 \includegraphics[width=8.9cm,height=6.0cm]{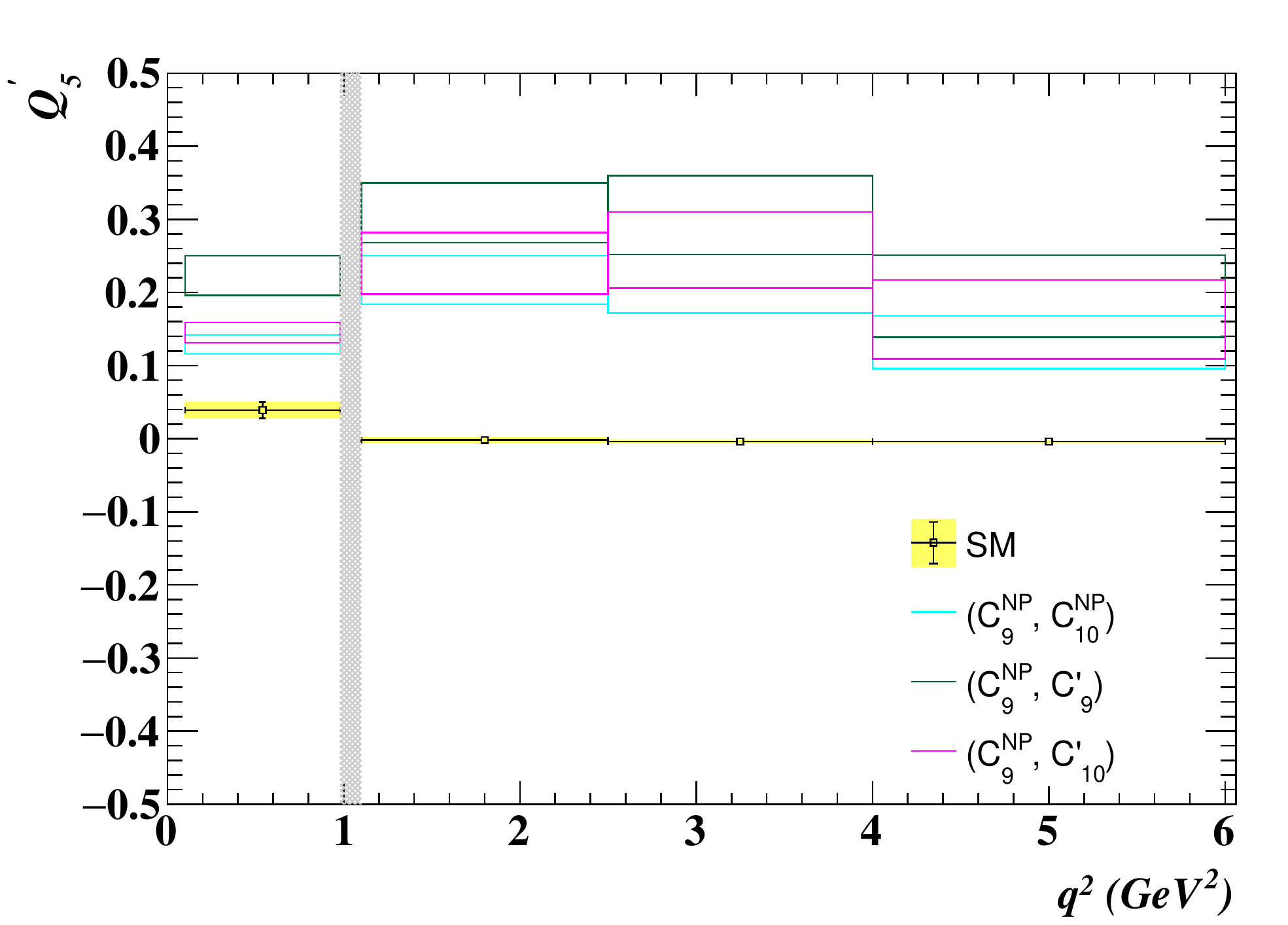}
 \includegraphics[width=8.9cm,height=6.0cm]{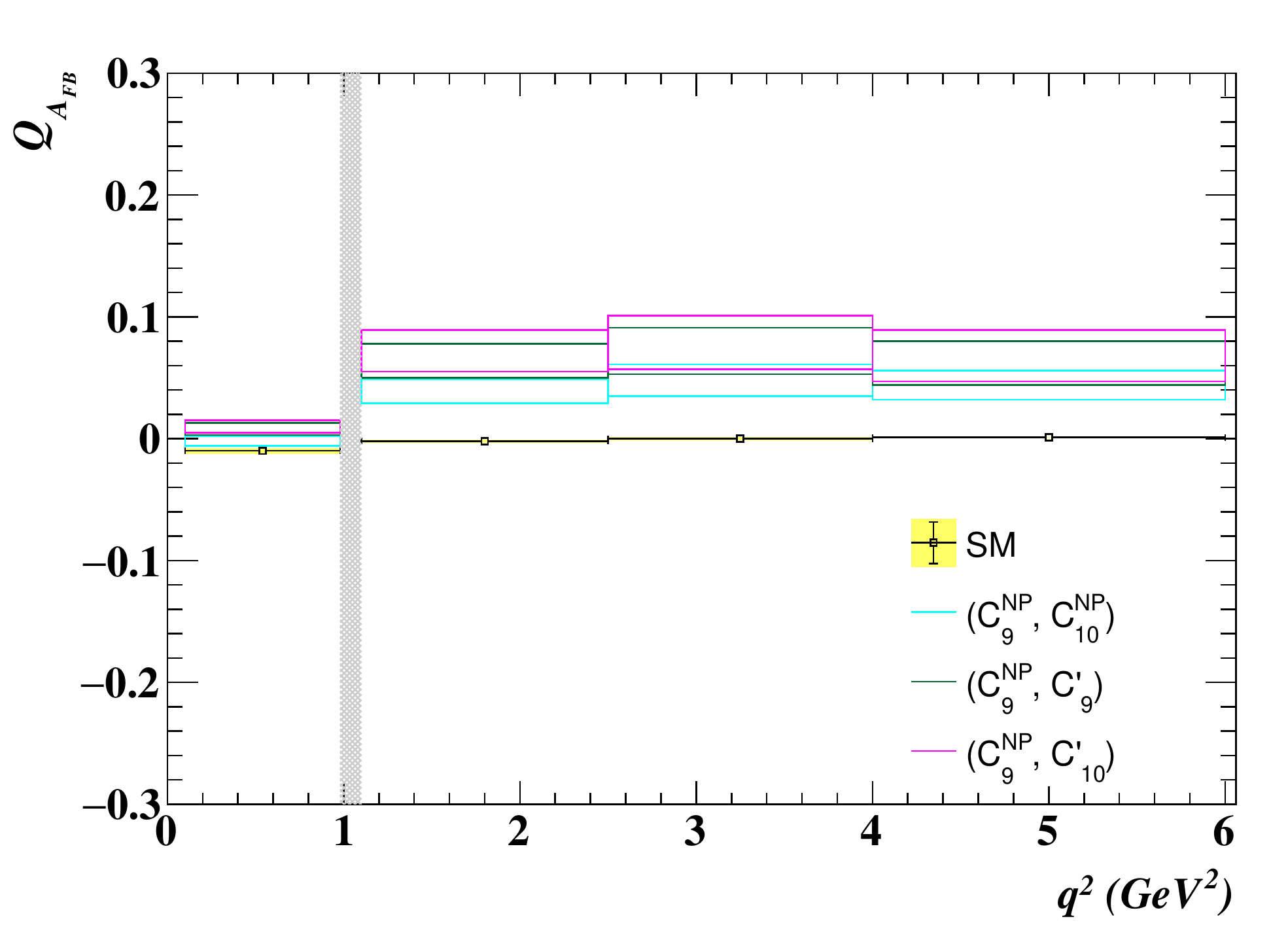}
 \includegraphics[width=8.9cm,height=6.0cm]{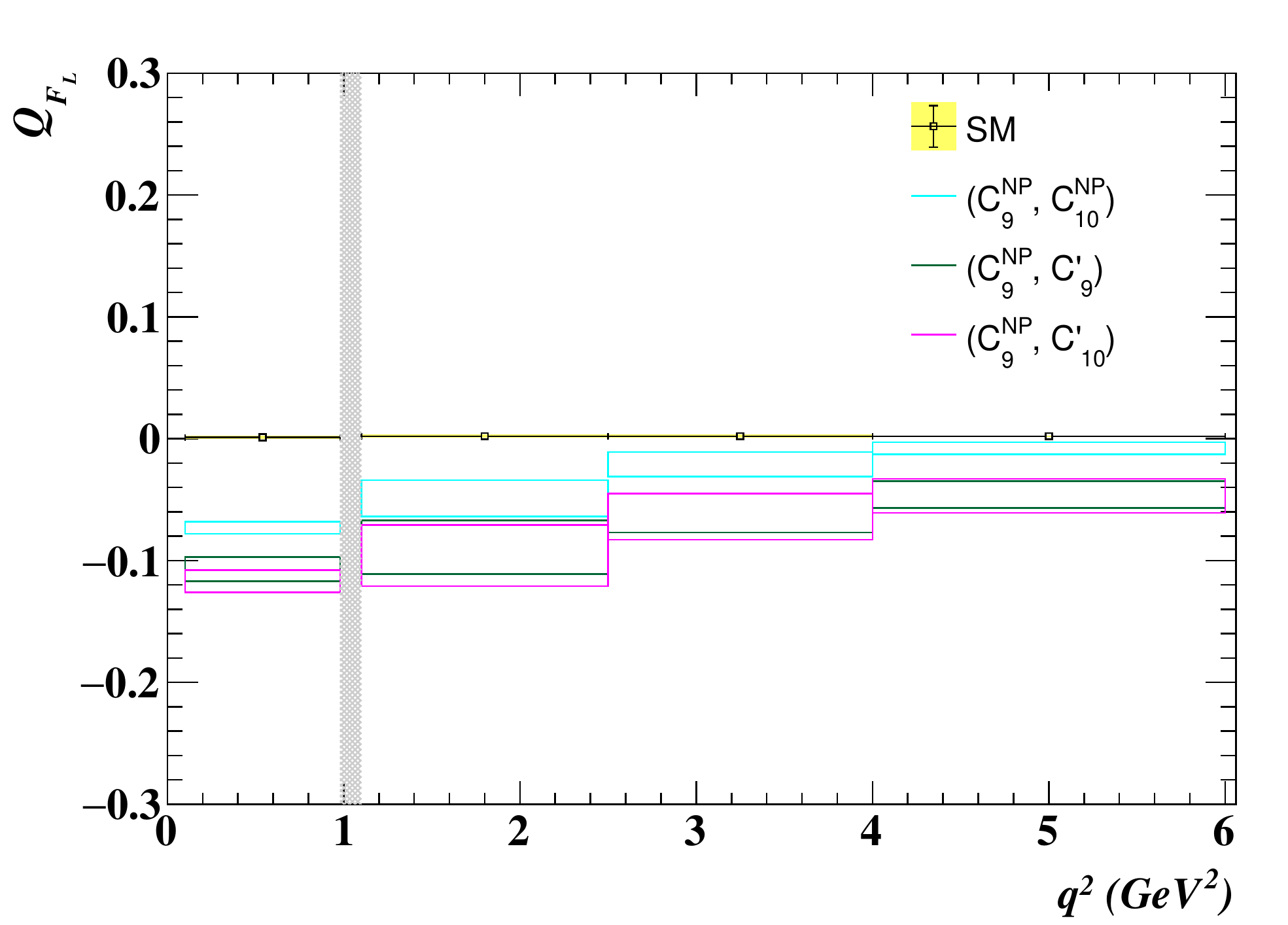}
 \caption{The central values and the corresponding $1\sigma$ error bands of various LFUV sensitive observables such as 
$\langle R_{f_{2}^{\prime}} \rangle$, $\langle Q_{i}^{(\prime)} \rangle$, $\langle Q_{A_{FB}} \rangle$, and $\langle Q_{F_L} \rangle$   
 in several $q^2$ bins in the SM and in the presence of three 2D NP scenarios.}
\label{fig_np2dlfuv}
\end{figure}

Study of LFUV in $B_s \to f_{2}^{\prime}(1525)(\to K^+\,K^-)\,\mu^+ \,\mu^-$ decays is interesting because
it is mediated via similar $b\to s\,l^+\,l^-$ quark level transition, and in principle, it can provide complementary information
regarding the anomalies present in $B\, \to\, (K\,,K^*)\, \mu^+\,\mu^-$ decay modes. 
We study the violation of LFU in two different 1D and 2D NP scenarios. We make a comparative study of the LFUV sensitive
observables such as $\langle R_{f_{2}^{\prime}} \rangle$, $\langle Q_{F_L} \rangle$, $\langle Q_{A_{FB}} \rangle$, and 
$\langle Q_{i}^{(\prime)} \rangle$ ($i\in1,2,4,5$) in the SM and in several $1D$ and $2D$ NP scenarios. 
We report in the Appendix in Tables~\ref{tab_npr},\ref{tab_npq1},
\ref{tab_npq2},\ref{tab_npq4},\ref{tab_npq5},\ref{tab_npqafb},\ref{tab_npqfl} the binned average values of each of the observables.
Similarly, the bin wise $q^2$ distribution plots for both $1D$ and $2D$ scenarios are shown in 
Figs.~\ref{fig_np1dlfuv} and \ref{fig_np2dlfuv}, respectively. Our observations are as follows:

\subsubsection{1D scenario}

\begin{itemize}
 
 \item $R_{f_{2}^{\prime}}$: Except in the low $q^2$ bin, all the NP scenarios are distinguishable
from the SM prediction at more than $5\sigma$ significance. Hence, a measurement of $R_{f_{2}^{\prime}}$ will be crucial to probe
NP in $b \to s\,l^+\,l^-$ transition decays.
 
 \item $Q_1$: The value of $Q_1$ obtained in case of ${C}_{9}^{NP}=-{C}_{9}^{\prime}$ NP scenario is distinguishable from the SM prediction 
at the level of $4-5\sigma$ significance in the $q^2 \in[0.045, 0.98]$ and $[1.1, 2.5]$ bins. 
 In the rest of the bins, although the central values obtained in each NP scenarios differ
 significantly from the SM, the SM band, however, overlaps with the NP band. 

\item $Q_2$: The value of $Q_2$ obtained in case of ${C}_{9}^{NP}$ and ${C}_{9}^{NP}=-{C}_{9}^{\prime}$ NP scenarios are distinguishable 
from the SM prediction at the level of more than $5\sigma$ significance in the region $q^2 \in[2.5, 6.0]$.
 
 \item $Q_4^{\prime}$: In the bin $q^2\in[1.1, 2.5]$, the ${C}_{10}^{NP}$ and ${C}_{9}^{NP}=-{C}_{10}^{NP}$ NP scenarios are distinguishable 
at $5-6\sigma$ from
 the SM. Although, the central values for ${C}_{9}^{NP}$ and ${C}_{9}^{NP}=-{C}_{9}^{\prime}$ differ significantly from the SM expectations,
 the associated error band is too large and the SM band overlaps with the NP band.
 Similarly, for $q^2 \geq 4$ the ${C}_{9}^{NP}=-{C}_{9}^{\prime}$ NP scenario is distinguishable at $4.8\sigma$ from the SM expectations.

 \item $Q_5^{\prime}$: In the bin $q^2\in[1.1, 2.5]$, the value of $Q_5^{\prime}$ obtained in case of ${C}_{9}^{NP}$, 
${C}_{9}^{NP}=-{C}_{10}^{NP}$ and ${C}_{9}^{NP}=-{C}_{9}^{\prime}$ NP scenarios are clearly distinguishable from the SM prediction at more 
than $5\sigma$ significance.
 Similarly, the ${C}_{9}^{NP}$ and ${C}_{9}^{NP}=-{C}_{9}^{\prime}$ NP scenarios
 are distinguishable at more than $3\sigma$ significance from the SM expectations for $q^2 \leq 4\,{\rm GeV^2}$. 
For $q^2 \geq 4\,{\rm GeV^2}$, the ${C}_{9}^{NP}$ 
and ${C}_{9}^{NP}=-{C}_{9}^{\prime}$ NP scenarios are clearly
distinguishable from the SM at the level of $4.4\sigma$ and $2.5\sigma$ significance, respectively.

 \item $Q_{A_{FB}}$: The value of $Q_{A_{FB}}$ obtained in case of ${C}_{9}^{NP}$, ${C}_{9}^{NP}=-{C}_{10}^{NP}$ and 
${C}_{9}^{NP}=-{C}_{9}^{\prime}$ NP scenarios are clearly distinguishable from the SM prediction
 at the level of more than $3\sigma$ significance, whereas, for the ${C}_{10}^{NP}$ NP scenario, it is SM 
like. 

 \item $Q_{F_L}$: In the low $q^2$ region, the value of $Q_{F_L}$ deviates significantly from the SM prediction for all the NP scenarios and
it is clearly distinguishable from the SM prediction at more than $5\sigma$ significance. Similarly, for $q^2 \geq 1$, except for 
${C}_{10}^{NP}$, the ${C}_{9}^{NP}$, ${C}_{9}^{NP}=-{C}_{10}^{NP}$ NP scenarios are distinguishable from the SM
at the level of $3\sigma$ significance.
 
\begin{figure}[htbp]
\centering
 \includegraphics[width=8.9cm,height=6.0cm]{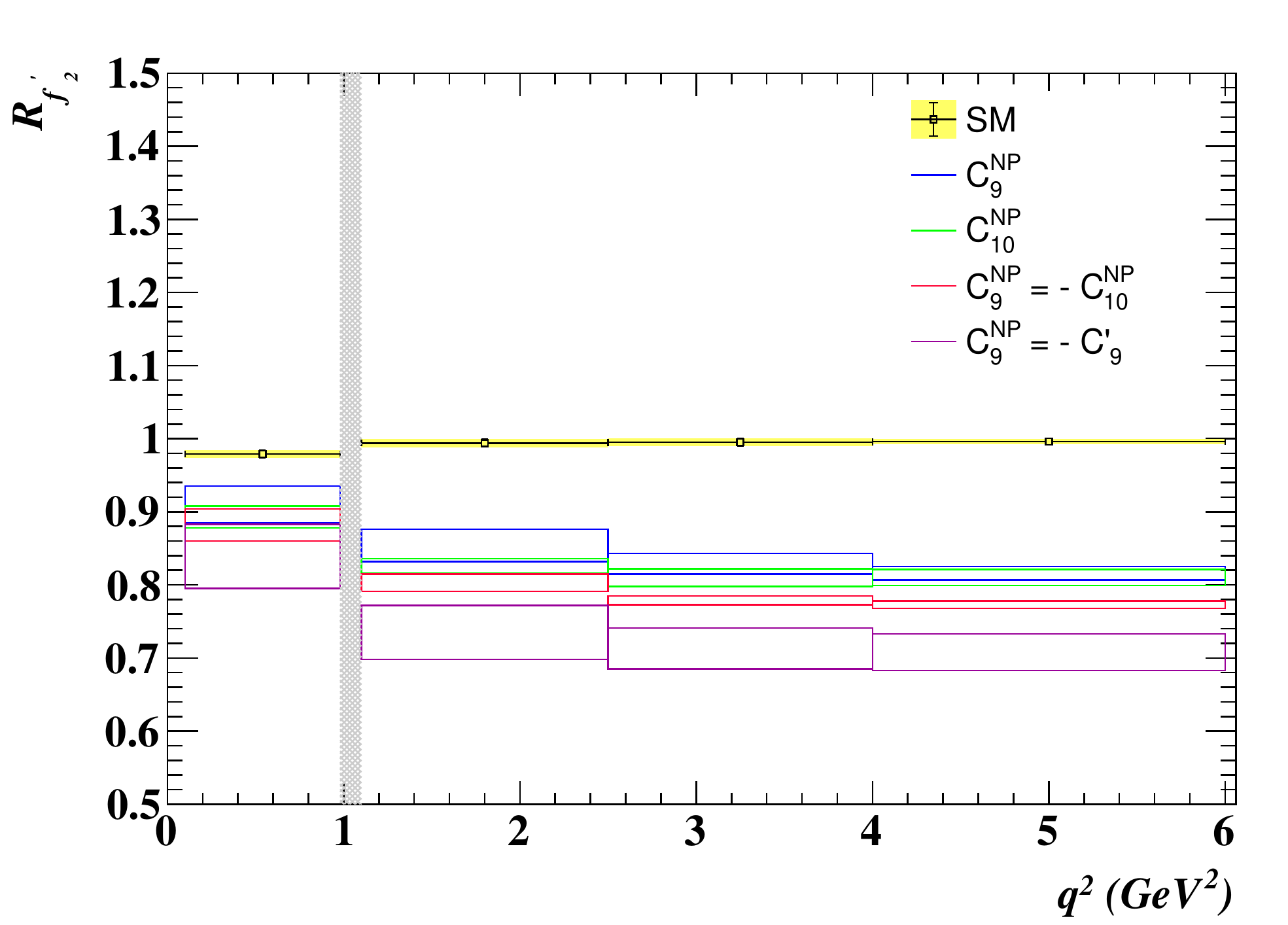}
 \includegraphics[width=8.9cm,height=6.0cm]{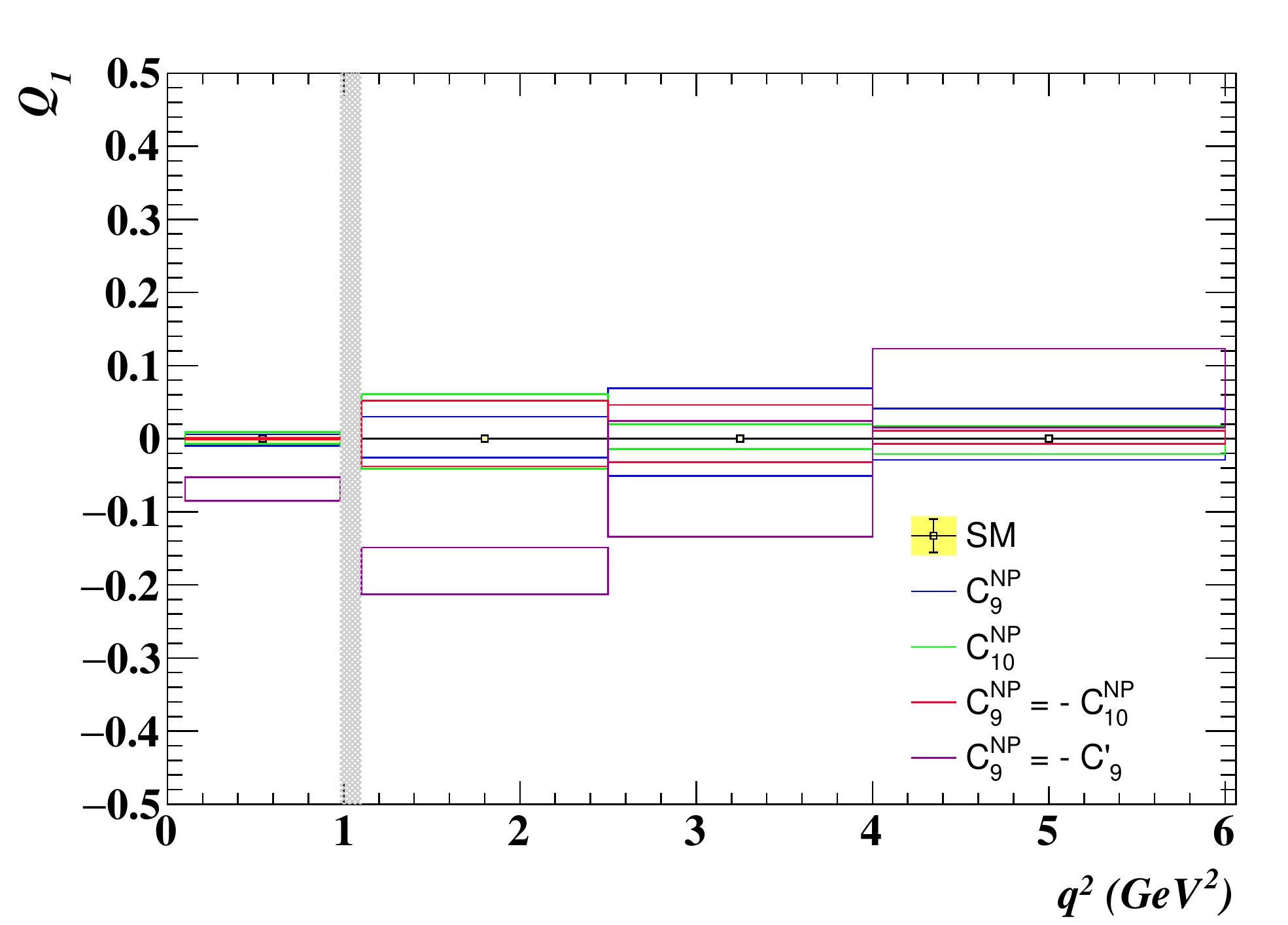}
 \includegraphics[width=8.9cm,height=6.0cm]{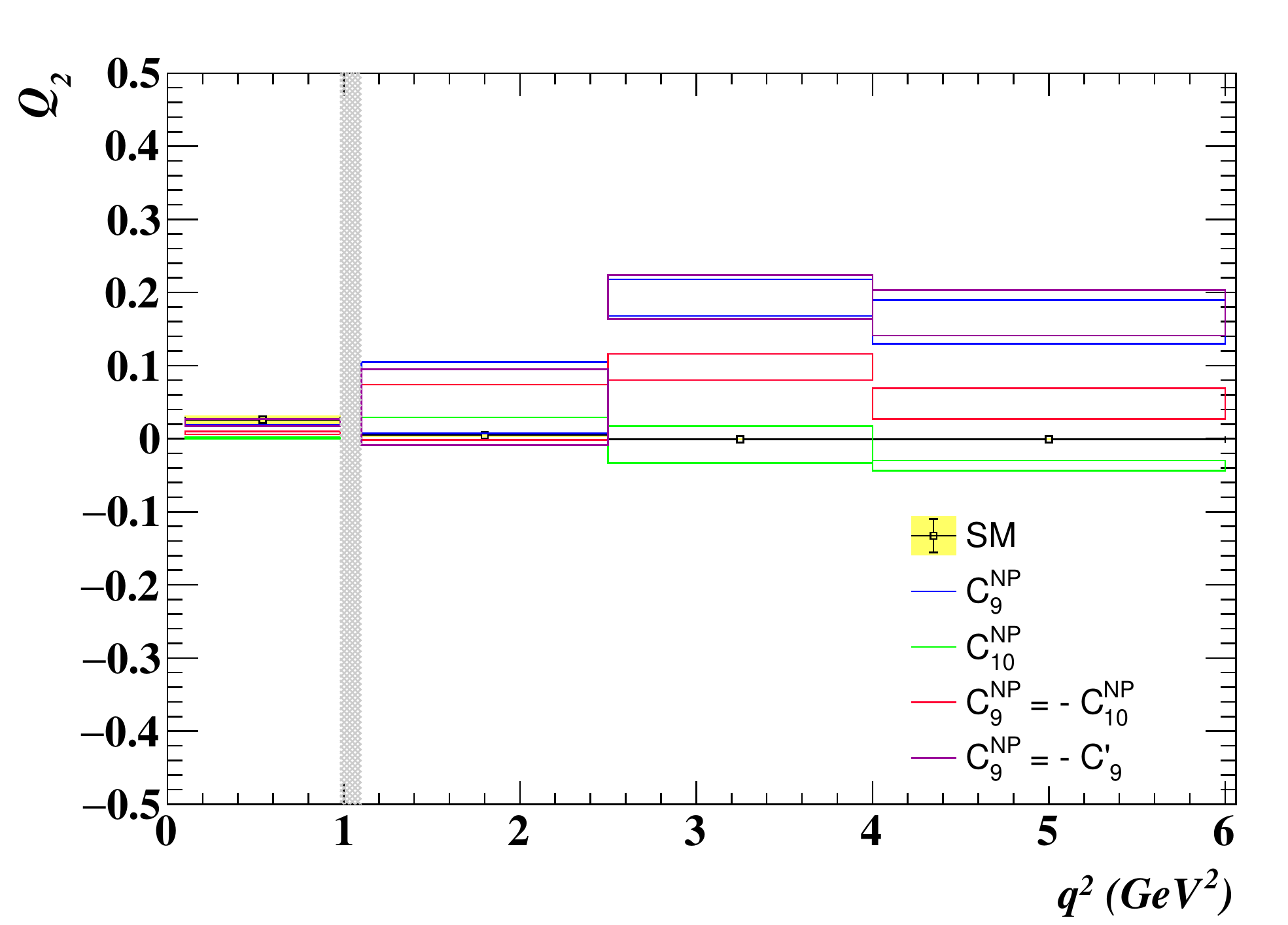}
 \includegraphics[width=8.9cm,height=6.0cm]{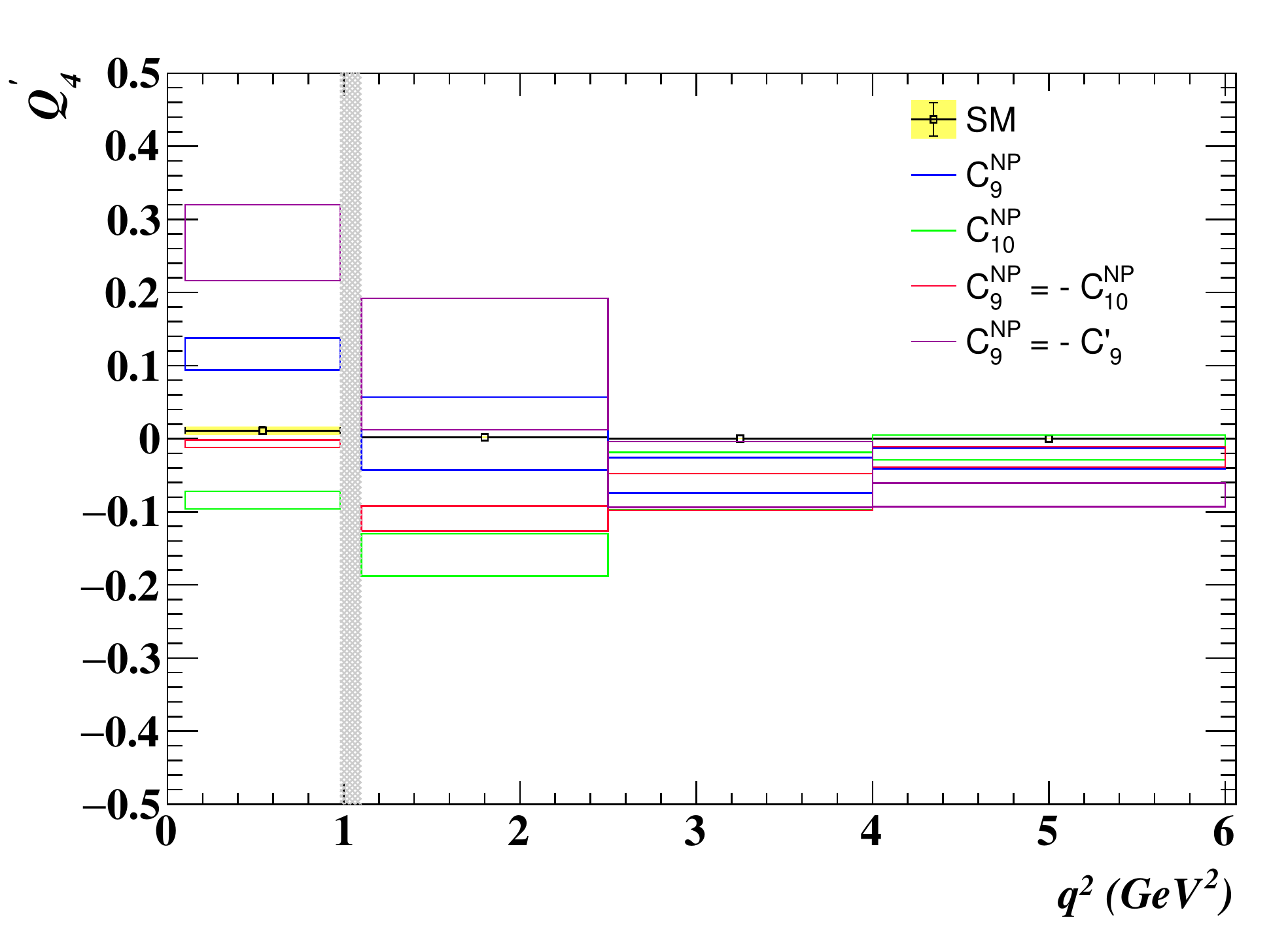}
 \includegraphics[width=8.9cm,height=6.0cm]{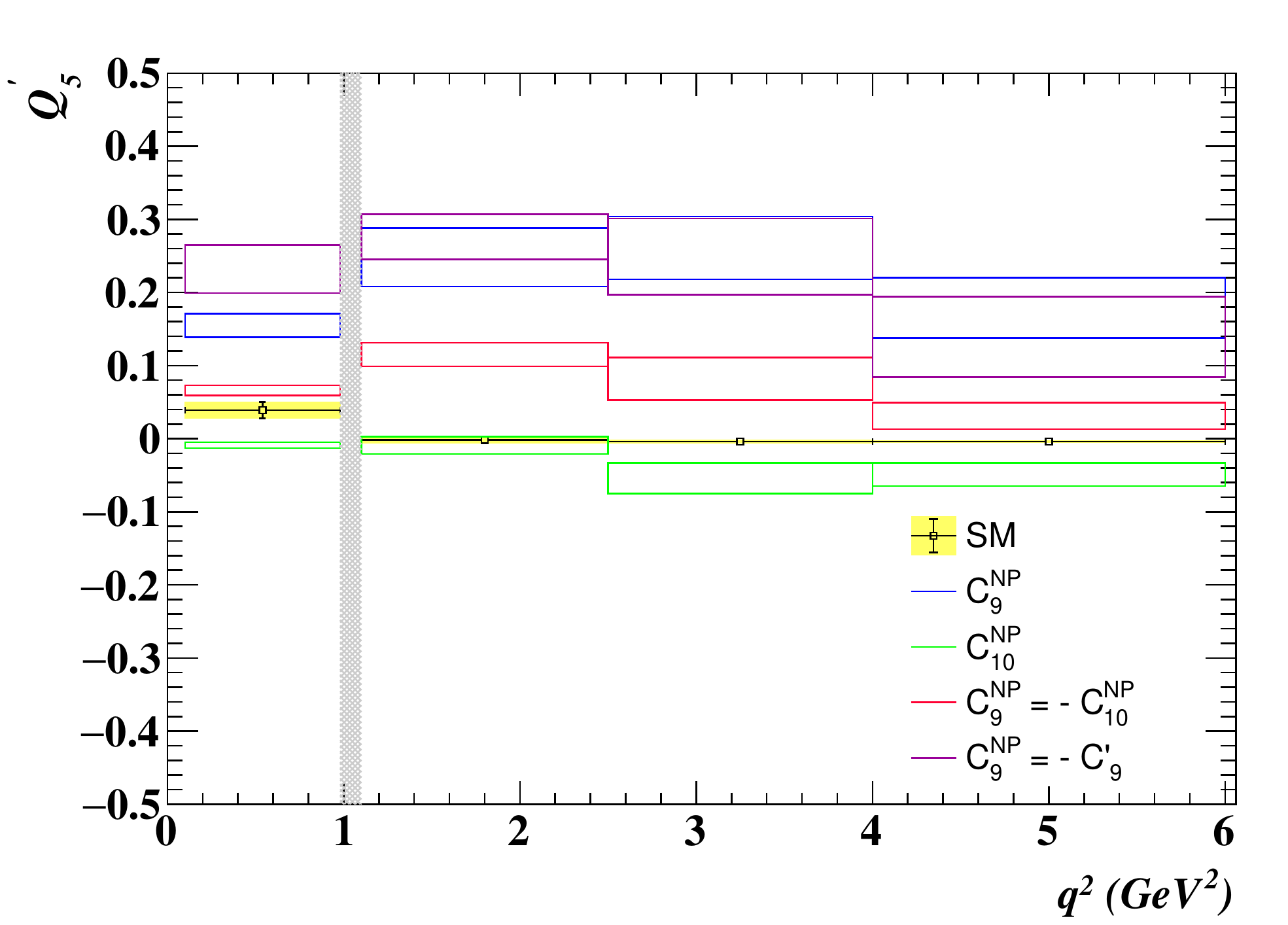}
 \includegraphics[width=8.9cm,height=6.0cm]{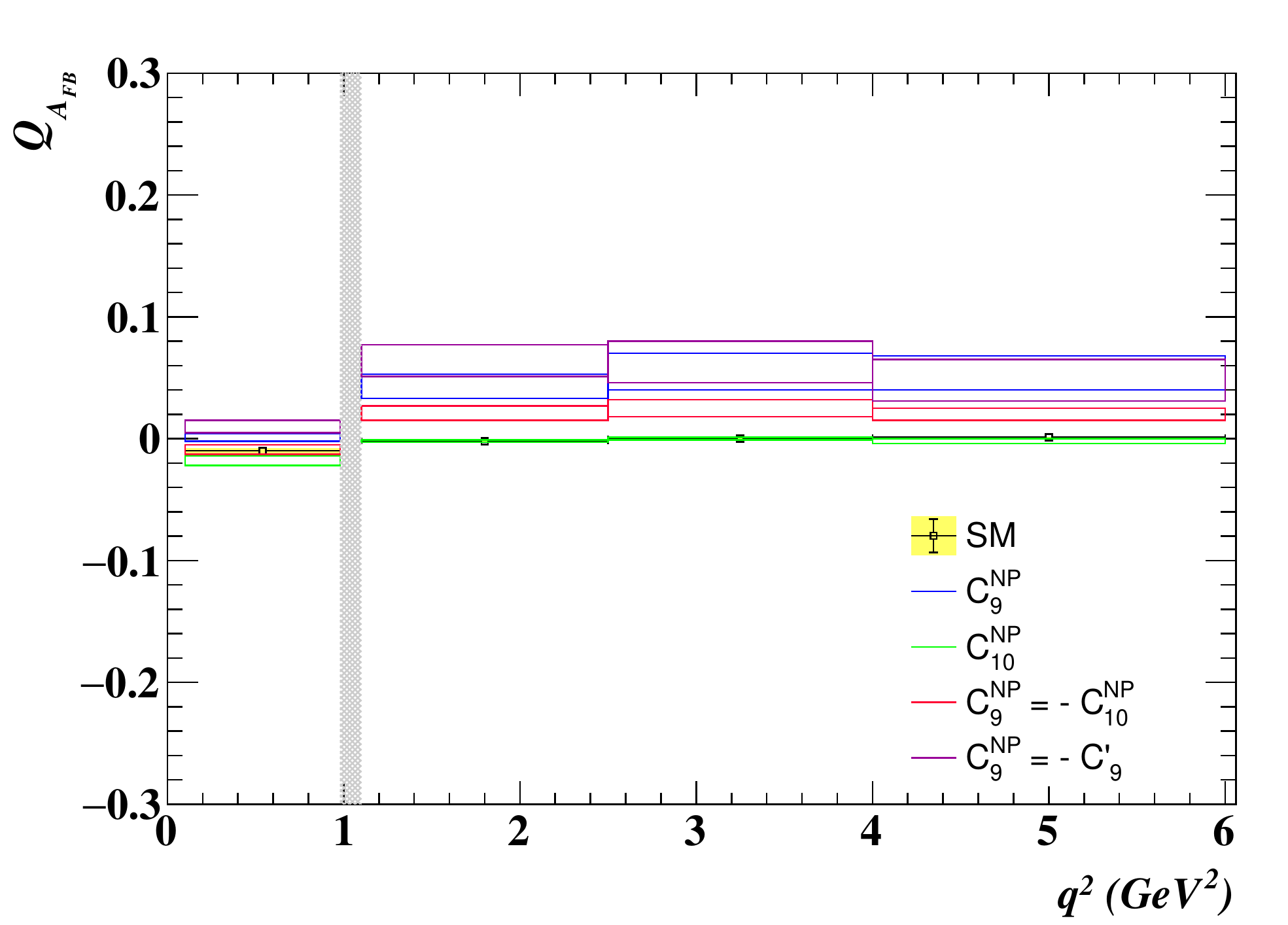}
 \includegraphics[width=8.9cm,height=6.0cm]{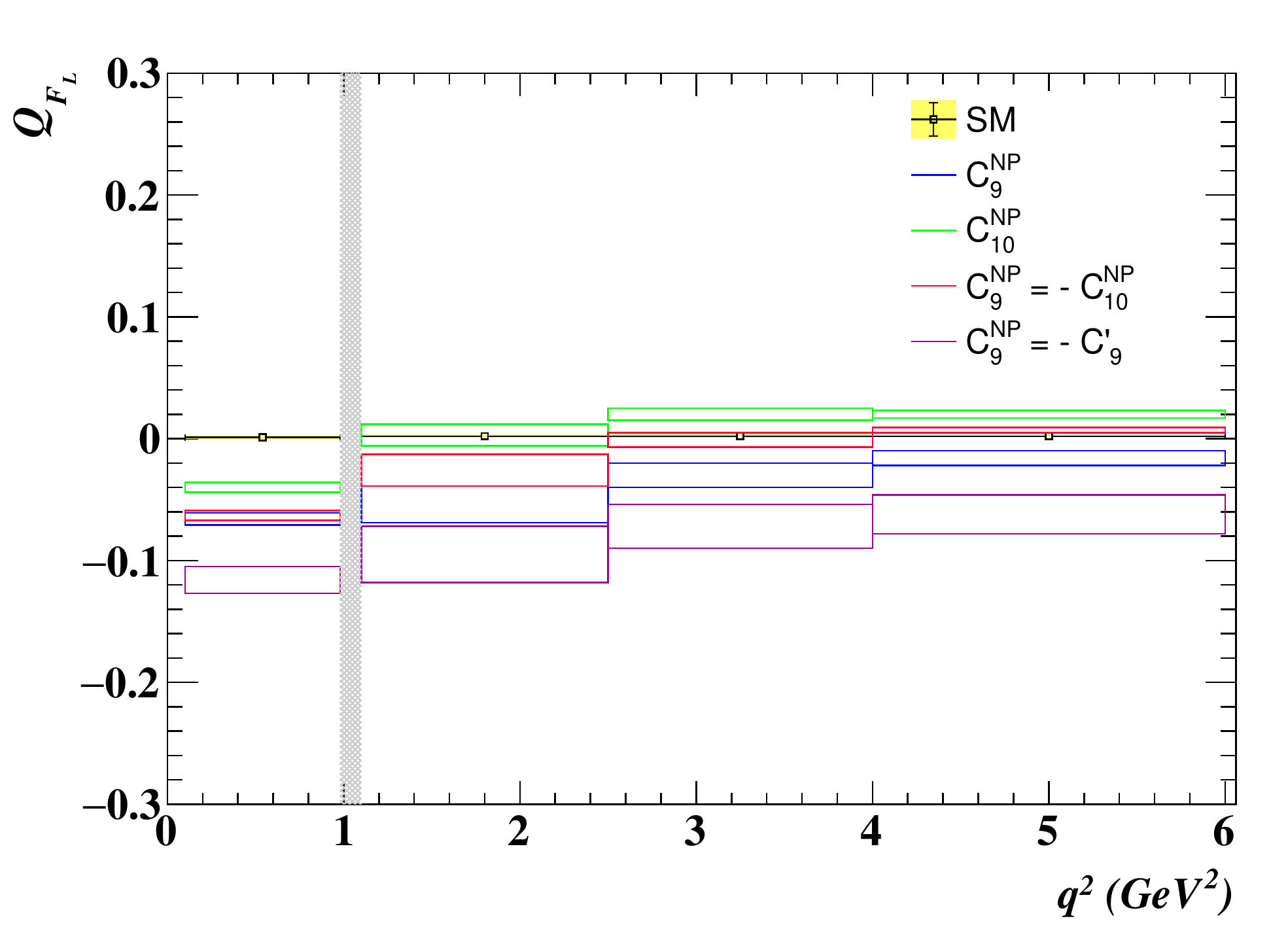}
\caption{The central values and the corresponding $1\sigma$ error bands of various LFUV sensitive observables such as 
$\langle R_{f_{2}^{\prime}} \rangle$, $\langle Q_{i}^{(\prime)} \rangle$, $\langle Q_{A_{FB}} \rangle$, and $\langle Q_{F_L} \rangle$   
 in several $q^2$ bins in the SM and in the presence of four 1D NP scenarios.}
\label{fig_np1dlfuv}
\end{figure}
 
\end{itemize}

\subsubsection{2D scenario}

\begin{itemize}

 \item $R_{f_{2}^{\prime}}$: All the NP scenarios are distinguishable at more than $3\sigma$ from the SM prediction
and in particular, the deviation of $R_{f_{2}^{\prime}}$ from the SM prediction in case of $({C}_{9}^{NP}, {C}_{9}^{\prime})$ and 
$({C}_{9}^{NP}, {C}_{10}^{\prime})$ NP scenarios are quite significant and it is clearly distinguishable from the SM prediction at more than 
$5\sigma$ significance.
 
 \item $Q_1$: The deviation observed in case of $({C}_{9}^{NP}, {C}_{10}^{\prime})$ NP scenario is clearly
 distinguishable from the SM prediction at more than $3\sigma$ significance in all $q^2$ bins. Again, for $({C}_{9}^{NP}, {C}_{9}^{\prime})$
NP Scenario, 
 although the central values differ significantly from the SM, the associated error band is too large in $q^2 \geq 2.5$ bins and 
 the SM value overlaps with the NP band.
 
 \item $Q_2$: No significant deviation is found in $q^2 \leq 2.5$ bins, whereas, for $q^2 \geq 2.5$ bin, the deviation observed in case of
$({C}_{9}^{NP}, {C}_{9}^{\prime})$ and $({C}_{9}^{NP}, {C}_{10}^{\prime})$ is quite significant and it is distinguishable from the SM
prediction at more than $5\sigma$ significance. 

 \item $Q_4^{\prime}$: In the low $q^2$ bin, the deviation observed in case of $({C}_{9}^{NP}, {C}_{9}^{\prime})$ is clearly distinguishable
from the SM prediction.
in $q^2 \in [2.5, 4.0]$ bin, the value of $Q_4^{\prime}$ obtained in case of $({C}_{9}^{NP}, {C}_{10}^{NP})$ is distinguishable from the 
SM prediction at $3\sigma$ significance, whereas, in case of $({C}_{9}^{NP}, {C}_{10}^{\prime})$ NP scenario, it
  is distinguishable at more than $5\sigma$ significance. Similarly, in $q^2 \geq 4$ bin, $({C}_{9}^{NP}, {C}_{9}^{\prime})$ and 
$({C}_{9}^{NP}, {C}_{10}^{\prime})$ NP scenarios are clearly distinguishable from the SM prediction at more than $4\sigma$ significance.

 \item $Q_5^{\prime}$: Although the deviation from the SM prediction is observed to be more pronounced in case of 
$({C}_{9}^{NP}, {C}_{9}^{\prime})$ NP scenario, the value of $Q_5^{\prime}$ obtained in each NP scenarios is clearly distinguishable 
from the SM prediction at more than $5\sigma$ significance.

\item $Q_{A_{FB}}$: We observe significant deviation from the SM prediction for each NP scenarios. It should be noted that the value of
$Q_{A_{FB}}$ obtained in each NP scenarios is clearly distinguishable from the SM prediction at more than $3\sigma$ significance.

 \item $Q_{F_L}$: In the low $q^2$ bin, all the three NP scenarios are clearly distinguishable from the SM at more than $5\sigma$ significance.
 Similarly, for $q^2 \geq 1$ bins, value of $Q_{F_L}$ obtained in case of $({C}_{9}^{NP}, {C}_{9}^{\prime})$ and 
$({C}_{9}^{NP}, {C}_{10}^{\prime})$ NP scenarios is distinguishable from the SM prediction at more than $3\sigma$ significance.
 
\end{itemize}

\section{Conclusion}
\label{conclusion}
In the light of the recent flavor anomalies reported in $B\, \to\, (K\,,K^*)\, \mu^+\,\mu^-$ and $B_s\, \to\, \phi\, \mu^+\,\mu^-$
decays, we analyze $B_s \to f_{2}^{\prime}(1525)\,\mu^+ \,\mu^-$ decays
mediated via similar $b\, \to\, s\, l^+\,l^-$ neutral current quark level transition. 
We perform a detailed angular study of the four body differential decay of $B_s \to f_{2}^{\prime}(1525)(\to K^+\,K^-)\,\mu^+ \,\mu^-$
within a model independent effective theory formalism. We give predictions of several observables in SM and in the presence of
various 1D and 2D NP scenarios proposed in several global fits. In the SM, we obtain the branching ratio of 
$B_s \to f_{2}^{\prime}(1525)(\to K^+\,K^-)\,\mu^+ \,\mu^-$ decays to be of the order of $\mathcal{O}(10^{-7})$.
We observe that the branching ratio is reduced at all $q^2$ for most of the NP cases. 
Except for $C_{10}^{NP}$, in all other NP scenarios, the zero crossing point for $A_{FB}(q^2)$ is shifted to the higher $q^2$ values than in 
the SM. In case of $F_{L}$, the peak seems to be reduced and shifted to the higher values of $q^2$ in comparison to the SM. 
It is worth mentioning that the zero crossing for 
$A_{FB}(q^2)$ is quite interesting and can, in principle, give useful information regarding lepton flavor universality violation in 
$b\, \to\, s\, l^+\,l^-$ transition decays. 
Importantly, we do observe significant contributions coming from ${C}_{9}^{NP}=-{C}_{9}^{\prime}$ in the 1D scenario and
$({C}_{9}^{NP}, {C}_{9}^{\prime})$ and $({C}_{9}^{NP}, {C}_{10}^{\prime})$ in the 2D scenario. 
Specially, these primed operators which corresponds to right handed currents seem to be very interesting.
As expected, the lepton flavor universal ratio $\langle R_{f_{2}^{\prime}} \rangle$, 
and other $Q$ observables such as $\langle Q_{i}^{(\prime)} \rangle$, $\langle Q_{A_{FB}} \rangle$, and $\langle Q_{F_L} \rangle$
are exceptionally clean observable with theoretical uncertainty of only 1$\%$ which makes them ideal candidates to probe NP in
$b\, \to\, s\, l^+\,l^-$ transition decays.
Although there have been several hints of NP reported in $b\, \to\, s\, l^+\,l^-$ transition decays, existence of NP is yet to
be confirmed.
Unlike $B\, \to\, (K\,,K^*)\, \mu^+\,\mu^-$ and $B_s\, \to\, \phi\, \mu^+\,\mu^-$ decays which have caught more attention of the theorist 
and experimentalists,
the $B_s \to f_{2}^{\prime}(1525)(\to K^+\,K^-)\,\mu^+ \,\mu^-$ decays mediated via the same quark level transitions has received less 
attention so far. Measurements of various observables for this decay mode in future can shed more light in identifying the exact
NP Lorentz structures. At the same time better theoretical understanding of the $B_s \to f_2^{\prime}$ transition form factors in future
will be crucial in disentangling genuine NP effects from the SM uncertainties.
More data samples are also needed in order to enhance the
significance of the various measurements and to reduce the statistical and systematic uncertainties to properly disentangle 
the NP effects.

\section*{Acknowledgements}
The author NS would like to thank Thomas Blake for several useful discussions.

%\FloatBarrier
 
\appendix
\label{appendix}
\section{Predictions of various physical observables in the SM and in the presence various 1D and 2D NP couplings for the 
$B_s \to f_{2}'(1525)(\to K^+\,K^-)\,\mu^+ \,\mu^-$ decays}
 
 %%%%%%%%%%%%%%%%%%%%%%%%%%%%%%%%%%%%%%%%%%%%%TABLE -DBR
 
 \begin{table}[ht!]
 \centering
 \resizebox{\columnwidth}{!}{
\begin{tabular}{|c|c|c|c|c|c|c|c|c|}
  \hline
  \hline
  \multirow{2}{*}{$q^{2}$ bins (GeV$^2$)}
  & \multicolumn{8}{|c|}{BR$\times 10^{-7}$} \\
  \cline{2-9} 
  & SM & $C_{9}^{NP}$ & $C_{10}^{NP}$ & $C_{9}^{NP}$=$-C_{10}^{NP}$ & $C_{9}^{NP}$=$-C_{9}^{\prime}$ & $(C_{9}^{NP}$, $C_{10}^{NP}$) & $(C_{9}^{NP}$, $C_{9}^{\prime}$) & $(\
C_{9}^{NP}$, $C_{10}^{\prime}$) \\
     \hline
     \hline
      [0.10, 0.98] & $0.114 \pm  0.021$ & $0.106 \pm  0.019$ & $0.104 \pm  0.019$ & $0.102 \pm  0.019$ & $0.097 \pm  0.018$ & $0.103 \pm  0.019$ & $0.099 \pm  0.018$ & $0.098 \pm  0.018$  \\
      \hline
      [1.1, 2.5]   & $0.105 \pm  0.025$ & $0.090 \pm  0.020$ & $0.087 \pm  0.021$ & $0.084 \pm  0.020$ & $0.077 \pm  0.016$ & $0.086 \pm  0.019$ & $0.080 \pm  0.017$ & $0.077 \pm  0.016$  \\
      \hline
      [2.5, 4.0]   & $0.110 \pm  0.026$ & $0.092 \pm  0.021$ & $0.090 \pm  0.022$ & $0.086 \pm  0.021$ & $0.078 \pm  0.017$  & $0.087 \pm  0.020$ & $0.081 \pm  0.018$ & $0.078 \pm  0.017$ \\
      \hline
      [4.0, 6.0]   & $0.153 \pm  0.035$ & $0.125 \pm  0.028$ & $0.125 \pm  0.030$ & $0.119 \pm  0.027$ & $0.108 \pm  0.023$ & $0.119 \pm  0.027$ & $0.111 \pm  0.024$ & $0.107 \pm  0.023$ \\
      \hline
%      [6.0, 8.0]   & $0.156 \pm  0.035$ & $0.126 \pm  0.028$ & $0.129 \pm  0.030$ & $0.121 \pm  0.028$ & $0.110 \pm  0.023$ & $0.121 \pm  0.027$ & $0.112 \pm  0.024$ & $0.108 \pm  0.024$  \\
%      \hline
      [1.1, 6.0]   & $0.368 \pm  0.085$ & $0.307 \pm  0.068$ & $0.302 \pm  0.071$ & $0.290 \pm  0.067$ &  $0.264 \pm  0.055$ & $0.292 \pm  0.065$ & $0.271 \pm  0.057$ & $0.262 \pm  0.056$ \\
      \hline
      [0.045, 6.0]   & $0.512 \pm  0.103$ & $0.440 \pm  0.083$ & $0.434 \pm  0.086$ & $0.420 \pm  0.081$ & $0.388 \pm  0.068$ & $0.423 \pm  0.080$ & $0.398 \pm  0.071$ & $0.387\pm  0.069$ \\
      \hline
 \hline
 \end{tabular}}
 \caption{The binned average central values and the corresponding $1\sigma$ uncertainties for the branching ratio BR$\times 10^{-7}$ 
of $B_s \to f_{2}'(1525)(\to K^+\,K^-)\,\mu^+ \,\mu^-$ decays in the SM and in several NP cases.}
 \label{tab_npdbr}
 \end{table}
% 
% 
% %%%%%%%%%%%%%%%%%%%%%%%%%%%%%%%%%%%%%%%%%%%%%TABLE -FL
% 
 \begin{table}[ht!]
 \centering
 \resizebox{\columnwidth}{!}{
 \begin{tabular}{|c|c|c|c|c|c|c|c|c|}
  \hline
  \hline
  \multirow{2}{*}{$q^{2}$ bins (GeV$^2$)}
  & \multicolumn{8}{|c|}{$\langle F_L \rangle$} \\
  \cline{2-9}
      & SM & $C_{9}^{NP}$ & $C_{10}^{NP}$ & $C_{9}^{NP}$=$-C_{10}^{NP}$ & $C_{9}^{NP}$=$-C_{9}^{\prime}$ & $(C_{9}^{NP}$, $C_{10}^{NP}$) & $(C_{9}^{NP}$, $C_{9}^{\prime}$) & $(C_{9}^{NP}$, $C_{10}^{\prime}$) \\
     \hline
     \hline

     [0.10, 0.98] & $ 0.503 \pm  0.108$ & $0.436 \pm  0.105$ & $0.462 \pm  0.110$ & $0.439 \pm  0.107$ & $0.386 \pm  0.100$ & $0.429 \pm  0.105$ & $0.394 \pm  0.101$  & $0.385 \pm  0.101$ \\
     \hline
     [1.1, 2.5]   & $0.855 \pm  0.047$ & $0.799 \pm  0.061$ & $0.856 \pm  0.051$ & $0.827 \pm  0.058$ & $0.758 \pm  0.069$ & $0.804 \pm  0.062$ & $0.764 \pm  0.068$ & $0.757 \pm  0.071$ \\
     \hline
     [2.5, 4.0]   & $0.843 \pm  0.045$ & $0.811 \pm  0.049$ & $0.861 \pm  0.041$ & $0.840 \pm  0.044$ & $0.769 \pm  0.058$ & $0.819 \pm  0.048$ & $0.780 \pm  0.055$ & $0.777 \pm  0.057$ \\
     \hline
      [4.0, 6.0]   & $0.762 \pm  0.062$ & $0.745 \pm  0.062$ & $0.780 \pm  0.061$ & $0.767 \pm  0.061$ & $0.698 \pm  0.070$ & $0.752 \pm  0.062$ & $0.714 \pm  0.067$ &  $0.713 \pm  0.068$ \\
     \hline
%      [6.0, 8.0]   & $0.673 \pm  0.069$ & $0.664 \pm  0.069$ & $0.687 \pm  0.070$ & $0.679 \pm  0.069$ & $0.615 \pm  0.075$  & $0.670 \pm  0.069$ & $0.634 \pm  0.072$  & $0.635 \pm  0.073$ \\
%     \hline
      [1.1, 6.0]   & $0.812 \pm  0.050$ & $0.779 \pm  0.054$ & $0.825 \pm  0.047$ & $0.805 \pm  0.050$ & $0.735 \pm  0.062$  & $0.786 \pm  0.053$ & $0.747 \pm  0.060$ & $0.744 \pm  0.061$ \\
     \hline
      [0.045, 6.0]   & $0.712 \pm  0.069$ & $0.662 \pm  0.077$ & $0.700 \pm  0.074$ & $0.677 \pm  0.077$ & $0.611 \pm  0.081$ & $0.662 \pm  0.078$ & $0.622 \pm  0.080$  & $0.615 \pm  0.083$ \\
     \hline
 \hline
 \end{tabular}}
 \caption{The binned average central values and the corresponding $1\sigma$ uncertainties for the normalized longitudinal polarization 
fraction $\langle F_L \rangle$ of $B_s \to f_{2}'(1525)(\to K^+\,K^-)\,\mu^+ \,\mu^-$ decays in the SM and in several NP cases.}
 \label{tab_npfl}
 \end{table}
% 
% %%%%%%%%%%%%%%%%%%%%%%%%%%%%%%%%%%%%%%%%%%%%%TABLE -AFB
% 
 \begin{table}[ht!]
 \centering
 \resizebox{\columnwidth}{!}{
 \begin{tabular}{|c|c|c|c|c|c|c|c|c|}
  \hline
  \hline
  \multirow{2}{*}{$q^{2}$ bins (GeV$^2$)}
  & \multicolumn{8}{|c|}{$\langle A_{FB} \rangle$} \\
  \cline{2-9}
  & SM & $C_{9}^{NP}$ & $C_{10}^{NP}$ & $C_{9}^{NP}$=$-C_{10}^{NP}$ & $C_{9}^{NP}$=$-C_{9}^{\prime}$ & $(C_{9}^{NP}$, $C_{10}^{NP}$) & $(C_{9}^{NP}$, $C_{9}^{\prime}$) & $(C_{9}^{NP}$, $C_{10}^{\prime}$) \\ 
     \hline
     \hline
      [0.10, 0.98] & $0.086 \pm  0.016$ & $0.098 \pm  0.016$ & $0.079 \pm  0.013$ & $0.087 \pm  0.014$ & $0.106 \pm  0.016$ & $0.094 \pm  0.015$ & $0.105 \pm  0.016$ & $0.106 \pm  0.017$ \\
      \hline
      [1.1, 2.5]   & $0.082 \pm  0.036$ & $0.127 \pm  0.042$ & $0.083 \pm  0.036$ & $0.105 \pm  0.039$ & $0.148 \pm  0.045$ & $0.123 \pm  0.041$ &  $0.149 \pm  0.045$  & $0.156 \pm  0.047$ \\
      \hline
      [2.5, 4.0]   & $-0.014 \pm  0.039$ & $0.040 \pm  0.042$ & $-0.015 \pm  0.040$ & $0.010 \pm  0.042$ & $0.049 \pm  0.048$ & $0.033 \pm  0.042$ & $0.057 \pm  0.046$ & $0.065 \pm  0.048$ \\
      \hline
      [4.0, 6.0]   & $-0.116 \pm  0.049$ & $-0.063 \pm  0.047$ & $-0.119 \pm  0.051$ & $-0.097 \pm  0.050$ & $-0.068 \pm  0.053$ & $-0.072 \pm  0.049$ & $-0.054 \pm  0.051$ &  $-0.049 \pm  0.051$  \\
      \hline
%      [6.0, 8.0]   & $-0.204 \pm  0.056$ & $-0.159 \pm  0.053$ & $-0.207 \pm  0.057$ & $-0.192 \pm  0.056$ & $-0.177 \pm  0.057$ & $-0.170 \pm  0.055$ & $-0.160 \pm  0.055$ & $-0.156 \pm  0.055$ \\
%      \hline
      [1.1, 6.0]   & $-0.030 \pm  0.040$ & $0.023 \pm  0.041$ & $-0.030 \pm  0.040$ & $-0.007 \pm  0.041$ & $0.029 \pm  0.046$  & $0.016 \pm  0.041$ & $0.038 \pm  0.044$  & $0.045 \pm  0.046$  \\
      \hline
      [0.045, 6.0]   & $-0.000 \pm  0.030$ & $0.042 \pm  0.030$ & $-0.000 \pm  0.030$ & $0.019 \pm  0.030$ & $0.048 \pm  0.032$  & $0.036 \pm  0.030$ & $0.054 \pm  0.031$ & $0.059 \pm  0.032$ \\
      \hline
 \hline
 \end{tabular}}
 \caption{The binned average central values and the corresponding $1\sigma$ uncertainties for the normalized forward-backward asymmetry 
 $\langle A_{FB} \rangle$ of $B_s \to f_{2}'(1525)(\to K^+\,K^-)\,\mu^+ \,\mu^-$ decays in the SM and in several NP cases.}
 \label{tab_npafb}
 \end{table}
% 
% %%%%%%%%%%%%%%%%%%%%%%%%%%%%%%%%%%%%%%%%%%%%%TABLE -P1
% 
 \begin{table}[ht!]
 \centering
 \resizebox{\columnwidth}{!}{
\begin{tabular}{|c|c|c|c|c|c|c|c|c|}
  \hline
  \hline
  \multirow{2}{*}{$q^{2}$ bins (GeV$^2$)}
  & \multicolumn{8}{|c|}{$\langle P_1 \rangle$} \\
      \cline{2-9}
      & SM & $C_{9}^{NP}$ & $C_{10}^{NP}$ & $C_{9}^{NP}$=$-C_{10}^{NP}$ & $C_{9}^{NP}$=$-C_{9}^{\prime}$ & $(C_{9}^{NP}$, $C_{10}^{NP}$) & $(C_{9}^{NP}$, $C_{9}^{\prime}$) & $(C_{9}^{NP}$, $C_{10}^{\prime}$) \\
     \hline
     \hline
     [0.10, 0.98] & $-0.008 \pm  0.267$ & $-0.010 \pm  0.259$ & $-0.007 \pm  0.270$ & $-0.009 \pm  0.265$ & $-0.077 \pm  0.258$ & $-0.010 \pm  0.261$ & $-0.051 \pm  0.257$ & $-0.004 \pm  0.257$ \\
      \hline
     [1.1, 2.5]   & $-0.043 \pm  0.197$ & $-0.041 \pm  0.203$ & $-0.034 \pm  0.224$ & $-0.036 \pm  0.217$ & $-0.224 \pm  0.192$ & $-0.039 \pm  0.209$ & $-0.156 \pm  0.198$ & $0.052 \pm  0.204$ \\
      \hline
     [2.5, 4.0]   & $-0.112 \pm  0.240$ & $-0.102 \pm  0.204$ & $-0.109 \pm  0.241$ & $-0.104 \pm  0.218$ & $-0.166 \pm  0.212$ & $-0.102 \pm  0.206$ & $-0.148 \pm  0.201$ & $0.069 \pm  0.198$ \\
      \hline
     [4.0, 6.0]   & $-0.159 \pm  0.282$ & $-0.154 \pm  0.256$ & $-0.161 \pm  0.297$ & $-0.158 \pm  0.277$ & $-0.090 \pm  0.264$  & $-0.155 \pm  0.261$ & $-0.120 \pm  0.256$ & $0.024 \pm  0.254$ \\
      \hline
%     [6.0, 8.0]   & $-0.211 \pm  0.279$ & $-0.209 \pm 0.267$ & $-0.212 \pm  0.291$ & $-0.211 \pm  0.282$ & $-0.074 \pm  0.276$ & $-0.209 \pm  0.272$ & $-0.130 \pm  0.271$ & $-0.047 \pm  0.273$  \\
%      \hline
     [1.1, 6.0]   & $-0.121 \pm  0.222$ & $-0.110 \pm  0.196$ & $-0.117 \pm  0.219$ & $-0.112 \pm  0.204$ & $-0.141 \pm  0.201$ & $-0.110 \pm  0.196$ & $-0.135 \pm  0.193$ & $0.043 \pm  0.193$ \\
      \hline
     [0.045, 6.0]   & $-0.074 \pm  0.175$ & $-0.067 \pm  0.177$ & $-0.066 \pm  0.178$ & $-0.064 \pm  0.179$ & $-0.106 \pm  0.176$  & $-0.065 \pm  0.178$ & $-0.094 \pm  0.176$ & $0.024 \pm  0.178$ \\      
      \hline
 \hline
 \end{tabular}}
 \caption{The binned average central values and the corresponding $1\sigma$ uncertainties for $\langle P_1 \rangle$ 
 of $B_s \to f_{2}'(1525)(\to K^+\,K^-)\,\mu^+ \,\mu^-$ decays in the SM and in several NP cases.}
 \label{tab_npp1}
 \end{table}
% 
% %%%%%%%%%%%%%%%%%%%%%%%%%%%%%%%%%%%%%%%%%%%%%TABLE -P2
% 
 \begin{table}[ht!]
 \centering
 \resizebox{\columnwidth}{!}{
\begin{tabular}{|c|c|c|c|c|c|c|c|c|}
  \hline
  \hline
  \multirow{2}{*}{$q^{2}$ bins (GeV$^2$)}
  & \multicolumn{8}{|c|}{$\langle P_2 \rangle$} \\
      \cline{2-9}
      & SM & $C_{9}^{NP}$ & $C_{10}^{NP}$ & $C_{9}^{NP}$=$-C_{10}^{NP}$ & $C_{9}^{NP}$=$-C_{9}^{\prime}$ & $(C_{9}^{NP}$, $C_{10}^{NP}$) & $(C_{9}^{NP}$, $C_{9}^{\prime}$) & $(C_{9}^{NP}$, $C_{10}^{\prime}$) \\
     \hline
     \hline
     [0.10, 0.98] & $0.158 \pm  0.029$ & $0.156 \pm  0.029$ & $0.133 \pm  0.025$ & $0.141 \pm  0.026$ & $0.155 \pm  0.028$ & $0.149 \pm  0.028$ & $0.155 \pm  0.028$ & $0.154 \pm  0.029$  \\
      \hline
     [1.1, 2.5]   & $0.378 \pm  0.081$ & $0.429 \pm  0.038$ & $0.383 \pm  0.070$ & $0.409 \pm  0.050$ & $0.416 \pm  0.042$ & $0.425 \pm  0.039$ & $0.428 \pm  0.035$ & $0.436 \pm  0.031$ \\
       \hline
     [2.5, 4.0]   & $-0.046 \pm  0.158$ & $0.148 \pm  0.143$ & $-0.054 \pm  0.181$ & $0.053 \pm  0.169$ & $0.148 \pm  0.136$ & $0.129 \pm  0.151$ & $0.179 \pm  0.135$ & $0.197\pm  0.134$ \\
      \hline
     [4.0, 6.0]   & $-0.315 \pm  0.082$ & $-0.154 \pm  0.104$ & $-0.351 \pm  0.086$ & $-0.266 \pm  0.101$ &  $-0.142 \pm  0.102$ & $-0.184 \pm  0.106$ & $-0.116 \pm  0.106$ & $-0.104 \pm  0.108$ \\
      \hline
%     [6.0, 8.0]   & $-0.413 \pm  0.048$ & $-0.312 \pm  0.062$ & $-0.440 \pm  0.048$ & $-0.395 \pm  0.055$ & $-0.303 \pm  0.060$  & $-0.338 \pm  0.061$ & $-0.286 \pm  0.064$ & $-0.281 \pm  0.065$ \\
%      \hline
     [1.1, 6.0]   & $-0.098 \pm  0.128$ & $0.072 \pm  0.119$ & $-0.109 \pm  0.141$ & $-0.019 \pm  0.135$ & $0.076 \pm  0.115$  & $0.052 \pm  0.125$ & $0.103 \pm  0.114$ & $0.118 \pm  0.113$ \\
      \hline
     [0.045, 6.0]   & $-0.003 \pm  0.085$ & $0.098 \pm  0.063$ & $-0.006 \pm  0.084$ & $0.044 \pm  0.074$ & $0.100 \pm  0.062$ & $0.085 \pm 0.066$ & $0.116 \pm  0.058$ & $0.124 \pm  0.056$ \\
      \hline
 \hline
 \end{tabular}}
 \caption{The binned average central values and the corresponding $1\sigma$ uncertainties for $\langle P_2 \rangle$ 
 of $B_s \to f_{2}'(1525)(\to K^+\,K^-)\,\mu^+ \,\mu^-$ decays in the SM and in several NP cases.}
 \label{tab_npp2}
 \end{table}
 
% %%%%%%%%%%%%%%%%%%%%%%%%%%%%%%%%%%%%%%%%%%%%%TABLE -P4
% 
 \begin{table}[ht!]
 \centering
 \resizebox{\columnwidth}{!}{
\begin{tabular}{|c|c|c|c|c|c|c|c|c|}
  \hline
  \hline
  \multirow{2}{*}{$q^{2}$ bins (GeV$^2$)}
  & \multicolumn{8}{|c|}{$\langle P^{\prime}_4 \rangle$} \\
      \cline{2-9}
      & SM & $C_{9}^{NP}$ & $C_{10}^{NP}$ & $C_{9}^{NP}$=$-C_{10}^{NP}$ & $C_{9}^{NP}$=$-C_{9}^{\prime}$ & $(C_{9}^{NP}$, $C_{10}^{NP}$) & $(C_{9}^{NP}$, $C_{9}^{\prime}$) & $(C_{9}^{NP}$, $C_{10}^{\prime}$) \\
     \hline
     \hline
     [0.10, 0.98] & $-0.457 \pm  0.089$ & $-0.351 \pm  0.069$ & $-0.552 \pm  0.095$ & $-0.475 \pm  0.083$ & $-0.199 \pm  0.050$ & $-0.394 \pm 0.073$ & $-0.239 \pm  0.054$  & $-0.372 \pm  0.065$ \\
      \hline
     [1.1, 2.5]   & $0.251 \pm  0.234$ & $0.256 \pm  0.186$ & $0.090 \pm  0.256$ & $0.140 \pm  0.221$ & $0.351 \pm  0.152$  & $0.207 \pm  0.195$ & $0.318 \pm  0.158$  & $0.166 \pm  0.169$ \\
      \hline
     [2.5, 4.0]   & $0.811 \pm  0.197$ & $0.760 \pm  0.181$ & $0.753 \pm  0.230$ & $0.737 \pm  0.211$ & $0.761 \pm  0.159$ & $0.743 \pm  0.191$ & $0.756 \pm  0.164$ & $0.643 \pm  0.182$  \\
       \hline
     [4.0, 6.0]   & $0.995 \pm  0.156$ & $0.968 \pm  0.150$ & $0.983 \pm  0.170$ & $0.970 \pm  0.163$ & $0.918 \pm  0.150$ & $0.964 \pm  0.155$ & $0.935 \pm  0.149$ & $0.867 \pm  0.161$ \\
      \hline
%     [6.0, 8.0]   & $1.067 \pm  0.138$ & $1.056 \pm  0.135$ & $1.065 \pm  0.145$ & $1.060 \pm  0.142$ &  $0.982 \pm  0.145$  & $1.056 \pm  0.137$ & $1.011 \pm  0.141$ & $0.973 \pm  0.148$  \\
%      \hline
     [1.1, 6.0]   & $0.739 \pm  0.188$ & $0.702 \pm  0.173$ & $0.668 \pm  0.215$ & $0.666 \pm  0.199$ & $0.707 \pm  0.152$ & $0.681 \pm  0.182$ &  $0.703 \pm  0.157$ &  $0.597 \pm  0.173$ \\
      \hline
     [0.045, 6.0]   & $0.405 \pm  0.183$ & $0.395 \pm  0.161$ & $0.297 \pm  0.198$ & $0.320 \pm  0.181$ & $0.439 \pm  0.135$ & $0.361 \pm  0.167$ & $0.423 \pm  0.141$  & $0.315 \pm  0.152$ \\
      \hline
 \hline
 \end{tabular}}
 \caption{The binned average central values and the corresponding $1\sigma$ uncertainties for $\langle P^{\prime}_4 \rangle$ 
 of $B_s \to f_{2}'(1525)(\to K^+\,K^-)\,\mu^+ \,\mu^-$ decays in the SM and in several NP cases.}
 \label{tab_npp4}
 \end{table}
% 
% %%%%%%%%%%%%%%%%%%%%%%%%%%%%%%%%%%%%%%%%%%%%%TABLE -P5
% 
 \begin{table}[ht!]
 \centering
 \resizebox{\columnwidth}{!}{
\begin{tabular}{|c|c|c|c|c|c|c|c|c|}
  \hline
  \hline
  \multirow{2}{*}{$q^{2}$ bins (GeV$^2$)}
  & \multicolumn{8}{|c|}{$\langle P^{\prime}_5 \rangle$} \\
      \cline{2-9}
      & SM & $C_{9}^{NP}$ & $C_{10}^{NP}$ & $C_{9}^{NP}$=$-C_{10}^{NP}$ & $C_{9}^{NP}$=$-C_{9}^{\prime}$ & $(C_{9}^{NP}$, $C_{10}^{NP}$) & $(C_{9}^{NP}$, $C_{9}^{\prime}$) & $(C_{9}^{NP}$, $C_{10}^{\prime}$) \\
     \hline
     \hline
     [0.10, 0.98] & $0.593 \pm  0.114$ & $0.709 \pm  0.116$ & $0.545 \pm  0.104$ & $0.620 \pm  0.109$ & $0.786 \pm  0.132$ & $0.684 \pm  0.114$ & $0.777 \pm  0.126$ & $0.700 \pm  0.115$ \\
      \hline
     [1.1, 2.5]   & $-0.079 \pm  0.267$ & $0.172 \pm  0.230$ & $-0.086 \pm  0.272$ & $0.038 \pm  0.253$ & $0.199 \pm  0.249$ & $0.141 \pm  0.236$ & $0.233 \pm  0.235$ & $0.163 \pm  0.226$ \\
      \hline
     [2.5, 4.0]   & $-0.620 \pm  0.236$ & $-0.354 \pm  0.23$ & $-0.670 \pm  0.255$ & $-0.534 \pm  0.253$ & $-0.366 \pm  0.247$ & $-0.403 \pm  0.243$ & $-0.310 \pm  0.242$ & $-0.358 \pm  0.233$ \\
      \hline
     [4.0, 6.0]   & $-0.797 \pm  0.187$ & $-0.615 \pm  0.187$ & $-0.842 \pm  0.196$ & $-0.763 \pm  0.197$ & $-0.655 \pm  0.189$  & $-0.662 \pm  0.193$ & $-0.599 \pm  0.190$ & $-0.631 \pm  0.183$ \\
      \hline
%     [6.0, 8.0]   & $-0.840 \pm  0.169$ & $-0.726 \pm  0.165$ & $-0.864 \pm  0.174$ & $-0.828 \pm  0.174$ & $-0.789 \pm  0.160$ & $-0.763 \pm  0.169$ & $-0.736 \pm  0.162$ & $-0.757 \pm  0.158$ \\
%      \hline
     [1.1, 6.0]   & $-0.552 \pm  0.220$ & $-0.316 \pm  0.219$ & $-0.580 \pm  0.232$ & $-0.469 \pm  0.233$ & $-0.329 \pm  0.233$  & $-0.358 \pm  0.226$ & $-0.277 \pm  0.227$ & $-0.322 \pm  0.219$  \\
      \hline
     [0.045, 6.0]   & $-0.238 \pm  0.204$ & $-0.036 \pm  0.194$ & $-0.237 \pm  0.204$ & $-0.148 \pm  0.203$ & $-0.029 \pm  0.211$ & $-0.064 \pm  0.198$ & $0.008 \pm  0.202$ & $-0.041 \pm  0.195$ \\      
      \hline
 \hline
 \end{tabular}}
 \caption{The binned average central values and the corresponding $1\sigma$ uncertainties for $\langle P^{\prime}_5 \rangle$ 
 of $B_s \to f_{2}'(1525)(\to K^+\,K^-)\,\mu^+ \,\mu^-$ decays in the SM and in several NP cases.}
 \label{tab_npp5}
 \end{table}
% 
% %%%%%%%%%%%%%%%%%%%%%%%%%%%%%%%%%%%%%%%%%%%%%TABLE -R
% 
 \begin{table}[ht!]
 \centering
 \resizebox{\columnwidth}{!}{
\begin{tabular}{|c|c|c|c|c|c|c|c|c|}
  \hline
  \hline
  \multirow{2}{*}{$q^{2}$ bins (GeV$^2$)}
  & \multicolumn{8}{|c|}{$\langle R_{f_{2}^{\prime}} \rangle$} \\
  \cline{2-9}
      & SM & $C_{9}^{NP}$ & $C_{10}^{NP}$ & $C_{9}^{NP}$=$-C_{10}^{NP}$ & $C_{9}^{NP}$=$-C_{9}^{\prime}$ & $(C_{9}^{NP}$, $C_{10}^{NP}$) & $(C_{9}^{NP}$, $C_{9}^{\prime}$) & $(C_{9}^{NP}$, $C_{10}^{\prime}$) \\
     \hline
     \hline
     [0.10, 0.98] & $0.979 \pm  0.005$ &  $0.910 \pm  0.025$  & $0.893 \pm  0.015$ & $0.882 \pm  0.022$ &  $0.839 \pm  0.044$ & $0.890 \pm  0.027$ & $0.857 \pm  0.040$ & $0.846 \pm  0.040$  \\
      \hline
     [1.1, 2.5]   & $0.994 \pm  0.005$ & $0.854 \pm  0.022$ & $0.826 \pm  0.010$ & $0.803 \pm  0.012$ & $0.735 \pm  0.037$ & $0.815 \pm  0.019$ & $0.761 \pm  0.035$  & $0.737 \pm  0.030$ \\
      \hline
     [2.5, 4.0]   & $0.995 \pm  0.005$ & $0.829 \pm  0.014$ & $0.810 \pm  0.012$ & $0.779 \pm  0.006$ & $0.713 \pm  0.028$ & $0.788 \pm  0.010$ & $0.734 \pm  0.024$ & $0.707 \pm  0.019$ \\
      \hline
     [4.0, 6.0]   & $0.996 \pm  0.003$ & $0.816 \pm  0.009$ & $0.810 \pm  0.011$ & $0.773 \pm  0.005$ & $0.708 \pm  0.025$ & $0.776 \pm  0.005$ & $0.723 \pm  0.019$ & $0.697 \pm  0.015$ \\
      \hline
%     [6.0, 8.0]   & $0.997 \pm  0.003$ & $0.806 \pm  0.006$ & $0.818 \pm  0.009$ & $0.774 \pm  0.005$ & $0.705 \pm  0.025$ & $0.770 \pm  0.003$ & $0.715 \pm  0.017$ & $0.693 \pm  0.014$ \\
%      \hline
     [1.1, 6.0]   & $0.995 \pm  0.002$ & $0.831 \pm  0.014$ & $0.814 \pm  0.010$ & $0.783 \pm  0.005$ & $0.717 \pm  0.028$ & $0.790 \pm  0.010$ & $0.737 \pm  0.024$ & $0.711 \pm  0.020$ \\
      \hline
     [0.045, 6.0]   & $0.976 \pm  0.005$ & $0.842 \pm  0.016$ & $0.827 \pm  0.007$ & $0.802 \pm  0.010$ & $0.744 \pm  0.033$ & $0.809 \pm  0.015$ & $0.762 \pm  0.029$  & $0.742 \pm  0.026$ \\       
      \hline
 \hline
 \end{tabular}}
 \caption{The binned average central values and the corresponding $1\sigma$ uncertainties for the ratio of branching ratio 
 $\langle R_{f_{2}^{\prime}} \rangle$ of $B_s \to f_{2}'(1525)(\to K^+\,K^-)\,\mu^+ \,\mu^-$ decays in the SM and in several NP cases.}
 \label{tab_npr}
 \end{table}

% 
% %%%%%%%%%%%%%%%%%%%%%%%%%%%%%%%%%%%%%%%%%%%%%TABLE -Q1 
% 
 
 \begin{table}[ht!]
 \centering
 \resizebox{\columnwidth}{!}{
\begin{tabular}{|c|c|c|c|c|c|c|c|c|}
  \hline
  \hline
  \multirow{2}{*}{$q^{2}$ bins (GeV$^2$)}
  & \multicolumn{8}{|c|}{$\langle Q_1 \rangle$} \\
      \cline{2-9}
      & SM & $C_{9}^{NP}$ & $C_{10}^{NP}$ & $C_{9}^{NP}$=$-C_{10}^{NP}$ & $C_{9}^{NP}$=$-C_{9}^{\prime}$ & $(C_{9}^{NP}$, $C_{10}^{NP}$) & $(C_{9}^{NP}$, $C_{9}^{\prime}$) & $(C_{9}^{NP}$, $C_{10}^{\prime}$) \\
     \hline
     \hline
     [0.10, 0.98] & $0.000 \pm  0.003$ & $-0.002 \pm  0.008$ & $0.001 \pm  0.008$ & $-0.000 \pm  0.001$ &  $-0.069 \pm  0.016$ & $-0.002 \pm  0.005$ & $-0.043 \pm  0.013$ & $0.004 \pm  0.011$ \\
     \hline
     [1.1, 2.5]   & $-0.000 \pm  0.001$ & $0.002 \pm  0.028$ & $0.010 \pm  0.051$ & $0.007 \pm  0.045$ & $-0.181 \pm  0.032$ & $0.004 \pm  0.036$ & $-0.113 \pm  0.032$ & $0.095 \pm  0.046$  \\
      \hline
     [2.5, 4.0]   & $-0.000 \pm  0.000$ & $0.009 \pm  0.060$ & $0.003 \pm  0.017$  & $0.007 \pm  0.039$ & $-0.055 \pm  0.079$ & $0.010 \pm  0.059$ & $-0.037 \pm  0.075$ & $0.181 \pm  0.076$ \\
      \hline
     [4.0, 6.0]   & $-0.000 \pm  0.000$ & $0.006 \pm  0.035$ & $-0.002 \pm  0.019$  & $0.002 \pm  0.009$ & $0.069 \pm  0.054$ & $0.005 \pm  0.028$ & $0.039 \pm  0.047$ & $0.183 \pm  0.046$ \\
      \hline
%     [6.0, 8.0]   & $-0.000 \pm  0.000$ & $0.002 \pm  0.016$ & $-0.002 \pm  0.016$ & $-0.000 \pm  0.004$ & $0.137 \pm  0.032$  & $0.001 \pm  0.010$ & $0.080 \pm  0.024$ & $0.163 \pm  0.024$ \\
%      \hline
     [1.1, 6.0]   & $-0.001 \pm  0.002$ & $0.010 \pm  0.048$ & $0.003 \pm  0.014$ & $0.008 \pm  0.035$ & $-0.021 \pm  0.066$ & $0.011 \pm  0.049$ & $-0.014 \pm  0.062$  & $0.163 \pm  0.062$\\
       \hline
     [0.045, 6.0]   & $-0.014 \pm  0.040$ & $-0.007 \pm  0.009$ & $-0.005 \pm  0.016$ & $-0.004 \pm  0.006$ & $-0.046 \pm  0.032$ & $-0.005 \pm  0.003$ & $-0.034 \pm  0.019$ & $0.084 \pm  0.021$ \\
      \hline
 \hline
 \end{tabular}}
 \caption{The binned average central values and the corresponding $1\sigma$ uncertainties for $\langle Q_1 \rangle$ 
 of $B_s \to f_{2}'(1525)(\to K^+\,K^-)\,\mu^+ \,\mu^-$ decays in the SM and in several NP cases.}
 \label{tab_npq1}
 \end{table}

 %
 % %%%%%%%%%%%%%%%%%%%%%%%%%%%%%%%%%%%%%%%%%%%%%TABLE -Q2
 %
 
 \begin{table}[ht!]
 \centering
 \resizebox{\columnwidth}{!}{
\begin{tabular}{|c|c|c|c|c|c|c|c|c|}
  \hline
  \hline
  \multirow{2}{*}{$q^{2}$ bins (GeV$^2$)}
  & \multicolumn{8}{|c|}{$\langle Q_2 \rangle$} \\
      \cline{2-9}
      & SM & $C_{9}^{NP}$ & $C_{10}^{NP}$ & $C_{9}^{NP}$=$-C_{10}^{NP}$ & $C_{9}^{NP}$=$-C_{9}^{\prime}$ & $(C_{9}^{NP}$, $C_{10}^{NP}$) & $(C_{9}^{NP}$, $C_{9}^{\prime}$) & $(C_{9}^{NP}$, $C_{10}^{\prime}$) \\
     \hline
     \hline
     [0.10, 0.98] & $0.026 \pm  0.005$ & $0.023 \pm  0.004$ & $0.001 \pm  0.001$ & $0.008 \pm  0.002$ & $0.022 \pm  0.005$ & $0.016 \pm  0.003$ & $0.022 \pm  0.004$ & $0.022 \pm  0.005$ \\
      \hline
     [1.1, 2.5]   & $0.005 \pm  0.002$ & $0.056 \pm  0.049$  & $0.010 \pm  0.019$ & $0.036 \pm  0.038$ & $0.043 \pm  0.052$ & $0.051 \pm  0.049$ & $0.055 \pm  0.057$ & $0.063 \pm  0.060$ \\
      \hline
     [2.5, 4.0]   & $-0.001 \pm  0.001$ & $0.193 \pm  0.025$ & $-0.008 \pm  0.025$ & $0.098 \pm  0.018$ & $0.194 \pm  0.030$ & $0.175 \pm  0.021$ & $0.225 \pm  0.033$ & $0.243 \pm  0.036$  \\
       \hline
     [4.0, 6.0]   & $-0.001 \pm  0.001$ & $0.160 \pm  0.030$ & $-0.037 \pm  0.007$ & $0.048 \pm  0.021$ & $0.172 \pm  0.031$ & $0.130 \pm  0.030$ & $0.198 \pm  0.034$ &  $0.210 \pm  0.040$ \\
      \hline
%     [6.0, 8.0]   & $-0.001 \pm  0.000$ & $0.101 \pm  0.025$  & $-0.028 \pm  0.008$ & $0.017 \pm  0.010$ & $0.109 \pm  0.035$ & $0.074 \pm  0.022$ & $0.127 \pm  0.034$ & $0.132 \pm  0.039$ \\
%      \hline
     [1.1, 6.0]   & $-0.003 \pm  0.001$ & $0.167 \pm  0.017$ & $-0.014 \pm  0.013$ & $0.076 \pm  0.012$  & $0.171 \pm  0.020$ & $0.147 \pm  0.015$ & $0.198 \pm  0.022$ & $0.213 \pm  0.027$ \\
      \hline
     [0.045, 6.0]   & $-0.007 \pm  0.019$ & $0.095 \pm  0.005$ & $-0.009 \pm  0.017$ & $0.041 \pm  0.008$ & $0.096 \pm  0.007$ & $0.082 \pm  0.004$ & $0.112 \pm  0.010$ & $0.121 \pm  0.014$ \\
      \hline
 \hline
 \end{tabular}}
 \caption{The binned average central values and the corresponding $1\sigma$ uncertainties for $\langle Q_2 \rangle$ 
 of $B_s \to f_{2}'(1525)(\to K^+\,K^-)\,\mu^+ \,\mu^-$ decays in the SM and in several NP cases.}
 \label{tab_npq2}
 \end{table}

 %
 % %%%%%%%%%%%%%%%%%%%%%%%%%%%%%%%%%%%%%%%%%%%%%TABLE -Q4'
 %
 
 \begin{table}[ht!]
 \centering
 \resizebox{\columnwidth}{!}{
\begin{tabular}{|c|c|c|c|c|c|c|c|c|}
  \hline
  \hline
  \multirow{2}{*}{$q^{2}$ bins (GeV$^2$)}
  & \multicolumn{8}{|c|}{$\langle Q^{\prime}_4 \rangle$} \\
      \cline{2-9}
      & SM & $C_{9}^{NP}$ & $C_{10}^{NP}$ & $C_{9}^{NP}$=$-C_{10}^{NP}$ & $C_{9}^{NP}$=$-C_{9}^{\prime}$ & $(C_{9}^{NP}$, $C_{10}^{NP}$) & $(C_{9}^{NP}$, $C_{9}^{\prime}$) & $(C_{9}^{NP}$, $C_{10}^{\prime}$) \\
     \hline
     \hline
     [0.10, 0.98] & $0.011 \pm  0.005$ & $0.116 \pm  0.022$ & $-0.084 \pm  0.012$ & $-0.007 \pm  0.005$ & $0.268 \pm  0.052$ & $0.073 \pm  0.016$ & $0.228 \pm  0.045$ & $0.096 \pm  0.026$ \\
      \hline
     [1.1, 2.5]   & $0.002 \pm  0.001$ & $0.007 \pm  0.050$ & $-0.159 \pm  0.029$ & $-0.109 \pm  0.017$ & $0.102 \pm  0.090$ & $-0.042 \pm  0.041$ & $0.070 \pm  0.082$ & $-0.082 \pm  0.068$ \\
      \hline
     [2.5, 4.0]   & $0.000 \pm  0.000$ & $-0.050 \pm  0.024$ & $-0.058 \pm  0.039$ & $-0.073 \pm  0.025$ & $-0.049 \pm  0.045$  & $-0.068 \pm  0.021$ & $-0.054 \pm  0.039$ & $-0.167 \pm  0.028$  \\
      \hline
     [4.0, 6.0]   & $0.000 \pm  0.000$ & $-0.027 \pm  0.014$ & $-0.012 \pm  0.017$ & $-0.025 \pm  0.014$ & $-0.077 \pm  0.016$ & $-0.031 \pm  0.014$ & $-0.060 \pm  0.014$ &  $-0.128 \pm  0.021$ \\
      \hline
%     [6.0, 8.0]   & $0.000 \pm  0.000$ & $-0.011 \pm  0.007$ & $-0.002 \pm  0.009$ & $-0.008 \pm  0.006$ & $-0.085 \pm  0.013$ & $-0.011 \pm  0.006$ & $-0.056 \pm  0.008$ & $-0.094 \pm  0.016$ \\
%      \hline
     [1.1, 6.0]   & $0.004 \pm  0.001$ & $-0.033 \pm  0.020$ & $-0.067 \pm  0.032$ & $-0.069 \pm  0.019$ & $-0.028 \pm  0.044$ & $-0.054 \pm  0.016$ & $-0.032 \pm  0.037$ & $-0.138 \pm  0.023$ \\
      \hline
     [0.045, 6.0]   & $0.098 \pm  0.015$ & $0.088 \pm  0.016$ & $-0.010 \pm  0.027$ & $0.013 \pm  0.010$ & $0.132 \pm  0.047$ & $0.054 \pm  0.010$ & $0.117 \pm  0.039$ & $0.008 \pm  0.023$  \\      
      \hline
 \hline
 \end{tabular}}
 \caption{The binned average central values and the corresponding $1\sigma$ uncertainties for $\langle Q^{\prime}_4 \rangle$ 
 of $B_s \to f_{2}'(1525)(\to K^+\,K^-)\,\mu^+ \,\mu^-$ decays in the SM and in several NP cases.}
 \label{tab_npq4}
 \end{table}

 %
 % %%%%%%%%%%%%%%%%%%%%%%%%%%%%%%%%%%%%%%%%%%%%%TABLE -Q5'
 %
 
 \begin{table}[ht!]
 \centering
 \resizebox{\columnwidth}{!}{
\begin{tabular}{|c|c|c|c|c|c|c|c|c|}
  \hline
  \hline
  \multirow{2}{*}{$q^{2}$ bins (GeV$^2$)}
  & \multicolumn{8}{|c|}{$\langle Q^{\prime}_5 \rangle$} \\
      \cline{2-9}
      & SM & $C_{9}^{NP}$ & $C_{10}^{NP}$ & $C_{9}^{NP}$=$-C_{10}^{NP}$ & $C_{9}^{NP}$=$-C_{9}^{\prime}$ & $(C_{9}^{NP}$, $C_{10}^{NP}$) & $(C_{9}^{NP}$, $C_{9}^{\prime}$) & $(C_{9}^{NP}$, $C_{10}^{\prime}$) \\
     \hline
     \hline
     [0.10, 0.98] & $0.039 \pm  0.011$ & $0.155 \pm  0.016$  & $-0.009 \pm  0.004$  & $0.066 \pm  0.007$ & $0.232 \pm  0.033$  & $0.129 \pm  0.013$ & $0.223 \pm  0.027$ & $0.145 \pm  0.014$ \\
      \hline
     [1.1, 2.5]   & $-0.002 \pm  0.004$ & $0.248 \pm  0.040$ & $-0.009 \pm  0.012$ & $0.115 \pm  0.016$ & $0.276 \pm  0.031$ & $0.217 \pm  0.033$ & $0.309 \pm  0.041$ & $0.240 \pm  0.042$ \\
      \hline
     [2.5, 4.0]   & $-0.004 \pm  0.002$ & $0.261 \pm  0.043$ & $-0.054 \pm  0.021$  & $0.082 \pm  0.029$ & $0.249 \pm  0.052$ & $0.212 \pm  0.040$ & $0.306 \pm  0.054$ &  $0.258 \pm  0.052$  \\
      \hline
     [4.0, 6.0]   & $-0.004 \pm  0.001$ & $0.179 \pm  0.041$ & $-0.049 \pm  0.016$ & $0.031 \pm  0.018$ & $0.139 \pm  0.055$  & $0.132 \pm  0.036$ & $0.195 \pm  0.056$ & $0.163 \pm  0.054$ \\
      \hline
%     [6.0, 8.0]   & $-0.003 \pm  0.001$ & $0.112 \pm  0.028$ & $-0.027 \pm  0.012$  & $0.009 \pm  0.009$ & $0.049 \pm  0.041$ & $0.075 \pm  0.022$ & $0.101 \pm  0.041$ & $0.080 \pm  0.038$   \\
%      \hline
     [1.1, 6.0]   & $-0.007 \pm  0.002$ & $0.229 \pm  0.035$ & $-0.035 \pm  0.014$ & $0.076 \pm  0.023$ & $0.217 \pm  0.044$ & $0.187 \pm  0.032$ & $0.269 \pm  0.045$  & $0.224 \pm  0.043$ \\
      \hline
     [0.045, 6.0]   & $-0.071 \pm  0.021$ & $0.131 \pm  0.015$ & $-0.070 \pm  0.020$ & $0.019 \pm  0.018$ & $0.139 \pm  0.029$ & $0.103 \pm  0.016$ & $0.175 \pm  0.023$ & $0.126 \pm  0.017$ \\      
      \hline
 \hline
 \end{tabular}}
 \caption{The binned average central values and the corresponding $1\sigma$ uncertainties for $\langle Q^{\prime}_5 \rangle$ 
 of $B_s \to f_{2}'(1525)(\to K^+\,K^-)\,\mu^+ \,\mu^-$ decays in the SM and in several NP cases.}
 \label{tab_npq5}
 \end{table}

 %
 % %%%%%%%%%%%%%%%%%%%%%%%%%%%%%%%%%%%%%%%%%%%%%TABLE -QAFB
 %

 \begin{table}[ht!]
 \centering
 \resizebox{\columnwidth}{!}{
\begin{tabular}{|c|c|c|c|c|c|c|c|c|}
  \hline
  \hline
  \multirow{2}{*}{$q^{2}$ bins (GeV$^2$)}
  & \multicolumn{8}{|c|}{$\langle Q_{A_{FB}} \rangle$} \\
      \cline{2-9}
      & SM & $C_{9}^{NP}$ & $C_{10}^{NP}$ & $C_{9}^{NP}$=$-C_{10}^{NP}$ & $C_{9}^{NP}$=$-C_{9}^{\prime}$ & $(C_{9}^{NP}$, $C_{10}^{NP}$) & $(C_{9}^{NP}$, $C_{9}^{\prime}$) & $(C_{9}^{NP}$, $C_{10}^{\prime}$) \\
     \hline
     \hline
     [0.10, 0.98] & $ -0.010 \pm  0.002 $ & $ 0.001 \pm  0.003 $ & $ -0.018 \pm  0.004 $ & $ -0.009 \pm  0.004 $  & $ 0.010 \pm  0.005 $ & $ -0.002 \pm  0.004 $ & $ 0.008 \pm 0.005 $  & $ 0.010 \pm  0.005 $  \\
      \hline
     [1.1, 2.5]   & $ -0.002 \pm  0.001 $ & $ 0.043 \pm  0.010 $ &  $ -0.002 \pm  0.001 $ & $ 0.021 \pm  0.006 $ & $ 0.064 \pm  0.013 $ & $ 0.039 \pm  0.010 $ & $ 0.064 \pm  0.014 $ & $ 0.072 \pm  0.017 $  \\
      \hline
     [2.5, 4.0]   & $ 0.000 \pm  0.001 $ & $ 0.055 \pm  0.015 $ & $ -0.000 \pm  0.001 $ & $ 0.025 \pm  0.007 $ & $ 0.063 \pm  0.017 $ & $ 0.048 \pm  0.013 $ & $ 0.072 \pm  0.019 $ & $ 0.079 \pm  0.022 $ \\
     \hline
     [4.0, 6.0]   & $ 0.001 \pm  0.000 $ & $ 0.054 \pm  0.014 $ & $ -0.002 \pm  0.002 $ & $ 0.020 \pm  0.005 $ & $ 0.048 \pm  0.017 $  & $ 0.044 \pm  0.012 $  & $ 0.062 \pm  0.018 $  & $ 0.068 \pm  0.021 $ \\
      \hline
%     [6.0, 8.0]   & $ 0.001 \pm  0.000 $ & $ 0.045 \pm  0.011 $ & $ -0.002 \pm  0.003 $  & $ 0.013 \pm  0.003 $ & $ 0.027 \pm  0.015 $ & $ 0.035 \pm  0.009 $ & $ 0.045 \pm  0.015 $ & $ 0.049 \pm  0.017 $ \\
%     \hline
     [1.1, 6.0]   & $ -0.000 \pm  0.001 $ & $ 0.052 \pm  0.013 $ & $ -0.001 \pm  0.001 $ & $ 0.023 \pm  0.006 $ & $ 0.058 \pm  0.016 $ & $ 0.045 \pm  0.012 $ & $ 0.067 \pm  0.017 $  & $ 0.074 \pm  0.020 $ \\
     \hline
     [0.045, 6.0]   & $ -0.004 \pm  0.001 $  & $ 0.038 \pm  0.008 $ & $ -0.004 \pm  0.001 $ & $ 0.015 \pm  0.003 $ & $ 0.045 \pm  0.009 $ & $ 0.032 \pm  0.007 $ & $ 0.050 \pm 0.010 $ & $ 0.056 \pm  0.012 $ \\     
     \hline
 \hline
\end{tabular}}
 \caption{The binned average central values and the corresponding $1\sigma$ uncertainties for $\langle Q_{A_{FB}} \rangle$ 
 of $B_s \to f_{2}'(1525)(\to K^+\,K^-)\,\mu^+ \,\mu^-$ decays in the SM and in several NP cases.}
 \label{tab_npqafb}
 \end{table}

 %
 % %%%%%%%%%%%%%%%%%%%%%%%%%%%%%%%%%%%%%%%%%%%%%TABLE -QFL
 %
 
 \begin{table}[ht!]
 \centering
 \resizebox{\columnwidth}{!}{
\begin{tabular}{|c|c|c|c|c|c|c|c|c|}
  \hline
  \hline
  \multirow{2}{*}{$q^{2}$ bins (GeV$^2$)}
  & \multicolumn{8}{|c|}{$\langle Q_{F_{L}} \rangle$} \\
      \cline{2-9}
      & SM & $C_{9}^{NP}$ & $C_{10}^{NP}$ & $C_{9}^{NP}$=$-C_{10}^{NP}$ & $C_{9}^{NP}$=$-C_{9}^{\prime}$ & $(C_{9}^{NP}$, $C_{10}^{NP}$) & $(C_{9}^{NP}$, $C_{9}^{\prime}$) & $(C_{9}^{NP}$, $C_{10}^{\prime}$) \\
  \hline
  \hline
      [0.10, 0.98] & $ 0.001 \pm  0.001 $ & $ -0.066 \pm  0.005 $ & $ -0.040 \pm  0.004 $ &  $ -0.063 \pm  0.004 $  & $ -0.116 \pm  0.011 $ & $ -0.073 \pm  0.005 $ & $ -0.107 \pm  0.010 $ & $ -0.117 \pm  0.009 $ \\
      \hline
      [1.1, 2.5]   & $ 0.002 \pm  0.001 $ & $ -0.054 \pm  0.015 $ & $ 0.003 \pm  0.009 $ & $ -0.026 \pm  0.013 $ & $ -0.095 \pm  0.023 $ & $ -0.049 \pm  0.015 $ & $ -0.089 \pm 0.022 $ & $ -0.096 \pm  0.025 $  \\
      \hline
      [2.5, 4.0]   & $ 0.002 \pm  0.001 $  & $ -0.030 \pm  0.010 $ & $ 0.020 \pm  0.005 $ &  $ -0.001 \pm  0.006 $ & $ -0.072 \pm  0.018 $ & $ -0.021 \pm  0.010 $ & $ -0.061 \pm  0.016 $ & $ -0.064 \pm  0.019 $ \\ 
      \hline
      [4.0, 6.0]   & $ 0.002 \pm  0.000 $  & $ -0.016 \pm  0.006 $  & $ 0.020 \pm  0.003 $ & $ 0.007 \pm  0.002 $ & $ -0.062 \pm  0.016 $ & $ -0.008 \pm  0.005 $ & $ -0.046 \pm0.011 $ & $ -0.047 \pm  0.014 $ \\
      \hline
%      [6.0, 8.0]   & $ 0.001 \pm  0.000 $ & $ -0.007 \pm  0.003 $ &  $ 0.015 \pm  0.003 $ & $ 0.008 \pm  0.001 $ & $ -0.002 \pm  0.002 $ &  $ -0.038 \pm  0.010 $ & $ -0.038 \pm0.010 $ & $ -0.038 \pm  0.010 $ \\
%      \hline
      [1.1, 6.0]   & $ 0.002 \pm  0.001 $ & $ -0.030 \pm  0.010 $ & $ 0.015 \pm  0.004 $ & $ -0.005 \pm  0.007 $ & $ -0.075 \pm  0.017 $ & $ -0.023 \pm  0.010 $ & $ -0.063 \pm 0.015 $   & $ -0.066 \pm  0.018 $ \\
      \hline
      [0.045, 6.0]   & $ 0.012 \pm  0.002 $ & $ -0.038 \pm  0.007 $ & $ -0.001 \pm  0.006 $ & $ -0.023 \pm  0.008 $ & $ -0.089 \pm  0.012 $ & $ -0.038 \pm  0.009 $ & $ -0.078 \pm  0.011 $ & $ -0.085 \pm  0.014 $ \\
      \hline
 \hline
\end{tabular}}
 \caption{The binned average central values and the corresponding $1\sigma$ uncertainties for $\langle Q_{F_{L}} \rangle$ 
 of $B_s \to f_{2}'(1525)(\to K^+\,K^-)\,\mu^+ \,\mu^-$ decays in the SM and in several NP cases.}
 \label{tab_npqfl}
 \end{table}

\label{appendix2}
\section{Helicity Amplitudes}

In the process of $B_s \to f_{2}'(1525)(\to K^+\,K^-)\,l^+ \,l^-$ decays, initially the $B_s$ meson decays into an on-shell 
strange meson along with the a pair of leptons. Further, the $f_{2}'(1525)$ decays strongly into $K^+\,K^-$. This multibody
decay evaluated in the helicity framework which uses the metric tensor,
\begin{equation}
 g_{\mu \nu}=- \sum_{\lambda} \epsilon_{\mu} (\lambda)\, \epsilon_{\nu}^* (\lambda)\, + \frac{q_{\mu}q_{\nu}}{q^2}
\end{equation}
where, $\epsilon$ is the polarization vector with the momentum $q$ and with $\lambda$ being the three kinds of polarizations.
Here, the metric tensor $g_{\mu \nu}$ is expressed as a summation of four polarizations where the last term is identified to be
the time like polarization $\epsilon_{\mu} (t)=q^{\mu}/q^2$. 
In SM the production of lepton pair in the final state is due to a $Z$ boson, an off-shell photon or any hadronic meson.
Although, the coupling strenghts of these states may be different but they processes the similar Lorentz structure of type
V-A, V+A or the combination of both. Hence, the decay amplitude for $B_s \to f_{2}'\,l^+ \,l^-$ is written as,
\begin{equation}
 \mathcal{A}(B_s \to f_{2}'\,l^+ \,l^-)=\mathcal{L}^{\mu}(L)\,\mathcal{H}_{\mu}(L)\,+\,\mathcal{L}^{\mu}(R)\,\mathcal{H}_{\mu}(R)
\end{equation}
where $\mathcal{L}^{\mu}(L)=\bar{l}\gamma_{\mu}(1-\gamma_5)l$ and $\mathcal{L}^{\mu}(R)=\bar{l}\gamma_{\mu}(1+\gamma_5)l$ are the 
lepton pair spinor products and similarly, $\mathcal{H}$ includes $B_s \to f_{2}'$. Further, the factarization of the decay amplitude
is obtained as
\begin{equation}
 \mathcal{A}(B_s \to f_{2}'\,l^+ \,l^-)=\mathcal{L}^{\mu}(L)\,\mathcal{H}_{\mu}(L)\,g_{\mu \nu}\,+\,
 \mathcal{L}^{\mu}(R)\,\mathcal{H}_{\mu}(R)\,g_{\mu \nu}=
 -\sum_{\lambda}\mathcal{L}_{L\lambda}\mathcal{H}_{L\lambda}-\sum_{\lambda}\mathcal{L}_{R\lambda}\mathcal{H}_{R\lambda}
\end{equation}
The Lorentz invariant amplitude for the lepton part $\mathcal{L}_{L\lambda}=\mathcal{L}^{\mu}(L) \epsilon_{\mu}(\lambda)$
and $\mathcal{L}_{R\lambda}=\mathcal{L}^{\mu}(R) \epsilon_{\mu}(\lambda)$ and similarly for the hadronic part. The timelike
polarization vanishes at $m_l=0$ for $l=e,\mu$; using the equation of motion, this term is proportional to the lepton mass.
Since the hadronic and leptonic amplitudes are Lorentz invariant, it allows us to evaluate in the different frames. 
Due to many similarities between $K^*$, $K_2^*$ and $f_{2}'$, the differential decay width for $B_s \to f_{2}'\,l^+ \,l^-$
are simply obtained in a comparative manner by multiplying the factors $\frac{\sqrt{\lambda}}{\sqrt{8}m_B m_{f_{2}'}}$ and
$\frac{\sqrt{\lambda}}{\sqrt{6}m_B m_{f_{2}'}}$ respectively with the longitudinal and transverse amplitudes.
This modification is because since we replace the polarization vector $\epsilon$ by $\epsilon_T$ in the definition of $B\to T$ 
form factors. Explicitly, the expressions for the hadronic amplitudes are written as~\cite{Li:2010ra}
\begin{eqnarray}
  H_{L0} &=& N_{f_{2}^{\prime}} \frac{\sqrt{\lambda}}{\sqrt{8}\,m_{B_s}m_{f_{2}^{\prime}}}\frac{1}{2m_{f_{2}^{\prime}}\sqrt{q^2}}
 \bigg\{(C_{9}^{eff} - C_{10}) \bigg[(m_{B_s}^2 - m_{f_{2}^{\prime}}^2 - q^2)(m_{B_s} + m_{f_{2}^{\prime}})A_1 - \frac{\lambda}{m_{B_s} + m_{f_{2}^{\prime}}}A_2\bigg]+ \nonumber \\  
 && 2\,m_b\, C_{7}^{eff}\, \bigg[(m_{B_s}^2 + 3m_{f_{2}^{\prime}}^2 - q^2)T_2 - \frac{\lambda}{m_{B_s}^2 - m_{f_{2}^{\prime}}^2}T_3 \bigg] \bigg\}\,, \nonumber \\
 H_{L\perp} &=& - N_{f_{2}^{\prime}}\sqrt{2} \frac{\sqrt{\lambda}}{\sqrt{6}\,m_{B_s}m_{f_{2}^{\prime}}} \bigg[(C_{9}^{eff} - C_{10})
 \frac{\sqrt{\lambda}}{m_{B_s} + m_{f_{2}^{\prime}}} V + \frac{\sqrt{\lambda}\,2\,m_b\,C_{7}^{eff}}{q^2} T_1 \bigg]\,, \nonumber \\
 H_{L\parallel} &=& N_{f_{2}^{\prime}}\sqrt{2} \frac{\sqrt{\lambda}}{\sqrt{6}\,m_{B_s}m_{f_{2}^{\prime}}} \bigg[(C_{9}^{eff} - C_{10})
 (m_{B_s} + m_{f_{2}^{\prime}}) A_1 + \frac{2\,m_b\,C_{7}^{eff}(m_{B_s}^2 - m_{f_{2}^{\prime}}^2)}{q^2} T_2 \bigg]\,, \nonumber \\
 H_{Lt} &=& N_{f_{2}^{\prime}} \frac{\sqrt{\lambda}}{\sqrt{8}\,m_{B_s}m_{f_{2}^{\prime}}} (C_{9}^{eff} - C_{10}) \frac{\sqrt{\lambda}}{\sqrt{q^2}}A_0\,,
\end{eqnarray}

The spherical harmonic functions for $K^*$ and ${f_{2}^{\prime}}$ are related as
\begin{eqnarray}
 \sqrt{\frac{3}{4\pi}}cos(\theta_K)\equiv C(K^*) \to \sqrt{\frac{5}{16\pi}}(3cos^2 \theta_K -1) \equiv C(f_{2}^{\prime}) \nonumber \\
 \sqrt{\frac{3}{8\pi}}sin(\theta_K)\equiv S(K^*) \to \sqrt{\frac{15}{32\pi}} sin(2\theta_K) \equiv S(f_{2}^{\prime})
\end{eqnarray}
The amplitude equations and the branching fraction formulas are compatible with the Ref.~\cite{Li:2010ra,Hatanaka:2009gb}.

\FloatBarrier
%%%%%%%%%%%%%%%%%%%%%%%%%%%%%%%%%%%%%%%%%%%%%%%%%%%%%%%%%%%%%%%%%%%%%%%%%%%%%%%%%%%%%%%

\end{document}